\documentclass[aps, prl, twocolumn, superscriptaddress]{revtex4-2}
\usepackage{bm, amsmath, amsfonts, amssymb, ascmac, mathtools, braket}
\usepackage{times}
\usepackage{multirow}
\usepackage{graphicx}
\usepackage{float, color, xcolor}
\usepackage{physics}

\usepackage{bbm}
\usepackage{tabularx}
\usepackage{enumerate}
\usepackage{comment}
\usepackage{makecell}

\usepackage[whole]{bxcjkjatype} 

\usepackage[
colorlinks=true, 
linkcolor={red!80!black},
citecolor={blue!70!black},
urlcolor={blue!80!black},
pdfstartview=FitV,
pdftitle={},
pdfauthor={},
pdfsubject={},
pdfkeywords={},
pdfpagemode=None,
bookmarksopen=true
]{hyperref}

\newcommand{\M}{U}
\newcommand{\nsg}[1]{{\textcolor{magenta}{#1}}}

\newcommand{\FP}{0}

\begin{document}

\title{Nonsymmorphic Topological Phases of Non-Hermitian Systems}

\author{Daichi Nakamura}
\email{daichi.nakamura@issp.u-tokyo.ac.jp}
\affiliation{Institute for Solid State Physics, University of Tokyo, Kashiwa, Chiba 277-8581, Japan}

\author{Yutaro Tanaka}
\email{yutaro.tanaka.ay@riken.jp}
\affiliation{RIKEN Center for Emergent Matter Science, Wako, Saitama, 351-0198, Japan}

\author{Ken Shiozaki}
\email{ken.shiozaki@yukawa.kyoto-u.ac.jp}
\affiliation{Center for Gravitational Physics and Quantum Information, Yukawa Institute for Theoretical Physics, Kyoto University, Kyoto 606-8502, Japan}

\author{Kohei Kawabata}
\email{kawabata@issp.u-tokyo.ac.jp}
\affiliation{Institute for Solid State Physics, University of Tokyo, Kashiwa, Chiba 277-8581, Japan}

\date{\today}

\begin{abstract}
Non-Hermiticity appears ubiquitously in various open classical and quantum systems and enriches classification of topological phases.
However, the role of nonsymmorphic symmetry, crystalline symmetry accompanying fractional lattice translations, has remained largely unexplored.
Here, we systematically classify non-Hermitian topological crystalline phases protected by nonsymmorphic symmetry and reveal unique phases that have no counterparts in either Hermitian topological crystalline phases or non-Hermitian topological phases protected solely by internal symmetry.
Specifically, we elucidate the $\mathbb{Z}_2$ and $\mathbb{Z}_4$ non-Hermitian topological phases and their associated anomalous boundary states characterized by distinctive complex-valued energy dispersions.
\end{abstract}

\maketitle

Topological insulators and superconductors play a pivotal role in the understanding of phases of matter~\cite{HK-review, QZ-review}.
Nontrivial topology of bulk wave functions results in the emergence of anomalous boundary states, influencing diverse condensed matter phenomena. 
A rich variety of topological materials appear depending on symmetry classes and spatial dimensions~\cite{CTSR-review}.
Beyond the internal symmetry classification, spatial symmetry also protects distinctive classes of topological phases---topological crystalline insulators~\cite{Fu-11, *Hsieh-12, Chiu-13, Morimoto-13, Shiozaki-PRB-2014, Benalcazar-17, Po-17, Bradlyn-17, Schindler-18}.
Among them, topological matter respecting nonsymmorphic symmetry constitutes a remarkable subclass~\cite{Parameswaran-13, Liu-14, Fang-Fu-15, Shiozaki-PRB-2015, Young-15, Watanabe-15, Dong-16, Lu-16, Shiozaki-PRB-2016, Wang-16, Chang-17}.
Unlike other symmetry, nonsymmorphic symmetry incorporates translation of the spatial origin by a fraction of the lattice period.
Even in the absence of time-reversal symmetry, nonsymmorphic symmetry stabilizes helical edge states spectrally detachable from the bulk bands~\cite{Fang-Fu-15, Shiozaki-PRB-2015}.
It also enables the $\mathbb{Z}_4$ topological classification and supports exotic surface states of hourglass fermions~\cite{Shiozaki-PRB-2016, Wang-16}.

Furthermore, topological phases of non-Hermitian systems have attracted widespread attention in both theory~\cite{Rudner-09, Sato-11, *Esaki-11, Hu-11, Schomerus-13, Longhi-15, Lee-16, Leykam-17, Xu-17, Shen-18, *Kozii-17, MartinezAlvarez-18, Gong-18, YW-18-SSH, *YSW-18-Chern, Kunst-18, McDonald-18, Lee-Thomale-19, Liu-19, Lee-Li-Gong-19, KSUS-19, ZL-19, Herviou-19, Zirnstein-19, Borgnia-19, Yokomizo-19, Zhang-20, OKSS-20, *KOS-20, Yi-Yang-20, Bessho-21, Denner-21, *Denner-23JPhysMater, KSR-21, Zhang-22, Sun-21, Nakamura-24, Wang-24, Nakai-24, Nakamura-23} and experiments~\cite{Poli-15, Zeuner-15, Zhen-15, *Zhou-18, Weimann-17, Xiao-17, St-Jean-17, Bahari-17, Harari-18, *Bandres-18, Zhao-19, Brandenbourger-19-skin-exp, *Ghatak-19-skin-exp, Helbig-19-skin-exp, *Hofmann-19-skin-exp, Xiao-19-skin-exp, Weidemann-20-skin-exp, Palacios-21, Zhao-25, Shen-25}.
Non-Hermiticity naturally arises from coupling with external environments and yields a wide range of phenomena and functionalities unique to open classical and quantum systems~\cite{Konotop-review, Christodoulides-review}.
A key characteristic of non-Hermitian topology lies in the unique gap structure of complex energy, known as a point gap~\cite{Gong-18, KSUS-19}.
A point gap is defined to be open when the spectrum does not intersect a reference point in the complex energy plane.
The bulk-boundary correspondence of point-gap topology manifests itself as anomalous boundary phenomena intrinsic to non-Hermitian systems, such as skin effects~\cite{Lee-16, YW-18-SSH, Kunst-18, Lee-Thomale-19, Yokomizo-19, Zhang-20, OKSS-20} and exceptional topological insulators~\cite{Bessho-21, Denner-21, KSR-21, Nakamura-24}.
More recently, beyond the regime of internal symmetry, topological crystalline phases of non-Hermitian systems have been investigated~\cite{Yoshida-20, Okugawa-20, KSS-20, Okugawa-21, Vecsei-21, Shiozaki-21, Schindler-23, Tanaka-24NSG, Ishikawa-24, YifanWang-24, Tanaka-24, WangBenalcazar-24}.
Nevertheless, the role of nonsymmorphic symmetry on non-Hermitian topological phases has remained largely unclear.
It has yet to be fully elucidated how the interplay of non-Hermiticity and nonsymmorphic symmetry enriches the topological classification and associated boundary phenomena.

In this Letter, we generalize nonsymmorphic symmetry to non-Hermitian systems and develop systematic classification of point-gap topological phases protected by nonsymmorphic symmetry, some of which are summarized in Table~\ref{tab:NSGmain} (see Ref.~\cite{supplement} for the entire classification).
Based on this classification, we uncover unconventional non-Hermitian topological crystalline phases that lack analogs in both Hermitian ones and non-Hermitian ones governed solely by internal symmetry.
Specifically, we clarify the $\mathbb{Z}_2$ and $\mathbb{Z}_4$ non-Hermitian topological phases and their associated anomalous boundary states with distinctive complex-valued energy spectra.
Our work broadens the landscape of topological materials in open systems.

\begin{table}[b]
    \caption{Classification of nonsymmorphic-symmetric point-gap topology for classes A, AII$^{\dag}$ (time-reversal symmetry$^{\dag}$), and AIII (chiral symmetry) with $d_{\parallel} \equiv 0, 1$ (mod $2$). 
    ${\cal \M}$ and $\tilde{\cal \M}$ denote nonsymmorphic symmetry and pseudo-nonsymmorphic symmetry (or equivalently, nonsymmorphic symmetry$^{\dag}$), respectively.
    The subscript of ${\cal \M}_\pm$ or $\tilde{\cal \M}_\pm$ specifies the commutation ($+$) or anticommutation ($-$) relation to time-reversal symmetry$^{\dag}$ or chiral symmetry. 
    The symmetry classes in parentheses represent those for Hermitized Hamiltonians $\sf H$, where $\sf U$ ($\bar{\sf U}$) represents nonsymmorphic symmetry (antisymmetry). 
    The first subscript of ${\sf U}_{\pm\pm}$ specifies the relation to time-reversal symmetry and the second one to particle-hole symmetry.
    The topological indices highlighted by $^{*}$ and color indicate classification intrinsic to nonsymmorphic symmetry.} 
        \label{tab:NSGmain}
    \centering
     \begin{tabular}{ccccc} \hline \hline
     ~~$d_\parallel$~~& ~~Symmetry class~~ & ~~$d=1$~~ & ~~$d=2$~~ & ~~$d=3$~~ \\ 
    \hline
     $0$ & \makecell{A + ${\cal \M}$ (AIII + ${\sf \M}_+$) \\ A + $\tilde{\cal \M}$ (AIII + ${\sf \M}_-$)} & \makecell{$\mathbb{Z}$ \\ $0$} & \makecell{$0$ \\ \nsg{$\mathbb{Z}_2^{*}$}} & \makecell{$\mathbb{Z}$ \\ $0$} \\
    \hline
     $0$ & \makecell{AII$^{\dag}$ + ${\cal \M}_+$ (DIII + ${\sf \M}_{++}$) \\ AII$^{\dag}$ + ${\cal \M}_-$ (DIII + ${\sf \M}_{--}$) \\ AII$^{\dag}$ + $\tilde{\cal \M}_+$ (DIII + ${\sf \M}_{+-}$) \\ AII$^{\dag}$ + $\tilde{\cal \M}_-$ (DIII + ${\sf \M}_{-+}$)} & \makecell{$\mathbb{Z}_2$ \\ $\mathbb{Z}_2$ \\ $\mathbb{Z}_2$ \\ $\mathbb{Z}_2$} & \makecell{~~$\mathbb{Z}_2 \oplus \nsg{\mathbb{Z}_2^{*}}$~~ \\ ~~$\mathbb{Z}_2 \oplus \nsg{\mathbb{Z}_2^{*}}$~~ \\ \nsg{$\mathbb{Z}_4^{*}$} \\ \nsg{$\mathbb{Z}_4^{*}$}} & \makecell{~~$\mathbb{Z} \oplus \nsg{\mathbb{Z}_2^{*}}$~~ \\ ~~$\mathbb{Z} \oplus \nsg{\mathbb{Z}_2^{*}}$~~ \\ \nsg{$\mathbb{Z}_2^{*}$} \\ \nsg{$\mathbb{Z}_2^{*}$}} \\
    \hline 
     $1$ & \makecell{AIII + ${\cal \M}_+$ (A + ${\mathsf \M}$) \\ AIII + ${\cal \M}_-$ (A + $\bar{\mathsf \M}$) \\ AIII + $\tilde{\cal \M}_+$ (A + $\bar{\mathsf \M}$) \\ AIII + $\tilde{\cal \M}_-$ (A + ${\mathsf \M}$)} & \makecell{\nsg{$\mathbb{Z}_2^{*}$} \\ $0$ \\ $0$ \\ \nsg{$\mathbb{Z}_2^{*}$}} & \makecell{$0$ \\ $\mathbb{Z}$ \\ $\mathbb{Z}$ \\ $0$} & \makecell{\nsg{$\mathbb{Z}_2^{*}$} \\ $0$ \\ $0$ \\ \nsg{$\mathbb{Z}_2^{*}$}} \\
    \hline \hline
  \end{tabular}
\end{table}

\textit{Classification}.---We investigate topological classification of $d$-dimensional non-Hermitian Hamiltonians $H(\bm{k})$ with nonsymmorphic symmetry.
In general, an operation of order-two nonsymmorphic symmetry accompanies a nonprimitive lattice half translation, $\left( x_1, x_2, \cdots, x_d \right) \mapsto \left( x_1 + 1/2, x_2, \cdots, x_{d-d_{\parallel}}, - x_{d-d_{\parallel}+1}, \cdots, - x_{d} \right)$, where the lattice constant is set to unity, the direction of the half translation is chosen as the $x_1$ axis, and $d_{\parallel}$ coordinates are flipped ($0 \leq d_\parallel < d$).
In momentum space, this operation leads to two types of nonsymmorphic symmetry, 
\begin{align}
    &\mathcal{U}(k_{1})H(\bm{k})\mathcal{V}^{-1}(k_{1})=H(\sigma\bm{k}),\label{sym:NSG} \\
    &\tilde{\mathcal{U}}(k_{1})H^\dagger (\bm{k})\tilde{\mathcal{V}}^{-1}(k_{1})= H (\sigma\bm{k}),\label{sym:pseudoNSG}
\end{align}
with 
\begin{equation}
\sigma\bm{k} \coloneqq (k_1,k_2,\dots,k_{d-d_{\parallel}},-k_{d-d_{\parallel}+1},\dots,-k_{d}).
\end{equation}
Here, $\mathcal{U}$, $\mathcal{V}$, $\tilde{\mathcal{U}}$, and $\tilde{\mathcal{V}}$ are $k_1$-dependent unitary matrices satisfying $[\mathcal{U}(k_1)]^2=[\mathcal{V}(k_1)]^2=e^{-ik_1}$ for Eq.~(\ref{sym:NSG}) and $\tilde{\mathcal{U}}(k_1)\tilde{\mathcal{V}}(k_1) =\tilde{\mathcal{V}}(k_1)\tilde{\mathcal{U}}(k_1) = e^{-ik_1}$ for Eq.~(\ref{sym:pseudoNSG}).
Notably, unlike in Hermitian systems, non-Hermiticity allows for $\mathcal{U} \neq \mathcal{V}$ and $\tilde{\mathcal{U}} \neq \tilde{\mathcal{V}}$~\cite{KSS-20, Shiozaki-21}. We refer to Eq.~(\ref{sym:NSG}) and Eq.~(\ref{sym:pseudoNSG}) as nonsymmorphic symmetry and pseudo-nonsymmorphic symmetry, respectively. 
According to the terminology of Ref.~\cite{KSUS-19}, the latter is also called nonsymmorphic symmetry$^{\dagger}$. 
Whereas both types of nonsymmorphic symmetry coincide in Hermitian systems, their distinction in non-Hermitian systems gives rise to the rich topological classification. 

We proceed to classify point-gap topological phases protected by nonsymmorphic symmetry. A non-Hermitian Hamiltonian $H$ is defined to have a point gap with respect to a reference energy $E \in \mathbb{C}$ for $\det \left[ H-E \right] \neq 0$~\cite{Gong-18, KSUS-19}. 
Non-Hermitian point-gap topology of $H$ is then equivalent to Hermitian topology of ${\sf H}$ defined by 
\begin{equation}
    {\sf H} \coloneqq \begin{pmatrix}
        0 & H - E \\
        H^{\dagger} - E^{\dag} & 0
    \end{pmatrix}.
        \label{eq: Hermitization}
\end{equation} 
Through this Hermitization mapping, we identify the relevant symmetry class of ${\sf H}$ for each internal symmetry class of $H$ in addition to nonsymmorphic symmetry, and thus classify nonsymmorphic-symmetric point-gap topology for arbitrary spatial dimensions~\cite{supplement}. 
While point-gap topology with order-two point-group symmetry was classified~\cite{YifanWang-24, Tanaka-24} in a similar manner to the Hermitian case~\cite{Shiozaki-PRB-2014}, we here extend the classification to encompass nonsymmorphic spatial symmetry.

Our classification uncovers a plethora of point-gap topological phases that are not present only with internal symmetry, some of which are highlighted in Table~\ref{tab:NSGmain}.
These phases have the potential to exhibit unique non-Hermitian topological phenomena inherent in nonsymmorphic symmetry. 
Below, we demonstrate that they indeed host anomalous non-Hermitian boundary states protected by nonsymmorphic symmetry.
We exemplify this for pseudo-nonsymmorphic-symmetric systems in two dimensions (2D classes A + $\tilde{\cal \M}$ and AII$^{\dag}$ + $\tilde{\cal \M}_{-}$) and a chiral-symmetric and nonsymmorphic-symmetric system in three dimensions (3D class AIII + ${\cal \M}_{+}$).
In Ref.~\cite{supplement}, we also discuss nonsymmorphic topological phases in one dimension.
Remarkably, we demonstrate that generic types of symmetry representations with $\mathcal{U}\neq\mathcal{V}$ ($\tilde{\mathcal{U}}\neq\tilde{\mathcal{V}}$), which are forbidden in Hermitian systems, yield boundary states with distinctive complex-valued energy dispersions. 
Hereafter, we identify $x_1$ as $x$ and may use $\mathcal{U}$ and $\mathcal{V}$ even for pseudo-nonsymmorphic cases when no ambiguity arises.

\begin{figure}[t]
    \centering
\includegraphics[width=0.8\columnwidth]{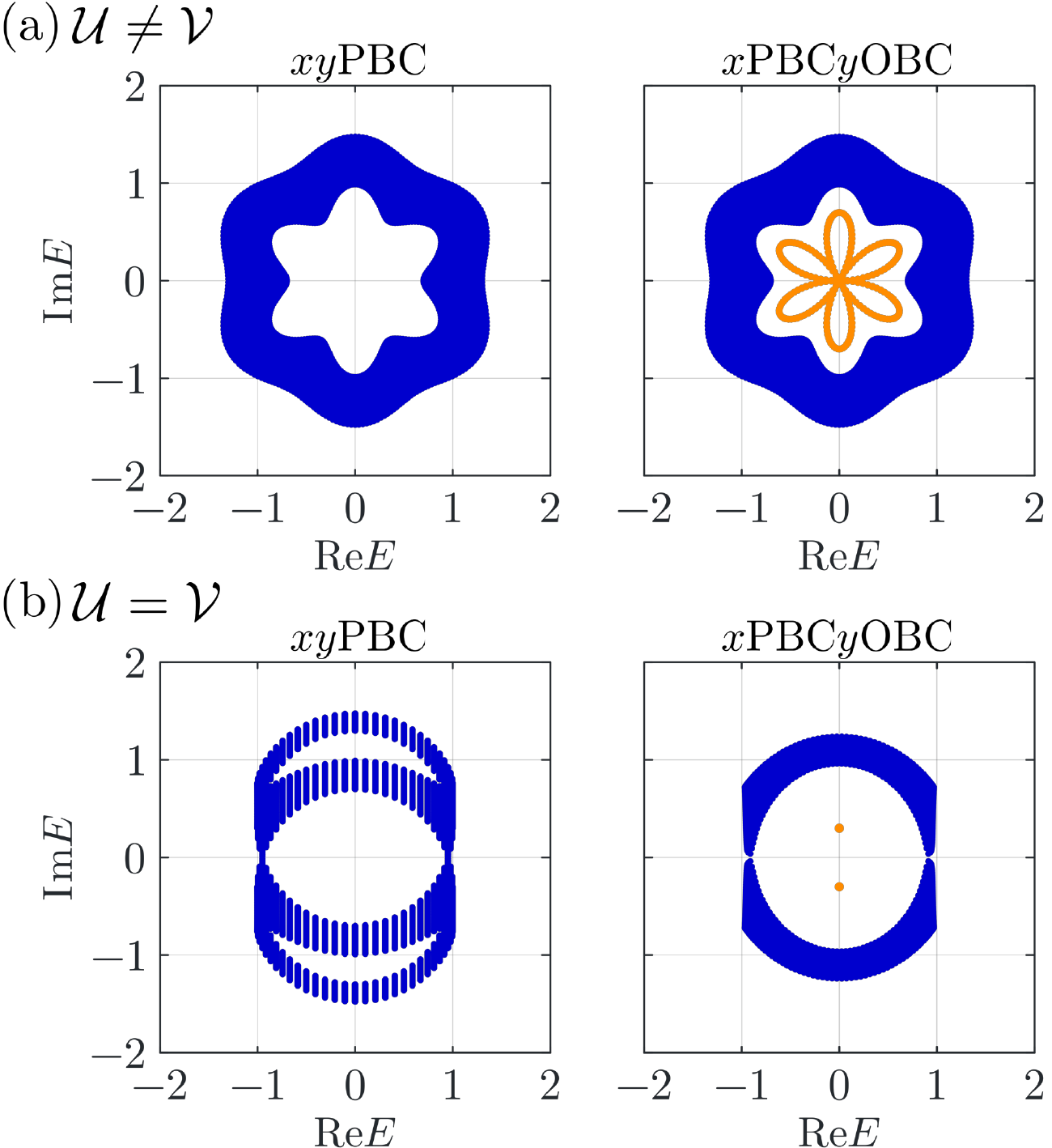}
\caption{Complex energy spectra of Eqs.~(\ref{model:A+Udag:UneqV}) and (\ref{model:A+Udag:U=V}) with $f(k_y)=-i(m+\cos k_y) + \sin k_y$ and $g(k_x)=\delta\sin (3k_x/2)$. 
$xy{\rm PBC}$ and $x{\rm PBC}y{\rm OBC}$ denote the periodic boundary conditions in both $x$ and $y$ directions and the open boundary conditions in the $y$ direction, respectively.
We set the parameters to $m=0.3$, $\delta=0.7$, $L_x=120$, and $L_y=50$, where $L_x$ and $L_y$ represent the system lengths in the $x$ and $y$ directions, respectively.
(a)~$\mathcal{U}\neq\mathcal{V}$. 
Edge modes (orange) described by $g(k_x)e^{ik_x/2}\sigma_z$ appear for $x{\rm PBC}y{\rm OBC}$. 
(b)~$\mathcal{U}=\mathcal{V}$. 
Skin modes (orange) from $f(k_y)$ appear for $x{\rm PBC}y{\rm OBC}$.}
    \label{fig:2DA+Udag}
\end{figure}

\textit{2D class A + $\tilde{\cal \M}$}.---We consider two-dimensional non-Hermitian systems only possessing pseudo-nonsymmorphic symmetry in Eq.~(\ref{sym:pseudoNSG}) with $d_\parallel=0$, 
which belong to class A + $\tilde{\cal \M}$ and thus host the $\mathbb{Z}_2$ topological phase (see Table~\ref{tab:NSGmain}). 
The associated $\mathbb{Z}_2$ topological invariant $\nu$ is given by the Berry phase of the Hermitian matrix $\mathcal{V}(k_x) H(\bm{k}) e^{i k_x/2}$~\cite{Shiozaki-PRB-2016, supplement}.
We first employ the generic symmetry representation,
\begin{equation}
    \mathcal{U}_{\rm I}(k_x)=\sigma_0,\quad
    \mathcal{V}_{\rm I}(k_x)=e^{-ik_x}\sigma_0,
\end{equation}
with the $2\times2$ identity matrix  $\sigma_0$.
A representative bulk Hamiltonian generating the $\mathbb{Z}_2$ phase takes the form of 
\begin{equation}
    H_{\rm I}(k_x,k_y)=
    \begin{pmatrix}
        g(k_x)e^{ik_x/2} & f(k_y) \\
        f^{*}(k_y)e^{ik_x} & -g(k_x)e^{ik_x/2}
    \end{pmatrix},\label{model:A+Udag:UneqV}
\end{equation}
where $f$ and $g$ are complex and real functions, respectively. 
Owing to the $2\pi$ periodicity of the Hamiltonian $H_{\rm I}(k_x,k_y)$, $f(k_x)$ is $2\pi$-periodic but $g(k_x)$ is $2\pi$-antiperiodic [i.e., $f(k_x+2\pi) = f(k_x)$, $g(k_x+2\pi) = - g(k_x)$].
Equation~(\ref{model:A+Udag:UneqV}) can be decomposed as
\begin{equation}
    H_{\rm I}(k_x,k_y)= e^{ik_x/2}H_{\rm SSH}(k_x,k_y) + H_{\rm edge}(k_x)\label{model:A+Udag:decomposition},
\end{equation}
with
\begin{align}
    H_{\rm SSH}(k_x,k_y)&\coloneqq
    \begin{pmatrix}
        0 & f(k_y)e^{-ik_x/2} \\
        f^{*}(k_y)e^{ik_x/2} & 0
    \end{pmatrix}\label{model:A+Udag:SSH},\\
    H_{\rm edge}(k_x)&\coloneqq g(k_x)e^{ik_x/2} \sigma_z
    \label{model:A+Udag:edge}.
\end{align}
Equation~(\ref{model:A+Udag:SSH}) gives the Su-Schrieffer-Heeger (SSH) model~\cite{SSH-79} along the $y$ direction for each $k_x$ and exhibits zero energy states if it has the nontrivial winding number of $f(k_y)e^{-ik_x/2}$: 
\begin{equation}
    W_{1}[f] \coloneqq -\int_{0}^{2\pi} \frac{dk_y}{2\pi i} \frac{d}{dk_y} \log f(k_y)\in\mathbb{Z}, \label{topology:A+Udag}
\end{equation}
where the factor $e^{-ik_x/2}$ is dropped.
We find that $W_{1}[f]$ also determines the $\mathbb{Z}_2$ point-gap topological invariant $\nu$ through its odd parity, $(-1)^{\nu} = (-1)^{W_{1}[f]}$~\cite{supplement}. 
Consequently, the entire non-Hermitian Hamiltonian $H_{\rm I}(k_x,k_y)$ in Eq.~(\ref{model:A+Udag:decomposition}) supports edge states described by $H_{\rm edge}(k_x)$ in Eq.~(\ref{model:A+Udag:edge}) under the open boundary conditions along the $y$ direction.

Remarkably, Eq.~(\ref{model:A+Udag:edge}) produces a loop-shaped edge spectrum in the complex energy plane, which is unique to non-Hermitian systems.
In Fig.~\ref{fig:2DA+Udag}\,(a), we provide the complex energy spectrum of Eq.~(\ref{model:A+Udag:UneqV}) for $f(k_y)=-i(m+\cos k_y) + \sin k_y$ and $g(k_x)=\delta\sin (3k_x/2)$ ($m, \delta \in \mathbb{R}$), demonstrating the emergence of the distinctive edge states.
While these unique edge states can be spectrally detached from the bulk bands in a similar manner to the Hermitian counterpart~\cite{Shiozaki-PRB-2015}, their robustness is protected by pseudo-nonsymmorphic symmetry.
Indeed, since $g(k_x)$ is $2\pi$-antiperiodic, it must vanish at some momenta $k_x=k_{\rm \FP}$ [i.e., $g(k_{\rm \FP}) = 0$], further necessitating the point-gap closing of $H_{\rm edge}(k_x)$.
Since this loop spectrum shows the nontrivial spectral winding number in terms of $k_x$, the corner skin effect can further occur under the open boundary conditions in both $x$ and $y$ directions. 

We also investigate an alternative choice of the symmetry representation, 
\begin{equation}
    \mathcal{U}_{\rm II}(k_x)=\mathcal{V}_{\rm II}(k_x)= 
    \begin{pmatrix}
        0 & e^{-ik_x} \\
        1 & 0
    \end{pmatrix},
\end{equation}
corresponding to 
\begin{align}
    H_{\rm II}(k_x,k_y)&=H_{\rm I}(k_x,k_y)\mathcal{U}_{\rm II}(k_x) \nonumber \\
    &=
    \begin{pmatrix}
        f(k_y) & g(k_x)e^{-ik_x/2} \\
        -g(k_x)e^{ik_x/2} & f^{*}(k_y)
    \end{pmatrix}.\label{model:A+Udag:U=V}
\end{align}
As in the previous case, $g(k_x)$ must accompany zeros at certain momenta $k_x=k_{\rm \FP}$, where the Hamiltonian becomes block-diagonalized:
$H_{\mathrm{II}}(k_{\rm \FP},k_y)={\rm diag}(f,f^*)$. 
Thus, in contrast to the generic case with $\mathcal{U}\neq\mathcal{V}$, the $\mathbb{Z}_2$ point-gap topology in Eq.~(\ref{model:A+Udag:U=V}) inevitably leads to the non-Hermitian skin effect of the individual subsectors $f$ and $f^{*}$ for $k_x=k_{\rm \FP}$, rather than the edge states in Eq.~(\ref{model:A+Udag:edge}). 
While this is reminiscent of the $\mathbb{Z}_2$ skin effect in Ref.~\cite{OKSS-20, *KOS-20}, it is protected by pseudo-nonsymmorphic symmetry instead of reciprocity.
We also confirm the $\mathbb{Z}_2$ skin effect for $\mathcal{U} = \mathcal{V}$ in Fig.~\ref{fig:2DA+Udag}\,(b).

\begin{figure}[t]
    \centering
\includegraphics[width=1.\columnwidth]{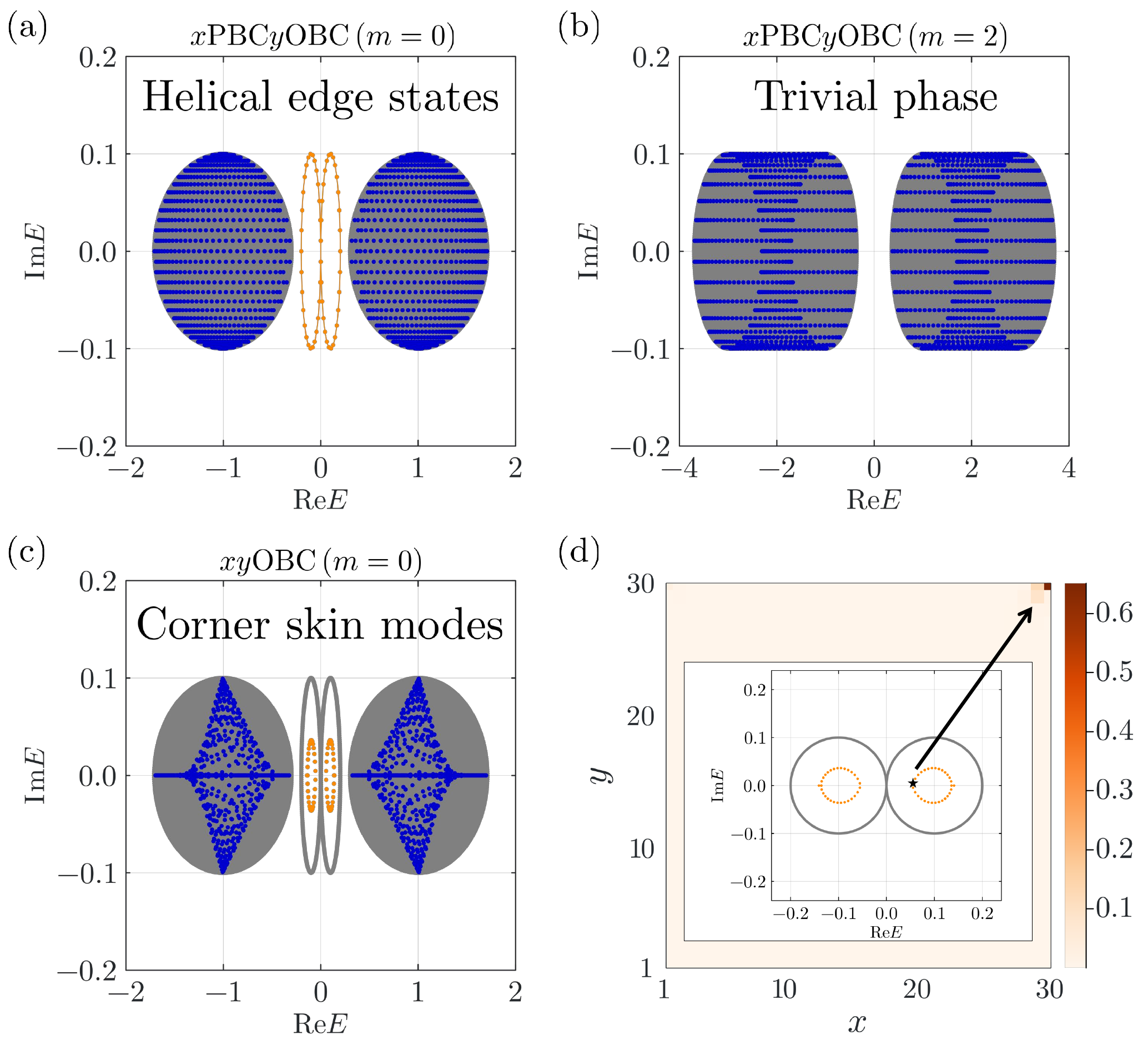}
\caption{Complex energy spectra of Eq.~(\ref{eq:nh_aIIdagger}) [$g(k_x)=c \sin(k_x/2)$; $t_x=0.7$, $t_y=1$, $\Delta=1$, and $c=0.2$] with (a,c,d)~$m=0$ and (b)~$m=2$. 
The system size is set to $L_x = L_y = 30$. 
(a, b)~Periodic boundary conditions (PBC) in the $x$ direction and open boundary conditions (OBC) in the $y$ direction for the bulk (blue dots) and edge (orange dots).
The gray region represents the spectrum under PBC in both $x$ and $y$ directions. 
For (a), the orange curve is $\pm ig(k_x)e^{\pm ik_x/2}$. 
(c)~OBC in both $x$ and $y$ directions for the bulk (blue dots) and corner (orange dots). 
The gray region and curve represent the spectrum for (a). 
(d)~Real-space distribution of the right eigenvector with the energy $E=0.0554+0.00458i$ (the star symbol in the inset). 
The inset shows the energy spectrum of (c) around $E=0$.}
    \label{fig:AIIdag}
\end{figure}

\textit{2D class AII$^{\dag}$ + $\tilde{\cal \M}_{-}$}.---An important feature of nonsymmorphic symmetry is the $\mathbb{Z}_4$ topological classification.
While $\mathbb{Z}_4$ Hermitian topology manifests as anomalous edge states with hourglass-like spectra~\cite{Shiozaki-PRB-2016, Wang-16}, we here elucidate its non-Hermitian counterpart.
Prime examples arise in 2D class AII$^{\dag}$ + $\tilde{\cal \M}_{-}$ with $d_{\parallel} = 0$, accompanying pseudo-nonsymmorphic symmetry in Eq.~(\ref{sym:pseudoNSG}) and time-reversal symmetry$^{\dag}$
\begin{equation}\label{eq:AIIdag_TRS}
    \mathcal{T}H^{T}(\boldsymbol{k})\mathcal{T}^{-1}=H(-\boldsymbol{k}), \quad \mathcal{T}\mathcal{T}^{*} = -1
\end{equation} 
with a unitary matrix $\mathcal{T}$ satisfying $\mathcal{T}\mathcal{V}^{*}(k_x)=-\mathcal{U}(-k_x)\mathcal{T}$.
The parity of the $\mathbb{Z}_4$ topological invariant $\nu$ reduces to the $\mathbb{Z}_2$ invariant protected solely by time-reversal symmetry$^{\dag}$~\cite{supplement, Fu-06}.
Consequently, $\nu \equiv 1, 3$ (mod $4$) merely yields the conventional $\mathbb{Z}_2$ skin effect~\cite{OKSS-20} and is thus irrelevant to nonsymmorphic symmetry.

Rather, we clarify that $\mathbb{Z}_4$ point-gap topology with $\nu \equiv 2$ is inherent in nonsymmorphic symmetry and gives rise to helical edge states characterized by distinctive complex-valued spectra, akin to class A + $\tilde{\cal U}$.
We identify their effective theory as
\begin{equation}\label{eq:AIIdagger_surface}
    H (k_x) = \begin{pmatrix}
     ie^{-ik_x/2} g(k_x) & f(k_x) \\
     f^{*}(k_x) & ie^{ik_x/2} g(-k_x)
    \end{pmatrix},
\end{equation}
where the symmetry operators are chosen as $\mathcal{U}(k_x)={\rm diag}({i},-{i}e^{-ik_x})$, $\mathcal{V}(k_x)={\rm diag}(-{i}e^{-ik_x},{i})$, and $\mathcal{T} = \sigma_y$.
Here, $f(k_x)$ is a complex $2\pi$-periodic function satisfying $f(-k_x)=-f(k_x)$, and $g(k_x)$ is a real $2\pi$-antiperiodic function.
Unlike the previous case of class A + $\tilde{\cal U}$, $H(k_x)$ effectively describes a Kramers pair of point-gapless edge states.
Indeed, for $f(k_x) = 0$ and $g(k_x)=\sin(k_x/2)$, the energy dispersions are given as the helical edge states $E=ie^{-ik_x/2}\sin(k_x/2)$, $-ie^{ik_x/2}\sin(k_x/2)$.

To realize this $\mathbb{Z}_4$ point-gap topological phase, we introduce the following tight-binding model on a square lattice:
\begin{align}
    &{H} (\boldsymbol{k}) \coloneqq (t_x\cos k_x + t_{y}\cos k_y+m)  \tau_x \sigma_x + \Delta \sin k_y \tau_0 \sigma_y \nonumber \\
    &\qquad +
    g(k_x)  \left[ 
    -i\cos (k_x/2) \tau_x \sigma_0 +\sin(k_x/2) \tau_x \sigma_z \right], \label{eq:nh_aIIdagger}
\end{align}
with $\mathcal{U}(\boldsymbol{k})= \tau_0 \otimes {\rm diag}({i}, -{i}e^{-ik_x})_{\sigma}$, $\mathcal{V}(\boldsymbol{k})= \tau_0 \otimes {\rm diag}(-{i}e^{-ik_x},{i})_{\sigma}$, and $\mathcal{T}=\tau_z \sigma_y$, as well as $g(k_x+2\pi)=-g(k_x)$ and $g(-k_x)=-g(k_x)$.
The corresponding Hermitized Hamiltonian in Eq.~(\ref{eq: Hermitization}) exhibits the nonsymmorphic topological phases supporting hourglass fermions at edges~\cite{supplement}. 
In Fig.~\ref{fig:AIIdag}\,(a, b), we provide the complex spectra under both periodic and open boundary conditions, confirming the non-Hermitian helical detachable edge states.
They originate from the zero modes of the SSH model in the $y$ direction comprised of the first two terms in Eq.~(\ref{eq:nh_aIIdagger}), as in the previous case.
As a result of their point-gap topology, they further exhibit the corner skin effect under the open boundary conditions along both directions [Fig.~\ref{fig:AIIdag}\,(c, d)].
In Ref.~\cite{Ishikawa-24}, the $\mathbb{Z}_4$ point-gap topology was argued to result in the skin effect for arbitrary $\nu$.
However, our results imply that this conclusion is specific to the model therein.
Instead, we demonstrate that while $\nu \equiv 1, 3$ leads to the skin effect, $\nu \equiv 2$ generally induces the emergence of helical edge states with unique complex-valued spectra.

\begin{figure}[t]
    \centering
\includegraphics[width=\columnwidth]{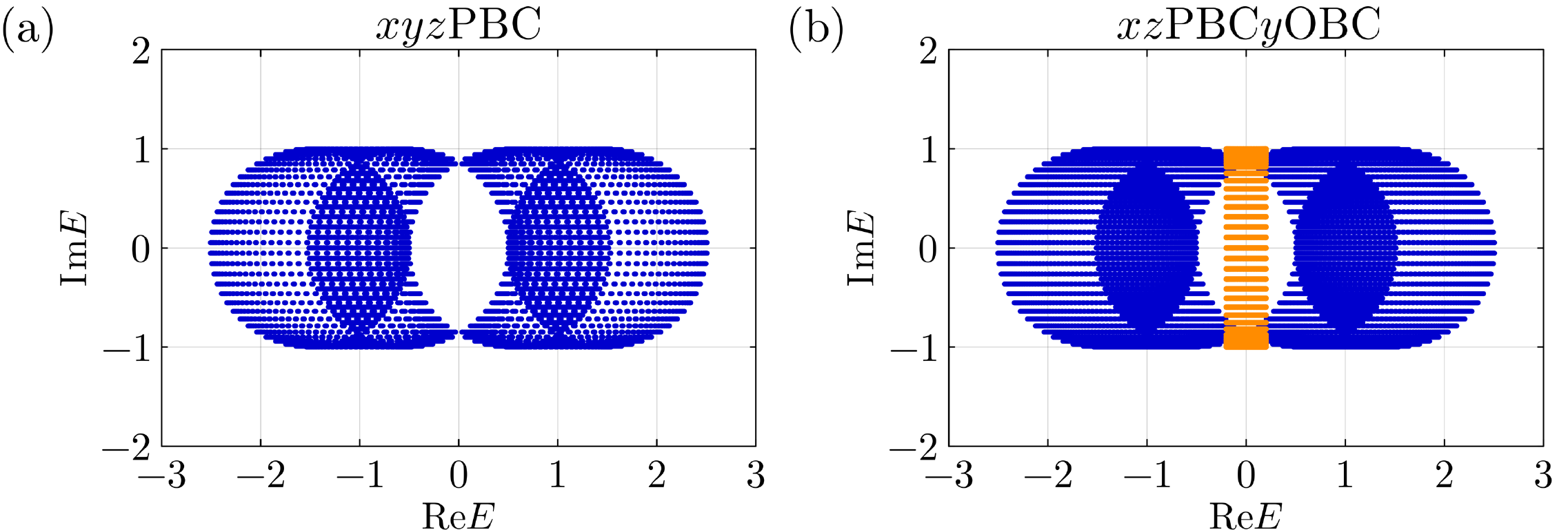}
\caption{Complex energy spectra of Eq.~(\ref{model:AIII+U+}) under the (a)~periodic boundary conditions in all directions ($xyz{\rm PBC}$) and (b)~open boundary conditions in the $y$ direction ($xz{\rm PBC}y{\rm OBC}$) 
[$a(k_x,k_z)=-\delta\sin (k_x/2)\sin (k_x/2)$ and $b(k_x,k_z)=\delta\cos (k_x/2)\sin (k_x/2)$;
$m=0.5, \delta=0.2$, and $L_x=L_y=L_z=60$].}
\label{fig:3DA+U+}
\end{figure}

\textit{3D class AIII + ${\cal \M}_{+}$}.---A nontrivial example in three spatial dimensions arises in the presence of both nonsymmorphic symmetry in Eq.~(\ref{sym:NSG}) and chiral symmetry 
\begin{equation}
    \Gamma H^{\dagger}(\bm{k})\Gamma^{-1} = -H(\bm{k}), \quad \Gamma^2 = 1
\end{equation}
with a unitary matrix $\Gamma$.
Specifically, we consider 3D class AIII + ${\cal \M}_{+}$ with $d_{\parallel} = 1$, which satisfies $\Gamma\mathcal{U}(k_x)=\mathcal{V}(k_x)\Gamma$ and hosts the $\mathbb{Z}_2$ point-gap classification (see Table~\ref{tab:NSGmain}). 
Non-Hermitian Hamiltonians $H(\bm{k})$ preserving these symmetries can be systematically constructed from Hermitian Hamiltonians ${\sf h}(\bm{k})$ in class A + $\sf U$ by $H(\bm{k})=-i{\sf h}(\bm{k})\Gamma$, 
where $\mathcal{U}(k_x)$ and $\mathcal{V}(k_x)$ are properly defined through a nonsymmorphic-symmetry operator of ${\sf h}(\bm{k})$~\cite{supplement}.
Following this procedure, we obtain a prototypical model as
\begin{align}
    &H(\bm{k}) = (m + \cos k_y + \cos k_z)\tau_x\sigma_z - \sin k_y\tau_y\sigma_0 \nonumber \\
    &\qquad + a\tau_x\sigma_x + b\tau_x\sigma_y -i\sin k_z\tau_0\sigma_0 \quad \left( m \in \mathbb{R} \right),\label{model:AIII+U+}
\end{align}
where $a=a(k_x,k_z)$ and $b=b(k_x,k_z)$ are real functions satisfying $[a(k_x,k_z) + ib(k_x,k_z)]e^{-ik_x} = -a(k_x,-k_z) + ib(k_x,-k_z)$. 
The symmetry operators are explicitly given as
\begin{equation}
    \Gamma = \tau_z\sigma_0,\quad
    \mathcal{U}(k_x)=-\mathcal{V}(k_x)=
    \tau_x\otimes
    \begin{pmatrix}
        0 & e^{-ik_x} \\
        1 & 0 \\
    \end{pmatrix}_\sigma.
\end{equation}
Similar to Eqs.~(\ref{model:A+Udag:decomposition}) and (\ref{eq:nh_aIIdagger}), this model is also decomposed into the SSH model and a non-Hermitian surface Hamiltonian, 
\begin{equation}
    H_{\rm surf}(k_x,k_z) = a\tau_x\sigma_x + b\tau_x\sigma_y -i\sin k_z\tau_0\sigma_0,
        \label{eq: 3D surface}
\end{equation}
within the surface Brillouin zone specified by $|m+\cos k_z|<1$ along the $z$ direction.

In Fig.~\ref{fig:3DA+U+}, we present the complex energy spectrum of Eq.~(\ref{model:AIII+U+}) for $a(k_x,k_z)=-\delta\sin (k_x/2)\sin (k_x/2)$ and $b(k_x,k_z)=\delta\cos (k_x/2)\sin (k_x/2)$ ($\delta \in \mathbb{R}$), 
exhibiting nonsymmorphic surface states consistent with Eq.~(\ref{eq: 3D surface}).
In contrast to the two-dimensional cases, which host the edge states completely detached from the bulk bands (see Figs.~\ref{fig:2DA+Udag} and \ref{fig:AIIdag}), the surface states in three dimensions are spectrally detached only along the real axis [Fig.~\ref{fig:3DA+U+}\,(b)]. 
In fact, the energy dispersion around $E=0$ reads $\pm\delta (k_x/2)-ik_z$. 
This is a unique feature of three-dimensional nonsymmorphic-symmetric systems, arising from the absence and presence of spectral flow in the $k_x$ and $k_z$ directions, respectively.

\textit{Conclusions}.---Nonsymmorphic symmetry serves as an important component of space groups.
While it yields unique topological classification and corresponding boundary phenomena in the Hermitian regime, its impact on non-Hermitian topology has remained largely elusive.
In this Letter, we have developed classification of nonsymmorphic-symmetry-protected point-gap topology.
Building upon this classification, we have identified the $\mathbb{Z}_2$ and $\mathbb{Z}_4$ non-Hermitian topological phases and their associated anomalous boundary states distinguished by the characteristic complex-valued energy dispersions.
In light of the recent experimental realization of non-Hermitian topological systems across diverse synthetic platforms, our nonsymmorphic-symmetric models should also be implemented in similar setups, potentially affecting the dynamics of open classical and quantum systems.
Our work thereby broadens the landscape of non-Hermitian topological materials and enriches the understanding of phases of matter.

\medskip
\begingroup
\renewcommand{\addcontentsline}[3]{}
\begin{acknowledgments}
We thank Ryo Okugawa, Masatoshi Sato, and Kazuki Yokomizo for helpful discussion. 
We appreciate the long-term workshop ``Recent Developments and Challenges in Topological Phases" (YITP-T-24-03) held at Yukawa Institute for Theoretical Physics (YITP), Kyoto University.
D.N., Y.T., and K.S. are supported by JST CREST Grant No.~JPMJCR19T2.
D.N. is supported by JSPS KAKENHI Grant No.~24K22857.
Y.T. is supported by RIKEN Special Postdoctoral Researchers Program and JSPS KAKENHI Grant No.~24K22868.
K.S. is supported by JSPS KAKENHI Grant No.~22H05118 and No.~23H01097.
K.K. is supported by MEXT KAKENHI Grant-in-Aid for Transformative Research Areas A ``Extreme Universe" No.~24H00945.
\end{acknowledgments}
\endgroup

\let\oldaddcontentsline\addcontentsline
\renewcommand{\addcontentsline}[3]{}
\bibliography{ref.bib}

\begin{thebibliography}{100}%
\makeatletter
\providecommand \@ifxundefined [1]{%
 \@ifx{#1\undefined}
}%
\providecommand \@ifnum [1]{%
 \ifnum #1\expandafter \@firstoftwo
 \else \expandafter \@secondoftwo
 \fi
}%
\providecommand \@ifx [1]{%
 \ifx #1\expandafter \@firstoftwo
 \else \expandafter \@secondoftwo
 \fi
}%
\providecommand \natexlab [1]{#1}%
\providecommand \enquote  [1]{``#1''}%
\providecommand \bibnamefont  [1]{#1}%
\providecommand \bibfnamefont [1]{#1}%
\providecommand \citenamefont [1]{#1}%
\providecommand \href@noop [0]{\@secondoftwo}%
\providecommand \href [0]{\begingroup \@sanitize@url \@href}%
\providecommand \@href[1]{\@@startlink{#1}\@@href}%
\providecommand \@@href[1]{\endgroup#1\@@endlink}%
\providecommand \@sanitize@url [0]{\catcode `\\12\catcode `\$12\catcode `\&12\catcode `\#12\catcode `\^12\catcode `\_12\catcode `\%12\relax}%
\providecommand \@@startlink[1]{}%
\providecommand \@@endlink[0]{}%
\providecommand \url  [0]{\begingroup\@sanitize@url \@url }%
\providecommand \@url [1]{\endgroup\@href {#1}{\urlprefix }}%
\providecommand \urlprefix  [0]{URL }%
\providecommand \Eprint [0]{\href }%
\providecommand \doibase [0]{https://doi.org/}%
\providecommand \selectlanguage [0]{\@gobble}%
\providecommand \bibinfo  [0]{\@secondoftwo}%
\providecommand \bibfield  [0]{\@secondoftwo}%
\providecommand \translation [1]{[#1]}%
\providecommand \BibitemOpen [0]{}%
\providecommand \bibitemStop [0]{}%
\providecommand \bibitemNoStop [0]{.\EOS\space}%
\providecommand \EOS [0]{\spacefactor3000\relax}%
\providecommand \BibitemShut  [1]{\csname bibitem#1\endcsname}%
\let\auto@bib@innerbib\@empty
\bibitem [{\citenamefont {Hasan}\ and\ \citenamefont {Kane}(2010)}]{HK-review}%
  \BibitemOpen
  \bibfield  {author} {\bibinfo {author} {\bibfnamefont {M.~Z.}\ \bibnamefont {Hasan}}\ and\ \bibinfo {author} {\bibfnamefont {C.~L.}\ \bibnamefont {Kane}},\ }\bibfield  {title} {\bibinfo {title} {{Colloquium: Topological insulators}},\ }\href {https://doi.org/10.1103/RevModPhys.82.3045} {\bibfield  {journal} {\bibinfo  {journal} {Rev. Mod. Phys.}\ }\textbf {\bibinfo {volume} {82}},\ \bibinfo {pages} {3045} (\bibinfo {year} {2010})}\BibitemShut {NoStop}%
\bibitem [{\citenamefont {Qi}\ and\ \citenamefont {Zhang}(2011)}]{QZ-review}%
  \BibitemOpen
  \bibfield  {author} {\bibinfo {author} {\bibfnamefont {X.-L.}\ \bibnamefont {Qi}}\ and\ \bibinfo {author} {\bibfnamefont {S.-C.}\ \bibnamefont {Zhang}},\ }\bibfield  {title} {\bibinfo {title} {{Topological insulators and superconductors}},\ }\href {https://doi.org/10.1103/RevModPhys.83.1057} {\bibfield  {journal} {\bibinfo  {journal} {{Rev. Mod. Phys.}}\ }\textbf {\bibinfo {volume} {83}},\ \bibinfo {pages} {1057} (\bibinfo {year} {2011})}\BibitemShut {NoStop}%
\bibitem [{\citenamefont {Chiu}\ \emph {et~al.}(2016)\citenamefont {Chiu}, \citenamefont {Teo}, \citenamefont {Schnyder},\ and\ \citenamefont {Ryu}}]{CTSR-review}%
  \BibitemOpen
  \bibfield  {author} {\bibinfo {author} {\bibfnamefont {C.-K.}\ \bibnamefont {Chiu}}, \bibinfo {author} {\bibfnamefont {J.~C.~Y.}\ \bibnamefont {Teo}}, \bibinfo {author} {\bibfnamefont {A.~P.}\ \bibnamefont {Schnyder}},\ and\ \bibinfo {author} {\bibfnamefont {S.}~\bibnamefont {Ryu}},\ }\bibfield  {title} {\bibinfo {title} {{Classification of topological quantum matter with symmetries}},\ }\href {https://doi.org/10.1103/RevModPhys.88.035005} {\bibfield  {journal} {\bibinfo  {journal} {Rev. Mod. Phys.}\ }\textbf {\bibinfo {volume} {88}},\ \bibinfo {pages} {035005} (\bibinfo {year} {2016})}\BibitemShut {NoStop}%
\bibitem [{\citenamefont {Fu}(2011)}]{Fu-11}%
  \BibitemOpen
  \bibfield  {author} {\bibinfo {author} {\bibfnamefont {L.}~\bibnamefont {Fu}},\ }\bibfield  {title} {\bibinfo {title} {{Topological Crystalline Insulators}},\ }\href {https://doi.org/10.1103/PhysRevLett.106.106802} {\bibfield  {journal} {\bibinfo  {journal} {Phys. Rev. Lett.}\ }\textbf {\bibinfo {volume} {106}},\ \bibinfo {pages} {106802} (\bibinfo {year} {2011})}\BibitemShut {NoStop}%
\bibitem [{\citenamefont {Hsieh}\ \emph {et~al.}(2012)\citenamefont {Hsieh}, \citenamefont {Lin}, \citenamefont {Liu}, \citenamefont {Duan}, \citenamefont {Bansil},\ and\ \citenamefont {Fu}}]{Hsieh-12}%
  \BibitemOpen
  \bibfield  {author} {\bibinfo {author} {\bibfnamefont {T.~H.}\ \bibnamefont {Hsieh}}, \bibinfo {author} {\bibfnamefont {H.}~\bibnamefont {Lin}}, \bibinfo {author} {\bibfnamefont {J.}~\bibnamefont {Liu}}, \bibinfo {author} {\bibfnamefont {W.}~\bibnamefont {Duan}}, \bibinfo {author} {\bibfnamefont {A.}~\bibnamefont {Bansil}},\ and\ \bibinfo {author} {\bibfnamefont {L.}~\bibnamefont {Fu}},\ }\bibfield  {title} {\bibinfo {title} {{Topological crystalline insulators in the SnTe material class}},\ }\href {https://doi.org/https://doi.org/10.1038/ncomms1969} {\bibfield  {journal} {\bibinfo  {journal} {Nat. Commun.}\ }\textbf {\bibinfo {volume} {3}},\ \bibinfo {pages} {982} (\bibinfo {year} {2012})}\BibitemShut {NoStop}%
\bibitem [{\citenamefont {Chiu}\ \emph {et~al.}(2013)\citenamefont {Chiu}, \citenamefont {Yao},\ and\ \citenamefont {Ryu}}]{Chiu-13}%
  \BibitemOpen
  \bibfield  {author} {\bibinfo {author} {\bibfnamefont {C.-K.}\ \bibnamefont {Chiu}}, \bibinfo {author} {\bibfnamefont {H.}~\bibnamefont {Yao}},\ and\ \bibinfo {author} {\bibfnamefont {S.}~\bibnamefont {Ryu}},\ }\bibfield  {title} {\bibinfo {title} {{Classification of topological insulators and superconductors in the presence of reflection symmetry}},\ }\href {https://doi.org/10.1103/PhysRevB.88.075142} {\bibfield  {journal} {\bibinfo  {journal} {Phys. Rev. B}\ }\textbf {\bibinfo {volume} {88}},\ \bibinfo {pages} {075142} (\bibinfo {year} {2013})}\BibitemShut {NoStop}%
\bibitem [{\citenamefont {Morimoto}\ and\ \citenamefont {Furusaki}(2013)}]{Morimoto-13}%
  \BibitemOpen
  \bibfield  {author} {\bibinfo {author} {\bibfnamefont {T.}~\bibnamefont {Morimoto}}\ and\ \bibinfo {author} {\bibfnamefont {A.}~\bibnamefont {Furusaki}},\ }\bibfield  {title} {\bibinfo {title} {{Topological classification with additional symmetries from Clifford algebras}},\ }\href {https://doi.org/10.1103/PhysRevB.88.125129} {\bibfield  {journal} {\bibinfo  {journal} {Phys. Rev. B}\ }\textbf {\bibinfo {volume} {88}},\ \bibinfo {pages} {125129} (\bibinfo {year} {2013})}\BibitemShut {NoStop}%
\bibitem [{\citenamefont {Shiozaki}\ and\ \citenamefont {Sato}(2014)}]{Shiozaki-PRB-2014}%
  \BibitemOpen
  \bibfield  {author} {\bibinfo {author} {\bibfnamefont {K.}~\bibnamefont {Shiozaki}}\ and\ \bibinfo {author} {\bibfnamefont {M.}~\bibnamefont {Sato}},\ }\bibfield  {title} {\bibinfo {title} {{Topology of crystalline insulators and superconductors}},\ }\href {https://doi.org/10.1103/PhysRevB.90.165114} {\bibfield  {journal} {\bibinfo  {journal} {Phys. Rev. B}\ }\textbf {\bibinfo {volume} {90}},\ \bibinfo {pages} {165114} (\bibinfo {year} {2014})}\BibitemShut {NoStop}%
\bibitem [{\citenamefont {Benalcazar}\ \emph {et~al.}(2017)\citenamefont {Benalcazar}, \citenamefont {Bernevig},\ and\ \citenamefont {Hughes}}]{Benalcazar-17}%
  \BibitemOpen
  \bibfield  {author} {\bibinfo {author} {\bibfnamefont {W.~A.}\ \bibnamefont {Benalcazar}}, \bibinfo {author} {\bibfnamefont {B.~A.}\ \bibnamefont {Bernevig}},\ and\ \bibinfo {author} {\bibfnamefont {T.~L.}\ \bibnamefont {Hughes}},\ }\bibfield  {title} {\bibinfo {title} {{Quantized electric multipole insulators}},\ }\href {https://doi.org/10.1126/science.aah6442} {\bibfield  {journal} {\bibinfo  {journal} {Science}\ }\textbf {\bibinfo {volume} {357}},\ \bibinfo {pages} {61} (\bibinfo {year} {2017})}\BibitemShut {NoStop}%
\bibitem [{\citenamefont {Po}\ \emph {et~al.}(2017)\citenamefont {Po}, \citenamefont {Vishwanath},\ and\ \citenamefont {Watanabe}}]{Po-17}%
  \BibitemOpen
  \bibfield  {author} {\bibinfo {author} {\bibfnamefont {H.~C.}\ \bibnamefont {Po}}, \bibinfo {author} {\bibfnamefont {A.}~\bibnamefont {Vishwanath}},\ and\ \bibinfo {author} {\bibfnamefont {H.}~\bibnamefont {Watanabe}},\ }\bibfield  {title} {\bibinfo {title} {{Symmetry-based indicators of band topology in the 230 space groups}},\ }\href {https://doi.org/https://doi.org/10.1038/s41467-017-00133-2} {\bibfield  {journal} {\bibinfo  {journal} {Nat. Commun.}\ }\textbf {\bibinfo {volume} {8}},\ \bibinfo {pages} {50} (\bibinfo {year} {2017})}\BibitemShut {NoStop}%
\bibitem [{\citenamefont {Bradlyn}\ \emph {et~al.}(2017)\citenamefont {Bradlyn}, \citenamefont {Elcoro}, \citenamefont {Cano}, \citenamefont {Vergniory}, \citenamefont {Wang}, \citenamefont {Felser}, \citenamefont {Aroyo},\ and\ \citenamefont {Bernevig}}]{Bradlyn-17}%
  \BibitemOpen
  \bibfield  {author} {\bibinfo {author} {\bibfnamefont {B.}~\bibnamefont {Bradlyn}}, \bibinfo {author} {\bibfnamefont {L.}~\bibnamefont {Elcoro}}, \bibinfo {author} {\bibfnamefont {J.}~\bibnamefont {Cano}}, \bibinfo {author} {\bibfnamefont {M.~G.}\ \bibnamefont {Vergniory}}, \bibinfo {author} {\bibfnamefont {Z.}~\bibnamefont {Wang}}, \bibinfo {author} {\bibfnamefont {C.}~\bibnamefont {Felser}}, \bibinfo {author} {\bibfnamefont {M.~I.}\ \bibnamefont {Aroyo}},\ and\ \bibinfo {author} {\bibfnamefont {B.~A.}\ \bibnamefont {Bernevig}},\ }\bibfield  {title} {\bibinfo {title} {{Topological quantum chemistry}},\ }\href {https://doi.org/https://doi.org/10.1038/nature23268} {\bibfield  {journal} {\bibinfo  {journal} {Nature}\ }\textbf {\bibinfo {volume} {547}},\ \bibinfo {pages} {298} (\bibinfo {year} {2017})}\BibitemShut {NoStop}%
\bibitem [{\citenamefont {Schindler}\ \emph {et~al.}(2018)\citenamefont {Schindler}, \citenamefont {Cook}, \citenamefont {Vergniory}, \citenamefont {Wang}, \citenamefont {Parkin}, \citenamefont {Bernevig},\ and\ \citenamefont {Neupert}}]{Schindler-18}%
  \BibitemOpen
  \bibfield  {author} {\bibinfo {author} {\bibfnamefont {F.}~\bibnamefont {Schindler}}, \bibinfo {author} {\bibfnamefont {A.~M.}\ \bibnamefont {Cook}}, \bibinfo {author} {\bibfnamefont {M.~G.}\ \bibnamefont {Vergniory}}, \bibinfo {author} {\bibfnamefont {Z.}~\bibnamefont {Wang}}, \bibinfo {author} {\bibfnamefont {S.~S.~P.}\ \bibnamefont {Parkin}}, \bibinfo {author} {\bibfnamefont {B.~A.}\ \bibnamefont {Bernevig}},\ and\ \bibinfo {author} {\bibfnamefont {T.}~\bibnamefont {Neupert}},\ }\bibfield  {title} {\bibinfo {title} {{Higher-order topological insulators}},\ }\href {https://doi.org/10.1126/sciadv.aat0346} {\bibfield  {journal} {\bibinfo  {journal} {Sci. Adv.}\ }\textbf {\bibinfo {volume} {4}},\ \bibinfo {pages} {eaat0346} (\bibinfo {year} {2018})}\BibitemShut {NoStop}%
\bibitem [{\citenamefont {Parameswaran}\ \emph {et~al.}(2013)\citenamefont {Parameswaran}, \citenamefont {Turner}, \citenamefont {Arovas},\ and\ \citenamefont {Vishwanath}}]{Parameswaran-13}%
  \BibitemOpen
  \bibfield  {author} {\bibinfo {author} {\bibfnamefont {S.~A.}\ \bibnamefont {Parameswaran}}, \bibinfo {author} {\bibfnamefont {A.~M.}\ \bibnamefont {Turner}}, \bibinfo {author} {\bibfnamefont {D.~P.}\ \bibnamefont {Arovas}},\ and\ \bibinfo {author} {\bibfnamefont {A.}~\bibnamefont {Vishwanath}},\ }\bibfield  {title} {\bibinfo {title} {{Topological order and absence of band insulators at integer filling in non-symmorphic crystals}},\ }\href {https://doi.org/https://doi.org/10.1038/nphys2600} {\bibfield  {journal} {\bibinfo  {journal} {Nat. Phys.}\ }\textbf {\bibinfo {volume} {9}},\ \bibinfo {pages} {299} (\bibinfo {year} {2013})}\BibitemShut {NoStop}%
\bibitem [{\citenamefont {Liu}\ \emph {et~al.}(2014)\citenamefont {Liu}, \citenamefont {Zhang},\ and\ \citenamefont {VanLeeuwen}}]{Liu-14}%
  \BibitemOpen
  \bibfield  {author} {\bibinfo {author} {\bibfnamefont {C.-X.}\ \bibnamefont {Liu}}, \bibinfo {author} {\bibfnamefont {R.-X.}\ \bibnamefont {Zhang}},\ and\ \bibinfo {author} {\bibfnamefont {B.~K.}\ \bibnamefont {VanLeeuwen}},\ }\bibfield  {title} {\bibinfo {title} {{Topological nonsymmorphic crystalline insulators}},\ }\href {https://doi.org/10.1103/PhysRevB.90.085304} {\bibfield  {journal} {\bibinfo  {journal} {Phys. Rev. B}\ }\textbf {\bibinfo {volume} {90}},\ \bibinfo {pages} {085304} (\bibinfo {year} {2014})}\BibitemShut {NoStop}%
\bibitem [{\citenamefont {Fang}\ and\ \citenamefont {Fu}(2015)}]{Fang-Fu-15}%
  \BibitemOpen
  \bibfield  {author} {\bibinfo {author} {\bibfnamefont {C.}~\bibnamefont {Fang}}\ and\ \bibinfo {author} {\bibfnamefont {L.}~\bibnamefont {Fu}},\ }\bibfield  {title} {\bibinfo {title} {{New classes of three-dimensional topological crystalline insulators: Nonsymmorphic and magnetic}},\ }\href {https://doi.org/10.1103/PhysRevB.91.161105} {\bibfield  {journal} {\bibinfo  {journal} {Phys. Rev. B}\ }\textbf {\bibinfo {volume} {91}},\ \bibinfo {pages} {161105} (\bibinfo {year} {2015})}\BibitemShut {NoStop}%
\bibitem [{\citenamefont {Shiozaki}\ \emph {et~al.}(2015)\citenamefont {Shiozaki}, \citenamefont {Sato},\ and\ \citenamefont {Gomi}}]{Shiozaki-PRB-2015}%
  \BibitemOpen
  \bibfield  {author} {\bibinfo {author} {\bibfnamefont {K.}~\bibnamefont {Shiozaki}}, \bibinfo {author} {\bibfnamefont {M.}~\bibnamefont {Sato}},\ and\ \bibinfo {author} {\bibfnamefont {K.}~\bibnamefont {Gomi}},\ }\bibfield  {title} {\bibinfo {title} {${Z}_{2}$ topology in nonsymmorphic crystalline insulators: M\"obius twist in surface states},\ }\href {https://doi.org/10.1103/PhysRevB.91.155120} {\bibfield  {journal} {\bibinfo  {journal} {Phys. Rev. B}\ }\textbf {\bibinfo {volume} {91}},\ \bibinfo {pages} {155120} (\bibinfo {year} {2015})}\BibitemShut {NoStop}%
\bibitem [{\citenamefont {Young}\ and\ \citenamefont {Kane}(2015)}]{Young-15}%
  \BibitemOpen
  \bibfield  {author} {\bibinfo {author} {\bibfnamefont {S.~M.}\ \bibnamefont {Young}}\ and\ \bibinfo {author} {\bibfnamefont {C.~L.}\ \bibnamefont {Kane}},\ }\bibfield  {title} {\bibinfo {title} {{Dirac Semimetals in Two Dimensions}},\ }\href {https://doi.org/10.1103/PhysRevLett.115.126803} {\bibfield  {journal} {\bibinfo  {journal} {Phys. Rev. Lett.}\ }\textbf {\bibinfo {volume} {115}},\ \bibinfo {pages} {126803} (\bibinfo {year} {2015})}\BibitemShut {NoStop}%
\bibitem [{\citenamefont {Watanabe}\ \emph {et~al.}(2015)\citenamefont {Watanabe}, \citenamefont {Po}, \citenamefont {Vishwanath},\ and\ \citenamefont {Zaletel}}]{Watanabe-15}%
  \BibitemOpen
  \bibfield  {author} {\bibinfo {author} {\bibfnamefont {H.}~\bibnamefont {Watanabe}}, \bibinfo {author} {\bibfnamefont {H.~C.}\ \bibnamefont {Po}}, \bibinfo {author} {\bibfnamefont {A.}~\bibnamefont {Vishwanath}},\ and\ \bibinfo {author} {\bibfnamefont {M.}~\bibnamefont {Zaletel}},\ }\bibfield  {title} {\bibinfo {title} {{Filling constraints for spin-orbit coupled insulators in symmorphic and nonsymmorphic crystals}},\ }\href {https://doi.org/https://doi.org/10.1073/pnas.1514665112} {\bibfield  {journal} {\bibinfo  {journal} {Proc. Natl. Acad. Sci. USA}\ }\textbf {\bibinfo {volume} {112}},\ \bibinfo {pages} {14551} (\bibinfo {year} {2015})}\BibitemShut {NoStop}%
\bibitem [{\citenamefont {Dong}\ and\ \citenamefont {Liu}(2016)}]{Dong-16}%
  \BibitemOpen
  \bibfield  {author} {\bibinfo {author} {\bibfnamefont {X.-Y.}\ \bibnamefont {Dong}}\ and\ \bibinfo {author} {\bibfnamefont {C.-X.}\ \bibnamefont {Liu}},\ }\bibfield  {title} {\bibinfo {title} {{Classification of topological crystalline insulators based on representation theory}},\ }\href {https://doi.org/10.1103/PhysRevB.93.045429} {\bibfield  {journal} {\bibinfo  {journal} {Phys. Rev. B}\ }\textbf {\bibinfo {volume} {93}},\ \bibinfo {pages} {045429} (\bibinfo {year} {2016})}\BibitemShut {NoStop}%
\bibitem [{\citenamefont {Lu}\ \emph {et~al.}(2016)\citenamefont {Lu}, \citenamefont {Fang}, \citenamefont {Fu}, \citenamefont {Johnson}, \citenamefont {Joannopoulos},\ and\ \citenamefont {Solja\v{c}i\'c}}]{Lu-16}%
  \BibitemOpen
  \bibfield  {author} {\bibinfo {author} {\bibfnamefont {L.}~\bibnamefont {Lu}}, \bibinfo {author} {\bibfnamefont {C.}~\bibnamefont {Fang}}, \bibinfo {author} {\bibfnamefont {L.}~\bibnamefont {Fu}}, \bibinfo {author} {\bibfnamefont {S.~G.}\ \bibnamefont {Johnson}}, \bibinfo {author} {\bibfnamefont {J.~D.}\ \bibnamefont {Joannopoulos}},\ and\ \bibinfo {author} {\bibfnamefont {M.}~\bibnamefont {Solja\v{c}i\'c}},\ }\bibfield  {title} {\bibinfo {title} {{Symmetry-protected topological photonic crystal in three dimensions}},\ }\href {https://doi.org/https://doi.org/10.1038/nphys3611} {\bibfield  {journal} {\bibinfo  {journal} {Nat. Phys.}\ }\textbf {\bibinfo {volume} {12}},\ \bibinfo {pages} {337} (\bibinfo {year} {2016})}\BibitemShut {NoStop}%
\bibitem [{\citenamefont {Shiozaki}\ \emph {et~al.}(2016)\citenamefont {Shiozaki}, \citenamefont {Sato},\ and\ \citenamefont {Gomi}}]{Shiozaki-PRB-2016}%
  \BibitemOpen
  \bibfield  {author} {\bibinfo {author} {\bibfnamefont {K.}~\bibnamefont {Shiozaki}}, \bibinfo {author} {\bibfnamefont {M.}~\bibnamefont {Sato}},\ and\ \bibinfo {author} {\bibfnamefont {K.}~\bibnamefont {Gomi}},\ }\bibfield  {title} {\bibinfo {title} {{Topology of nonsymmorphic crystalline insulators and superconductors}},\ }\href {https://doi.org/10.1103/PhysRevB.93.195413} {\bibfield  {journal} {\bibinfo  {journal} {Phys. Rev. B}\ }\textbf {\bibinfo {volume} {93}},\ \bibinfo {pages} {195413} (\bibinfo {year} {2016})}\BibitemShut {NoStop}%
\bibitem [{\citenamefont {Wang}\ \emph {et~al.}(2016)\citenamefont {Wang}, \citenamefont {Alexandradinata}, \citenamefont {Cava},\ and\ \citenamefont {Bernevig}}]{Wang-16}%
  \BibitemOpen
  \bibfield  {author} {\bibinfo {author} {\bibfnamefont {Z.}~\bibnamefont {Wang}}, \bibinfo {author} {\bibfnamefont {A.}~\bibnamefont {Alexandradinata}}, \bibinfo {author} {\bibfnamefont {R.~J.}\ \bibnamefont {Cava}},\ and\ \bibinfo {author} {\bibfnamefont {B.~A.}\ \bibnamefont {Bernevig}},\ }\bibfield  {title} {\bibinfo {title} {{Hourglass fermions}},\ }\href {https://doi.org/10.1038/nature17410} {\bibfield  {journal} {\bibinfo  {journal} {Nature}\ }\textbf {\bibinfo {volume} {532}},\ \bibinfo {pages} {189} (\bibinfo {year} {2016})}\BibitemShut {NoStop}%
\bibitem [{\citenamefont {Chang}\ \emph {et~al.}(2017)\citenamefont {Chang}, \citenamefont {Erten},\ and\ \citenamefont {Coleman}}]{Chang-17}%
  \BibitemOpen
  \bibfield  {author} {\bibinfo {author} {\bibfnamefont {P.-Y.}\ \bibnamefont {Chang}}, \bibinfo {author} {\bibfnamefont {O.}~\bibnamefont {Erten}},\ and\ \bibinfo {author} {\bibfnamefont {P.}~\bibnamefont {Coleman}},\ }\bibfield  {title} {\bibinfo {title} {{M\"obius Kondo insulators}},\ }\href {https://doi.org/https://doi.org/10.1038/nphys4092} {\bibfield  {journal} {\bibinfo  {journal} {Nat. Phys.}\ }\textbf {\bibinfo {volume} {13}},\ \bibinfo {pages} {794} (\bibinfo {year} {2017})}\BibitemShut {NoStop}%
\bibitem [{\citenamefont {Rudner}\ and\ \citenamefont {Levitov}(2009)}]{Rudner-09}%
  \BibitemOpen
  \bibfield  {author} {\bibinfo {author} {\bibfnamefont {M.~S.}\ \bibnamefont {Rudner}}\ and\ \bibinfo {author} {\bibfnamefont {L.~S.}\ \bibnamefont {Levitov}},\ }\bibfield  {title} {\bibinfo {title} {{Topological Transition in a Non-Hermitian Quantum Walk}},\ }\href {https://doi.org/10.1103/PhysRevLett.102.065703} {\bibfield  {journal} {\bibinfo  {journal} {Phys. Rev. Lett.}\ }\textbf {\bibinfo {volume} {102}},\ \bibinfo {pages} {065703} (\bibinfo {year} {2009})}\BibitemShut {NoStop}%
\bibitem [{\citenamefont {Sato}\ \emph {et~al.}(2012)\citenamefont {Sato}, \citenamefont {Hasebe}, \citenamefont {Esaki},\ and\ \citenamefont {Kohmoto}}]{Sato-11}%
  \BibitemOpen
  \bibfield  {author} {\bibinfo {author} {\bibfnamefont {M.}~\bibnamefont {Sato}}, \bibinfo {author} {\bibfnamefont {K.}~\bibnamefont {Hasebe}}, \bibinfo {author} {\bibfnamefont {K.}~\bibnamefont {Esaki}},\ and\ \bibinfo {author} {\bibfnamefont {M.}~\bibnamefont {Kohmoto}},\ }\bibfield  {title} {\bibinfo {title} {{Time-Reversal Symmetry in Non-Hermitian Systems}},\ }\href {https://doi.org/10.1143/PTP.127.937} {\bibfield  {journal} {\bibinfo  {journal} {Prog. Theor. Phys.}\ }\textbf {\bibinfo {volume} {127}},\ \bibinfo {pages} {937} (\bibinfo {year} {2012})}\BibitemShut {NoStop}%
\bibitem [{\citenamefont {Esaki}\ \emph {et~al.}(2011)\citenamefont {Esaki}, \citenamefont {Sato}, \citenamefont {Hasebe},\ and\ \citenamefont {Kohmoto}}]{Esaki-11}%
  \BibitemOpen
  \bibfield  {author} {\bibinfo {author} {\bibfnamefont {K.}~\bibnamefont {Esaki}}, \bibinfo {author} {\bibfnamefont {M.}~\bibnamefont {Sato}}, \bibinfo {author} {\bibfnamefont {K.}~\bibnamefont {Hasebe}},\ and\ \bibinfo {author} {\bibfnamefont {M.}~\bibnamefont {Kohmoto}},\ }\bibfield  {title} {\bibinfo {title} {{Edge states and topological phases in non-Hermitian systems}},\ }\href {https://doi.org/10.1103/PhysRevB.84.205128} {\bibfield  {journal} {\bibinfo  {journal} {Phys. Rev. B}\ }\textbf {\bibinfo {volume} {84}},\ \bibinfo {pages} {205128} (\bibinfo {year} {2011})}\BibitemShut {NoStop}%
\bibitem [{\citenamefont {Hu}\ and\ \citenamefont {Hughes}(2011)}]{Hu-11}%
  \BibitemOpen
  \bibfield  {author} {\bibinfo {author} {\bibfnamefont {Y.~C.}\ \bibnamefont {Hu}}\ and\ \bibinfo {author} {\bibfnamefont {T.~L.}\ \bibnamefont {Hughes}},\ }\bibfield  {title} {\bibinfo {title} {{Absence of topological insulator phases in non-Hermitian $\textit{PT}$-symmetric Hamiltonians}},\ }\href {https://doi.org/10.1103/PhysRevB.84.153101} {\bibfield  {journal} {\bibinfo  {journal} {Phys. Rev. B}\ }\textbf {\bibinfo {volume} {84}},\ \bibinfo {pages} {153101} (\bibinfo {year} {2011})}\BibitemShut {NoStop}%
\bibitem [{\citenamefont {Schomerus}(2013)}]{Schomerus-13}%
  \BibitemOpen
  \bibfield  {author} {\bibinfo {author} {\bibfnamefont {H.}~\bibnamefont {Schomerus}},\ }\bibfield  {title} {\bibinfo {title} {{Topologically protected midgap states in complex photonic lattices}},\ }\href {https://doi.org/10.1364/OL.38.001912} {\bibfield  {journal} {\bibinfo  {journal} {Opt. Lett.}\ }\textbf {\bibinfo {volume} {38}},\ \bibinfo {pages} {1912} (\bibinfo {year} {2013})}\BibitemShut {NoStop}%
\bibitem [{\citenamefont {Longhi}\ \emph {et~al.}(2015)\citenamefont {Longhi}, \citenamefont {Gatti},\ and\ \citenamefont {Valle}}]{Longhi-15}%
  \BibitemOpen
  \bibfield  {author} {\bibinfo {author} {\bibfnamefont {S.}~\bibnamefont {Longhi}}, \bibinfo {author} {\bibfnamefont {D.}~\bibnamefont {Gatti}},\ and\ \bibinfo {author} {\bibfnamefont {G.~D.}\ \bibnamefont {Valle}},\ }\bibfield  {title} {\bibinfo {title} {{Robust light transport in non-Hermitian photonic lattices}},\ }\href {https://doi.org/10.1038/srep13376} {\bibfield  {journal} {\bibinfo  {journal} {Sci. Rep.}\ }\textbf {\bibinfo {volume} {5}},\ \bibinfo {pages} {13376} (\bibinfo {year} {2015})}\BibitemShut {NoStop}%
\bibitem [{\citenamefont {Lee}(2016)}]{Lee-16}%
  \BibitemOpen
  \bibfield  {author} {\bibinfo {author} {\bibfnamefont {T.~E.}\ \bibnamefont {Lee}},\ }\bibfield  {title} {\bibinfo {title} {{Anomalous Edge State in a Non-Hermitian Lattice}},\ }\href {https://doi.org/10.1103/PhysRevLett.116.133903} {\bibfield  {journal} {\bibinfo  {journal} {Phys. Rev. Lett.}\ }\textbf {\bibinfo {volume} {116}},\ \bibinfo {pages} {133903} (\bibinfo {year} {2016})}\BibitemShut {NoStop}%
\bibitem [{\citenamefont {Leykam}\ \emph {et~al.}(2017)\citenamefont {Leykam}, \citenamefont {Bliokh}, \citenamefont {Huang}, \citenamefont {Chong},\ and\ \citenamefont {Nori}}]{Leykam-17}%
  \BibitemOpen
  \bibfield  {author} {\bibinfo {author} {\bibfnamefont {D.}~\bibnamefont {Leykam}}, \bibinfo {author} {\bibfnamefont {K.~Y.}\ \bibnamefont {Bliokh}}, \bibinfo {author} {\bibfnamefont {C.}~\bibnamefont {Huang}}, \bibinfo {author} {\bibfnamefont {Y.~D.}\ \bibnamefont {Chong}},\ and\ \bibinfo {author} {\bibfnamefont {F.}~\bibnamefont {Nori}},\ }\bibfield  {title} {\bibinfo {title} {{Edge Modes, Degeneracies, and Topological Numbers in Non-Hermitian Systems}},\ }\href {https://doi.org/10.1103/PhysRevLett.118.040401} {\bibfield  {journal} {\bibinfo  {journal} {Phys. Rev. Lett.}\ }\textbf {\bibinfo {volume} {118}},\ \bibinfo {pages} {040401} (\bibinfo {year} {2017})}\BibitemShut {NoStop}%
\bibitem [{\citenamefont {Xu}\ \emph {et~al.}(2017)\citenamefont {Xu}, \citenamefont {Wang},\ and\ \citenamefont {Duan}}]{Xu-17}%
  \BibitemOpen
  \bibfield  {author} {\bibinfo {author} {\bibfnamefont {Y.}~\bibnamefont {Xu}}, \bibinfo {author} {\bibfnamefont {S.-T.}\ \bibnamefont {Wang}},\ and\ \bibinfo {author} {\bibfnamefont {L.-M.}\ \bibnamefont {Duan}},\ }\bibfield  {title} {\bibinfo {title} {{Weyl Exceptional Rings in a Three-Dimensional Dissipative Cold Atomic Gas}},\ }\href {https://doi.org/10.1103/PhysRevLett.118.045701} {\bibfield  {journal} {\bibinfo  {journal} {Phys. Rev. Lett.}\ }\textbf {\bibinfo {volume} {118}},\ \bibinfo {pages} {045701} (\bibinfo {year} {2017})}\BibitemShut {NoStop}%
\bibitem [{\citenamefont {Shen}\ \emph {et~al.}(2018)\citenamefont {Shen}, \citenamefont {Zhen},\ and\ \citenamefont {Fu}}]{Shen-18}%
  \BibitemOpen
  \bibfield  {author} {\bibinfo {author} {\bibfnamefont {H.}~\bibnamefont {Shen}}, \bibinfo {author} {\bibfnamefont {B.}~\bibnamefont {Zhen}},\ and\ \bibinfo {author} {\bibfnamefont {L.}~\bibnamefont {Fu}},\ }\bibfield  {title} {\bibinfo {title} {{Topological Band Theory for Non-Hermitian Hamiltonians}},\ }\href {https://doi.org/10.1103/PhysRevLett.120.146402} {\bibfield  {journal} {\bibinfo  {journal} {Phys. Rev. Lett.}\ }\textbf {\bibinfo {volume} {120}},\ \bibinfo {pages} {146402} (\bibinfo {year} {2018})}\BibitemShut {NoStop}%
\bibitem [{\citenamefont {Kozii}\ and\ \citenamefont {Fu}(2024)}]{Kozii-17}%
  \BibitemOpen
  \bibfield  {author} {\bibinfo {author} {\bibfnamefont {V.}~\bibnamefont {Kozii}}\ and\ \bibinfo {author} {\bibfnamefont {L.}~\bibnamefont {Fu}},\ }\bibfield  {title} {\bibinfo {title} {{Non-Hermitian topological theory of finite-lifetime quasiparticles: Prediction of bulk Fermi arc due to exceptional point}},\ }\href {https://doi.org/10.1103/PhysRevB.109.235139} {\bibfield  {journal} {\bibinfo  {journal} {Phys. Rev. B}\ }\textbf {\bibinfo {volume} {109}},\ \bibinfo {pages} {235139} (\bibinfo {year} {2024})}\BibitemShut {NoStop}%
\bibitem [{\citenamefont {Martinez~Alvarez}\ \emph {et~al.}(2018)\citenamefont {Martinez~Alvarez}, \citenamefont {Barrios~Vargas},\ and\ \citenamefont {Foa~Torres}}]{MartinezAlvarez-18}%
  \BibitemOpen
  \bibfield  {author} {\bibinfo {author} {\bibfnamefont {V.~M.}\ \bibnamefont {Martinez~Alvarez}}, \bibinfo {author} {\bibfnamefont {J.~E.}\ \bibnamefont {Barrios~Vargas}},\ and\ \bibinfo {author} {\bibfnamefont {L.~E.~F.}\ \bibnamefont {Foa~Torres}},\ }\bibfield  {title} {\bibinfo {title} {{Non-Hermitian robust edge states in one dimension: Anomalous localization and eigenspace condensation at exceptional points}},\ }\href {https://doi.org/10.1103/PhysRevB.97.121401} {\bibfield  {journal} {\bibinfo  {journal} {Phys. Rev. B}\ }\textbf {\bibinfo {volume} {97}},\ \bibinfo {pages} {121401(R)} (\bibinfo {year} {2018})}\BibitemShut {NoStop}%
\bibitem [{\citenamefont {Gong}\ \emph {et~al.}(2018)\citenamefont {Gong}, \citenamefont {Ashida}, \citenamefont {Kawabata}, \citenamefont {Takasan}, \citenamefont {Higashikawa},\ and\ \citenamefont {Ueda}}]{Gong-18}%
  \BibitemOpen
  \bibfield  {author} {\bibinfo {author} {\bibfnamefont {Z.}~\bibnamefont {Gong}}, \bibinfo {author} {\bibfnamefont {Y.}~\bibnamefont {Ashida}}, \bibinfo {author} {\bibfnamefont {K.}~\bibnamefont {Kawabata}}, \bibinfo {author} {\bibfnamefont {K.}~\bibnamefont {Takasan}}, \bibinfo {author} {\bibfnamefont {S.}~\bibnamefont {Higashikawa}},\ and\ \bibinfo {author} {\bibfnamefont {M.}~\bibnamefont {Ueda}},\ }\bibfield  {title} {\bibinfo {title} {{Topological Phases of Non-Hermitian Systems}},\ }\href {https://doi.org/10.1103/PhysRevX.8.031079} {\bibfield  {journal} {\bibinfo  {journal} {Phys. Rev. X}\ }\textbf {\bibinfo {volume} {8}},\ \bibinfo {pages} {031079} (\bibinfo {year} {2018})}\BibitemShut {NoStop}%
\bibitem [{\citenamefont {Yao}\ and\ \citenamefont {Wang}(2018)}]{YW-18-SSH}%
  \BibitemOpen
  \bibfield  {author} {\bibinfo {author} {\bibfnamefont {S.}~\bibnamefont {Yao}}\ and\ \bibinfo {author} {\bibfnamefont {Z.}~\bibnamefont {Wang}},\ }\bibfield  {title} {\bibinfo {title} {{Edge States and Topological Invariants of Non-Hermitian Systems}},\ }\href {https://doi.org/10.1103/PhysRevLett.121.086803} {\bibfield  {journal} {\bibinfo  {journal} {Phys. Rev. Lett.}\ }\textbf {\bibinfo {volume} {121}},\ \bibinfo {pages} {086803} (\bibinfo {year} {2018})}\BibitemShut {NoStop}%
\bibitem [{\citenamefont {Yao}\ \emph {et~al.}(2018)\citenamefont {Yao}, \citenamefont {Song},\ and\ \citenamefont {Wang}}]{YSW-18-Chern}%
  \BibitemOpen
  \bibfield  {author} {\bibinfo {author} {\bibfnamefont {S.}~\bibnamefont {Yao}}, \bibinfo {author} {\bibfnamefont {F.}~\bibnamefont {Song}},\ and\ \bibinfo {author} {\bibfnamefont {Z.}~\bibnamefont {Wang}},\ }\bibfield  {title} {\bibinfo {title} {{Non-Hermitian Chern Bands}},\ }\href {https://doi.org/10.1103/PhysRevLett.121.136802} {\bibfield  {journal} {\bibinfo  {journal} {{Phys. Rev. Lett.}}\ }\textbf {\bibinfo {volume} {121}},\ \bibinfo {pages} {136802} (\bibinfo {year} {2018})}\BibitemShut {NoStop}%
\bibitem [{\citenamefont {Kunst}\ \emph {et~al.}(2018)\citenamefont {Kunst}, \citenamefont {Edvardsson}, \citenamefont {Budich},\ and\ \citenamefont {Bergholtz}}]{Kunst-18}%
  \BibitemOpen
  \bibfield  {author} {\bibinfo {author} {\bibfnamefont {F.~K.}\ \bibnamefont {Kunst}}, \bibinfo {author} {\bibfnamefont {E.}~\bibnamefont {Edvardsson}}, \bibinfo {author} {\bibfnamefont {J.~C.}\ \bibnamefont {Budich}},\ and\ \bibinfo {author} {\bibfnamefont {E.~J.}\ \bibnamefont {Bergholtz}},\ }\bibfield  {title} {\bibinfo {title} {{Biorthogonal Bulk-Boundary Correspondence in Non-Hermitian Systems}},\ }\href {https://doi.org/10.1103/PhysRevLett.121.026808} {\bibfield  {journal} {\bibinfo  {journal} {Phys. Rev. Lett.}\ }\textbf {\bibinfo {volume} {121}},\ \bibinfo {pages} {026808} (\bibinfo {year} {2018})}\BibitemShut {NoStop}%
\bibitem [{\citenamefont {McDonald}\ \emph {et~al.}(2018)\citenamefont {McDonald}, \citenamefont {Pereg-Barnea},\ and\ \citenamefont {Clerk}}]{McDonald-18}%
  \BibitemOpen
  \bibfield  {author} {\bibinfo {author} {\bibfnamefont {A.}~\bibnamefont {McDonald}}, \bibinfo {author} {\bibfnamefont {T.}~\bibnamefont {Pereg-Barnea}},\ and\ \bibinfo {author} {\bibfnamefont {A.~A.}\ \bibnamefont {Clerk}},\ }\bibfield  {title} {\bibinfo {title} {{Phase-Dependent Chiral Transport and Effective Non-Hermitian Dynamics in a Bosonic Kitaev-Majorana Chain}},\ }\href {https://doi.org/10.1103/PhysRevX.8.041031} {\bibfield  {journal} {\bibinfo  {journal} {Phys. Rev. X}\ }\textbf {\bibinfo {volume} {8}},\ \bibinfo {pages} {041031} (\bibinfo {year} {2018})}\BibitemShut {NoStop}%
\bibitem [{\citenamefont {Lee}\ and\ \citenamefont {Thomale}(2019)}]{Lee-Thomale-19}%
  \BibitemOpen
  \bibfield  {author} {\bibinfo {author} {\bibfnamefont {C.~H.}\ \bibnamefont {Lee}}\ and\ \bibinfo {author} {\bibfnamefont {R.}~\bibnamefont {Thomale}},\ }\bibfield  {title} {\bibinfo {title} {{Anatomy of skin modes and topology in non-Hermitian systems}},\ }\href {https://doi.org/10.1103/PhysRevB.99.201103} {\bibfield  {journal} {\bibinfo  {journal} {Phys. Rev. B}\ }\textbf {\bibinfo {volume} {99}},\ \bibinfo {pages} {201103(R)} (\bibinfo {year} {2019})}\BibitemShut {NoStop}%
\bibitem [{\citenamefont {Liu}\ \emph {et~al.}(2019)\citenamefont {Liu}, \citenamefont {Zhang}, \citenamefont {Ai}, \citenamefont {Gong}, \citenamefont {Kawabata}, \citenamefont {Ueda},\ and\ \citenamefont {Nori}}]{Liu-19}%
  \BibitemOpen
  \bibfield  {author} {\bibinfo {author} {\bibfnamefont {T.}~\bibnamefont {Liu}}, \bibinfo {author} {\bibfnamefont {Y.-R.}\ \bibnamefont {Zhang}}, \bibinfo {author} {\bibfnamefont {Q.}~\bibnamefont {Ai}}, \bibinfo {author} {\bibfnamefont {Z.}~\bibnamefont {Gong}}, \bibinfo {author} {\bibfnamefont {K.}~\bibnamefont {Kawabata}}, \bibinfo {author} {\bibfnamefont {M.}~\bibnamefont {Ueda}},\ and\ \bibinfo {author} {\bibfnamefont {F.}~\bibnamefont {Nori}},\ }\bibfield  {title} {\bibinfo {title} {{Second-Order Topological Phases in Non-Hermitian Systems}},\ }\href {https://doi.org/10.1103/PhysRevLett.122.076801} {\bibfield  {journal} {\bibinfo  {journal} {Phys. Rev. Lett.}\ }\textbf {\bibinfo {volume} {122}},\ \bibinfo {pages} {076801} (\bibinfo {year} {2019})}\BibitemShut {NoStop}%
\bibitem [{\citenamefont {Lee}\ \emph {et~al.}(2019)\citenamefont {Lee}, \citenamefont {Li},\ and\ \citenamefont {Gong}}]{Lee-Li-Gong-19}%
  \BibitemOpen
  \bibfield  {author} {\bibinfo {author} {\bibfnamefont {C.~H.}\ \bibnamefont {Lee}}, \bibinfo {author} {\bibfnamefont {L.}~\bibnamefont {Li}},\ and\ \bibinfo {author} {\bibfnamefont {J.}~\bibnamefont {Gong}},\ }\bibfield  {title} {\bibinfo {title} {{Hybrid Higher-Order Skin-Topological Modes in Nonreciprocal Systems}},\ }\href {https://doi.org/10.1103/PhysRevLett.123.016805} {\bibfield  {journal} {\bibinfo  {journal} {Phys. Rev. Lett.}\ }\textbf {\bibinfo {volume} {123}},\ \bibinfo {pages} {016805} (\bibinfo {year} {2019})}\BibitemShut {NoStop}%
\bibitem [{\citenamefont {Kawabata}\ \emph {et~al.}(2019)\citenamefont {Kawabata}, \citenamefont {Shiozaki}, \citenamefont {Ueda},\ and\ \citenamefont {Sato}}]{KSUS-19}%
  \BibitemOpen
  \bibfield  {author} {\bibinfo {author} {\bibfnamefont {K.}~\bibnamefont {Kawabata}}, \bibinfo {author} {\bibfnamefont {K.}~\bibnamefont {Shiozaki}}, \bibinfo {author} {\bibfnamefont {M.}~\bibnamefont {Ueda}},\ and\ \bibinfo {author} {\bibfnamefont {M.}~\bibnamefont {Sato}},\ }\bibfield  {title} {\bibinfo {title} {{Symmetry and Topology in Non-Hermitian Physics}},\ }\href {https://doi.org/10.1103/PhysRevX.9.041015} {\bibfield  {journal} {\bibinfo  {journal} {Phys. Rev. X}\ }\textbf {\bibinfo {volume} {9}},\ \bibinfo {pages} {041015} (\bibinfo {year} {2019})}\BibitemShut {NoStop}%
\bibitem [{\citenamefont {Zhou}\ and\ \citenamefont {Lee}(2019)}]{ZL-19}%
  \BibitemOpen
  \bibfield  {author} {\bibinfo {author} {\bibfnamefont {H.}~\bibnamefont {Zhou}}\ and\ \bibinfo {author} {\bibfnamefont {J.~Y.}\ \bibnamefont {Lee}},\ }\bibfield  {title} {\bibinfo {title} {{Periodic table for topological bands with non-Hermitian symmetries}},\ }\href {https://doi.org/10.1103/PhysRevB.99.235112} {\bibfield  {journal} {\bibinfo  {journal} {Phys. Rev. B}\ }\textbf {\bibinfo {volume} {99}},\ \bibinfo {pages} {235112} (\bibinfo {year} {2019})}\BibitemShut {NoStop}%
\bibitem [{\citenamefont {Herviou}\ \emph {et~al.}(2019)\citenamefont {Herviou}, \citenamefont {Bardarson},\ and\ \citenamefont {Regnault}}]{Herviou-19}%
  \BibitemOpen
  \bibfield  {author} {\bibinfo {author} {\bibfnamefont {L.}~\bibnamefont {Herviou}}, \bibinfo {author} {\bibfnamefont {J.~H.}\ \bibnamefont {Bardarson}},\ and\ \bibinfo {author} {\bibfnamefont {N.}~\bibnamefont {Regnault}},\ }\bibfield  {title} {\bibinfo {title} {{Defining a bulk-edge correspondence for non-Hermitian Hamiltonians via singular-value decomposition}},\ }\href {https://doi.org/10.1103/PhysRevA.99.052118} {\bibfield  {journal} {\bibinfo  {journal} {Phys. Rev. A}\ }\textbf {\bibinfo {volume} {99}},\ \bibinfo {pages} {052118} (\bibinfo {year} {2019})}\BibitemShut {NoStop}%
\bibitem [{\citenamefont {Zirnstein}\ \emph {et~al.}(2021)\citenamefont {Zirnstein}, \citenamefont {Refael},\ and\ \citenamefont {Rosenow}}]{Zirnstein-19}%
  \BibitemOpen
  \bibfield  {author} {\bibinfo {author} {\bibfnamefont {H.-G.}\ \bibnamefont {Zirnstein}}, \bibinfo {author} {\bibfnamefont {G.}~\bibnamefont {Refael}},\ and\ \bibinfo {author} {\bibfnamefont {B.}~\bibnamefont {Rosenow}},\ }\bibfield  {title} {\bibinfo {title} {{Bulk-Boundary Correspondence for Non-Hermitian Hamiltonians via Green Functions}},\ }\href {https://doi.org/10.1103/PhysRevLett.126.216407} {\bibfield  {journal} {\bibinfo  {journal} {Phys. Rev. Lett.}\ }\textbf {\bibinfo {volume} {126}},\ \bibinfo {pages} {216407} (\bibinfo {year} {2021})}\BibitemShut {NoStop}%
\bibitem [{\citenamefont {Borgnia}\ \emph {et~al.}(2020)\citenamefont {Borgnia}, \citenamefont {Kruchkov},\ and\ \citenamefont {Slager}}]{Borgnia-19}%
  \BibitemOpen
  \bibfield  {author} {\bibinfo {author} {\bibfnamefont {D.~S.}\ \bibnamefont {Borgnia}}, \bibinfo {author} {\bibfnamefont {A.~J.}\ \bibnamefont {Kruchkov}},\ and\ \bibinfo {author} {\bibfnamefont {R.-J.}\ \bibnamefont {Slager}},\ }\bibfield  {title} {\bibinfo {title} {{Non-Hermitian Boundary Modes and Topology}},\ }\href {https://doi.org/10.1103/PhysRevLett.124.056802} {\bibfield  {journal} {\bibinfo  {journal} {Phys. Rev. Lett.}\ }\textbf {\bibinfo {volume} {124}},\ \bibinfo {pages} {056802} (\bibinfo {year} {2020})}\BibitemShut {NoStop}%
\bibitem [{\citenamefont {Yokomizo}\ and\ \citenamefont {Murakami}(2019)}]{Yokomizo-19}%
  \BibitemOpen
  \bibfield  {author} {\bibinfo {author} {\bibfnamefont {K.}~\bibnamefont {Yokomizo}}\ and\ \bibinfo {author} {\bibfnamefont {S.}~\bibnamefont {Murakami}},\ }\bibfield  {title} {\bibinfo {title} {{Non-Bloch Band Theory of Non-Hermitian Systems}},\ }\href {https://doi.org/10.1103/PhysRevLett.123.066404} {\bibfield  {journal} {\bibinfo  {journal} {Phys. Rev. Lett.}\ }\textbf {\bibinfo {volume} {123}},\ \bibinfo {pages} {066404} (\bibinfo {year} {2019})}\BibitemShut {NoStop}%
\bibitem [{\citenamefont {Zhang}\ \emph {et~al.}(2020)\citenamefont {Zhang}, \citenamefont {Yang},\ and\ \citenamefont {Fang}}]{Zhang-20}%
  \BibitemOpen
  \bibfield  {author} {\bibinfo {author} {\bibfnamefont {K.}~\bibnamefont {Zhang}}, \bibinfo {author} {\bibfnamefont {Z.}~\bibnamefont {Yang}},\ and\ \bibinfo {author} {\bibfnamefont {C.}~\bibnamefont {Fang}},\ }\bibfield  {title} {\bibinfo {title} {{Correspondence between Winding Numbers and Skin Modes in Non-Hermitian Systems}},\ }\href {https://doi.org/10.1103/PhysRevLett.125.126402} {\bibfield  {journal} {\bibinfo  {journal} {Phys. Rev. Lett.}\ }\textbf {\bibinfo {volume} {125}},\ \bibinfo {pages} {126402} (\bibinfo {year} {2020})}\BibitemShut {NoStop}%
\bibitem [{\citenamefont {Okuma}\ \emph {et~al.}(2020)\citenamefont {Okuma}, \citenamefont {Kawabata}, \citenamefont {Shiozaki},\ and\ \citenamefont {Sato}}]{OKSS-20}%
  \BibitemOpen
  \bibfield  {author} {\bibinfo {author} {\bibfnamefont {N.}~\bibnamefont {Okuma}}, \bibinfo {author} {\bibfnamefont {K.}~\bibnamefont {Kawabata}}, \bibinfo {author} {\bibfnamefont {K.}~\bibnamefont {Shiozaki}},\ and\ \bibinfo {author} {\bibfnamefont {M.}~\bibnamefont {Sato}},\ }\bibfield  {title} {\bibinfo {title} {{Topological Origin of Non-Hermitian Skin Effects}},\ }\href {https://doi.org/10.1103/PhysRevLett.124.086801} {\bibfield  {journal} {\bibinfo  {journal} {Phys. Rev. Lett.}\ }\textbf {\bibinfo {volume} {124}},\ \bibinfo {pages} {086801} (\bibinfo {year} {2020})}\BibitemShut {NoStop}%
\bibitem [{\citenamefont {Kawabata}\ \emph {et~al.}(2020{\natexlab{a}})\citenamefont {Kawabata}, \citenamefont {Okuma},\ and\ \citenamefont {Sato}}]{KOS-20}%
  \BibitemOpen
  \bibfield  {author} {\bibinfo {author} {\bibfnamefont {K.}~\bibnamefont {Kawabata}}, \bibinfo {author} {\bibfnamefont {N.}~\bibnamefont {Okuma}},\ and\ \bibinfo {author} {\bibfnamefont {M.}~\bibnamefont {Sato}},\ }\bibfield  {title} {\bibinfo {title} {{Non-Bloch band theory of non-Hermitian Hamiltonians in the symplectic class}},\ }\href {https://doi.org/10.1103/PhysRevB.101.195147} {\bibfield  {journal} {\bibinfo  {journal} {Phys. Rev. B}\ }\textbf {\bibinfo {volume} {101}},\ \bibinfo {pages} {195147} (\bibinfo {year} {2020}{\natexlab{a}})}\BibitemShut {NoStop}%
\bibitem [{\citenamefont {Yi}\ and\ \citenamefont {Yang}(2020)}]{Yi-Yang-20}%
  \BibitemOpen
  \bibfield  {author} {\bibinfo {author} {\bibfnamefont {Y.}~\bibnamefont {Yi}}\ and\ \bibinfo {author} {\bibfnamefont {Z.}~\bibnamefont {Yang}},\ }\bibfield  {title} {\bibinfo {title} {{Non-Hermitian Skin Modes Induced by On-Site Dissipations and Chiral Tunneling Effect}},\ }\href {https://doi.org/10.1103/PhysRevLett.125.186802} {\bibfield  {journal} {\bibinfo  {journal} {Phys. Rev. Lett.}\ }\textbf {\bibinfo {volume} {125}},\ \bibinfo {pages} {186802} (\bibinfo {year} {2020})}\BibitemShut {NoStop}%
\bibitem [{\citenamefont {Bessho}\ and\ \citenamefont {Sato}(2021)}]{Bessho-21}%
  \BibitemOpen
  \bibfield  {author} {\bibinfo {author} {\bibfnamefont {T.}~\bibnamefont {Bessho}}\ and\ \bibinfo {author} {\bibfnamefont {M.}~\bibnamefont {Sato}},\ }\bibfield  {title} {\bibinfo {title} {{Nielsen-Ninomiya Theorem with Bulk Topology: Duality in Floquet and Non-Hermitian Systems}},\ }\href {https://doi.org/10.1103/PhysRevLett.127.196404} {\bibfield  {journal} {\bibinfo  {journal} {Phys. Rev. Lett.}\ }\textbf {\bibinfo {volume} {127}},\ \bibinfo {pages} {196404} (\bibinfo {year} {2021})}\BibitemShut {NoStop}%
\bibitem [{\citenamefont {Denner}\ \emph {et~al.}(2021)\citenamefont {Denner}, \citenamefont {Skurativska}, \citenamefont {Schindler}, \citenamefont {Fischer}, \citenamefont {Thomale}, \citenamefont {Bzdu\v{s}ek},\ and\ \citenamefont {Neupert}}]{Denner-21}%
  \BibitemOpen
  \bibfield  {author} {\bibinfo {author} {\bibfnamefont {M.~M.}\ \bibnamefont {Denner}}, \bibinfo {author} {\bibfnamefont {A.}~\bibnamefont {Skurativska}}, \bibinfo {author} {\bibfnamefont {F.}~\bibnamefont {Schindler}}, \bibinfo {author} {\bibfnamefont {M.~H.}\ \bibnamefont {Fischer}}, \bibinfo {author} {\bibfnamefont {R.}~\bibnamefont {Thomale}}, \bibinfo {author} {\bibfnamefont {T.}~\bibnamefont {Bzdu\v{s}ek}},\ and\ \bibinfo {author} {\bibfnamefont {T.}~\bibnamefont {Neupert}},\ }\bibfield  {title} {\bibinfo {title} {{Exceptional topological insulators}},\ }\href {https://doi.org/10.1038/s41467-021-25947-z} {\bibfield  {journal} {\bibinfo  {journal} {Nat. Commun.}\ }\textbf {\bibinfo {volume} {12}},\ \bibinfo {pages} {5681} (\bibinfo {year} {2021})}\BibitemShut {NoStop}%
\bibitem [{\citenamefont {Denner}\ \emph {et~al.}(2023)\citenamefont {Denner}, \citenamefont {Neupert},\ and\ \citenamefont {Schindler}}]{Denner-23JPhysMater}%
  \BibitemOpen
  \bibfield  {author} {\bibinfo {author} {\bibfnamefont {M.~M.}\ \bibnamefont {Denner}}, \bibinfo {author} {\bibfnamefont {T.}~\bibnamefont {Neupert}},\ and\ \bibinfo {author} {\bibfnamefont {F.}~\bibnamefont {Schindler}},\ }\bibfield  {title} {\bibinfo {title} {{Infernal and exceptional edge modes: non-Hermitian topology beyond the skin effect}},\ }\href {https://doi.org/10.1088/2515-7639/acf2ca} {\bibfield  {journal} {\bibinfo  {journal} {J. Phys. Mater.}\ }\textbf {\bibinfo {volume} {6}},\ \bibinfo {pages} {045006} (\bibinfo {year} {2023})}\BibitemShut {NoStop}%
\bibitem [{\citenamefont {Kawabata}\ \emph {et~al.}(2021)\citenamefont {Kawabata}, \citenamefont {Shiozaki},\ and\ \citenamefont {Ryu}}]{KSR-21}%
  \BibitemOpen
  \bibfield  {author} {\bibinfo {author} {\bibfnamefont {K.}~\bibnamefont {Kawabata}}, \bibinfo {author} {\bibfnamefont {K.}~\bibnamefont {Shiozaki}},\ and\ \bibinfo {author} {\bibfnamefont {S.}~\bibnamefont {Ryu}},\ }\bibfield  {title} {\bibinfo {title} {{Topological Field Theory of Non-Hermitian Systems}},\ }\href {https://doi.org/10.1103/PhysRevLett.126.216405} {\bibfield  {journal} {\bibinfo  {journal} {Phys. Rev. Lett.}\ }\textbf {\bibinfo {volume} {126}},\ \bibinfo {pages} {216405} (\bibinfo {year} {2021})}\BibitemShut {NoStop}%
\bibitem [{\citenamefont {Zhang}\ \emph {et~al.}(2022)\citenamefont {Zhang}, \citenamefont {Yang},\ and\ \citenamefont {Fang}}]{Zhang-22}%
  \BibitemOpen
  \bibfield  {author} {\bibinfo {author} {\bibfnamefont {K.}~\bibnamefont {Zhang}}, \bibinfo {author} {\bibfnamefont {Z.}~\bibnamefont {Yang}},\ and\ \bibinfo {author} {\bibfnamefont {C.}~\bibnamefont {Fang}},\ }\bibfield  {title} {\bibinfo {title} {{Universal non-Hermitian skin effect in two and higher dimensions}},\ }\href {https://doi.org/10.1038/s41467-022-30161-6} {\bibfield  {journal} {\bibinfo  {journal} {Nat. Commun.}\ }\textbf {\bibinfo {volume} {13}},\ \bibinfo {pages} {2496} (\bibinfo {year} {2022})}\BibitemShut {NoStop}%
\bibitem [{\citenamefont {Sun}\ \emph {et~al.}(2021)\citenamefont {Sun}, \citenamefont {Zhu},\ and\ \citenamefont {Hughes}}]{Sun-21}%
  \BibitemOpen
  \bibfield  {author} {\bibinfo {author} {\bibfnamefont {X.-Q.}\ \bibnamefont {Sun}}, \bibinfo {author} {\bibfnamefont {P.}~\bibnamefont {Zhu}},\ and\ \bibinfo {author} {\bibfnamefont {T.~L.}\ \bibnamefont {Hughes}},\ }\bibfield  {title} {\bibinfo {title} {{Geometric Response and Disclination-Induced Skin Effects in Non-Hermitian Systems}},\ }\href {https://doi.org/10.1103/PhysRevLett.127.066401} {\bibfield  {journal} {\bibinfo  {journal} {Phys. Rev. Lett.}\ }\textbf {\bibinfo {volume} {127}},\ \bibinfo {pages} {066401} (\bibinfo {year} {2021})}\BibitemShut {NoStop}%
\bibitem [{\citenamefont {Nakamura}\ \emph {et~al.}(2024)\citenamefont {Nakamura}, \citenamefont {Bessho},\ and\ \citenamefont {Sato}}]{Nakamura-24}%
  \BibitemOpen
  \bibfield  {author} {\bibinfo {author} {\bibfnamefont {D.}~\bibnamefont {Nakamura}}, \bibinfo {author} {\bibfnamefont {T.}~\bibnamefont {Bessho}},\ and\ \bibinfo {author} {\bibfnamefont {M.}~\bibnamefont {Sato}},\ }\bibfield  {title} {\bibinfo {title} {{Bulk-Boundary Correspondence in Point-Gap Topological Phases}},\ }\href {https://doi.org/10.1103/PhysRevLett.132.136401} {\bibfield  {journal} {\bibinfo  {journal} {Phys. Rev. Lett.}\ }\textbf {\bibinfo {volume} {132}},\ \bibinfo {pages} {136401} (\bibinfo {year} {2024})}\BibitemShut {NoStop}%
\bibitem [{\citenamefont {Wang}\ \emph {et~al.}(2024)\citenamefont {Wang}, \citenamefont {Song},\ and\ \citenamefont {Wang}}]{Wang-24}%
  \BibitemOpen
  \bibfield  {author} {\bibinfo {author} {\bibfnamefont {H.-Y.}\ \bibnamefont {Wang}}, \bibinfo {author} {\bibfnamefont {F.}~\bibnamefont {Song}},\ and\ \bibinfo {author} {\bibfnamefont {Z.}~\bibnamefont {Wang}},\ }\bibfield  {title} {\bibinfo {title} {{Amoeba Formulation of Non-Bloch Band Theory in Arbitrary Dimensions}},\ }\href {https://doi.org/10.1103/PhysRevX.14.021011} {\bibfield  {journal} {\bibinfo  {journal} {Phys. Rev. X}\ }\textbf {\bibinfo {volume} {14}},\ \bibinfo {pages} {021011} (\bibinfo {year} {2024})}\BibitemShut {NoStop}%
\bibitem [{\citenamefont {Nakai}\ \emph {et~al.}(2024)\citenamefont {Nakai}, \citenamefont {Okuma}, \citenamefont {Nakamura}, \citenamefont {Shimomura},\ and\ \citenamefont {Sato}}]{Nakai-24}%
  \BibitemOpen
  \bibfield  {author} {\bibinfo {author} {\bibfnamefont {Y.~O.}\ \bibnamefont {Nakai}}, \bibinfo {author} {\bibfnamefont {N.}~\bibnamefont {Okuma}}, \bibinfo {author} {\bibfnamefont {D.}~\bibnamefont {Nakamura}}, \bibinfo {author} {\bibfnamefont {K.}~\bibnamefont {Shimomura}},\ and\ \bibinfo {author} {\bibfnamefont {M.}~\bibnamefont {Sato}},\ }\bibfield  {title} {\bibinfo {title} {{Topological enhancement of nonnormality in non-Hermitian skin effects}},\ }\href {https://doi.org/10.1103/PhysRevB.109.144203} {\bibfield  {journal} {\bibinfo  {journal} {Phys. Rev. B}\ }\textbf {\bibinfo {volume} {109}},\ \bibinfo {pages} {144203} (\bibinfo {year} {2024})}\BibitemShut {NoStop}%
\bibitem [{\citenamefont {Nakamura}\ \emph {et~al.}(2023)\citenamefont {Nakamura}, \citenamefont {Inaka}, \citenamefont {Okuma},\ and\ \citenamefont {Sato}}]{Nakamura-23}%
  \BibitemOpen
  \bibfield  {author} {\bibinfo {author} {\bibfnamefont {D.}~\bibnamefont {Nakamura}}, \bibinfo {author} {\bibfnamefont {K.}~\bibnamefont {Inaka}}, \bibinfo {author} {\bibfnamefont {N.}~\bibnamefont {Okuma}},\ and\ \bibinfo {author} {\bibfnamefont {M.}~\bibnamefont {Sato}},\ }\bibfield  {title} {\bibinfo {title} {{Universal Platform of Point-Gap Topological Phases from Topological Materials}},\ }\href {https://doi.org/10.1103/PhysRevLett.131.256602} {\bibfield  {journal} {\bibinfo  {journal} {Phys. Rev. Lett.}\ }\textbf {\bibinfo {volume} {131}},\ \bibinfo {pages} {256602} (\bibinfo {year} {2023})}\BibitemShut {NoStop}%
\bibitem [{\citenamefont {Poli}\ \emph {et~al.}(2015)\citenamefont {Poli}, \citenamefont {Bellec}, \citenamefont {Kuhl}, \citenamefont {Mortessagne},\ and\ \citenamefont {Schomerus}}]{Poli-15}%
  \BibitemOpen
  \bibfield  {author} {\bibinfo {author} {\bibfnamefont {C.}~\bibnamefont {Poli}}, \bibinfo {author} {\bibfnamefont {M.}~\bibnamefont {Bellec}}, \bibinfo {author} {\bibfnamefont {U.}~\bibnamefont {Kuhl}}, \bibinfo {author} {\bibfnamefont {F.}~\bibnamefont {Mortessagne}},\ and\ \bibinfo {author} {\bibfnamefont {H.}~\bibnamefont {Schomerus}},\ }\bibfield  {title} {\bibinfo {title} {{Selective enhancement of topologically induced interface states in a dielectric resonator chain}},\ }\href {https://doi.org/10.1038/ncomms7710} {\bibfield  {journal} {\bibinfo  {journal} {Nat. Commun.}\ }\textbf {\bibinfo {volume} {6}},\ \bibinfo {pages} {6710} (\bibinfo {year} {2015})}\BibitemShut {NoStop}%
\bibitem [{\citenamefont {Zeuner}\ \emph {et~al.}(2015)\citenamefont {Zeuner}, \citenamefont {Rechtsman}, \citenamefont {Plotnik}, \citenamefont {Lumer}, \citenamefont {Nolte}, \citenamefont {Rudner}, \citenamefont {Segev},\ and\ \citenamefont {Szameit}}]{Zeuner-15}%
  \BibitemOpen
  \bibfield  {author} {\bibinfo {author} {\bibfnamefont {J.~M.}\ \bibnamefont {Zeuner}}, \bibinfo {author} {\bibfnamefont {M.~C.}\ \bibnamefont {Rechtsman}}, \bibinfo {author} {\bibfnamefont {Y.}~\bibnamefont {Plotnik}}, \bibinfo {author} {\bibfnamefont {Y.}~\bibnamefont {Lumer}}, \bibinfo {author} {\bibfnamefont {S.}~\bibnamefont {Nolte}}, \bibinfo {author} {\bibfnamefont {M.~S.}\ \bibnamefont {Rudner}}, \bibinfo {author} {\bibfnamefont {M.}~\bibnamefont {Segev}},\ and\ \bibinfo {author} {\bibfnamefont {A.}~\bibnamefont {Szameit}},\ }\bibfield  {title} {\bibinfo {title} {{Observation of a Topological Transition in the Bulk of a Non-Hermitian System}},\ }\href {https://doi.org/10.1103/PhysRevLett.115.040402} {\bibfield  {journal} {\bibinfo  {journal} {{Phys. Rev. Lett.}}\ }\textbf {\bibinfo {volume} {115}},\ \bibinfo {pages} {040402} (\bibinfo {year} {2015})}\BibitemShut {NoStop}%
\bibitem [{\citenamefont {Zhen}\ \emph {et~al.}(2015)\citenamefont {Zhen}, \citenamefont {Hsu}, \citenamefont {Igarashi}, \citenamefont {Lu}, \citenamefont {Kaminer}, \citenamefont {Pick}, \citenamefont {Chua}, \citenamefont {Joannopoulos},\ and\ \citenamefont {Solja\v{c}i\'c}}]{Zhen-15}%
  \BibitemOpen
  \bibfield  {author} {\bibinfo {author} {\bibfnamefont {B.}~\bibnamefont {Zhen}}, \bibinfo {author} {\bibfnamefont {C.~W.}\ \bibnamefont {Hsu}}, \bibinfo {author} {\bibfnamefont {Y.}~\bibnamefont {Igarashi}}, \bibinfo {author} {\bibfnamefont {L.}~\bibnamefont {Lu}}, \bibinfo {author} {\bibfnamefont {I.}~\bibnamefont {Kaminer}}, \bibinfo {author} {\bibfnamefont {A.}~\bibnamefont {Pick}}, \bibinfo {author} {\bibfnamefont {S.-L.}\ \bibnamefont {Chua}}, \bibinfo {author} {\bibfnamefont {J.~D.}\ \bibnamefont {Joannopoulos}},\ and\ \bibinfo {author} {\bibfnamefont {M.}~\bibnamefont {Solja\v{c}i\'c}},\ }\bibfield  {title} {\bibinfo {title} {{Spawning rings of exceptional points out of Dirac cones}},\ }\href {https://doi.org/10.1038/nature14889} {\bibfield  {journal} {\bibinfo  {journal} {Nature}\ }\textbf {\bibinfo {volume} {525}},\ \bibinfo {pages} {354} (\bibinfo {year} {2015})}\BibitemShut {NoStop}%
\bibitem [{\citenamefont {Zhou}\ \emph {et~al.}(2018)\citenamefont {Zhou}, \citenamefont {Peng}, \citenamefont {Yoon}, \citenamefont {Hsu}, \citenamefont {Nelson}, \citenamefont {Fu}, \citenamefont {Joannopoulos}, \citenamefont {Solja\v{c}i\'c},\ and\ \citenamefont {Zhen}}]{Zhou-18}%
  \BibitemOpen
  \bibfield  {author} {\bibinfo {author} {\bibfnamefont {H.}~\bibnamefont {Zhou}}, \bibinfo {author} {\bibfnamefont {C.}~\bibnamefont {Peng}}, \bibinfo {author} {\bibfnamefont {Y.}~\bibnamefont {Yoon}}, \bibinfo {author} {\bibfnamefont {C.~W.}\ \bibnamefont {Hsu}}, \bibinfo {author} {\bibfnamefont {K.~A.}\ \bibnamefont {Nelson}}, \bibinfo {author} {\bibfnamefont {L.}~\bibnamefont {Fu}}, \bibinfo {author} {\bibfnamefont {J.~D.}\ \bibnamefont {Joannopoulos}}, \bibinfo {author} {\bibfnamefont {M.}~\bibnamefont {Solja\v{c}i\'c}},\ and\ \bibinfo {author} {\bibfnamefont {B.}~\bibnamefont {Zhen}},\ }\bibfield  {title} {\bibinfo {title} {{Observation of bulk Fermi arc and polarization half charge from paired exceptional points}},\ }\href {https://doi.org/10.1126/science.aap9859} {\bibfield  {journal} {\bibinfo  {journal} {Science}\ }\textbf {\bibinfo {volume} {359}},\ \bibinfo {pages} {1009} (\bibinfo {year} {2018})}\BibitemShut {NoStop}%
\bibitem [{\citenamefont {Weimann}\ \emph {et~al.}(2017)\citenamefont {Weimann}, \citenamefont {Kremer}, \citenamefont {Plotnik}, \citenamefont {Lumer}, \citenamefont {Nolte}, \citenamefont {Makris}, \citenamefont {Segev}, \citenamefont {Rechtsman},\ and\ \citenamefont {Szameit}}]{Weimann-17}%
  \BibitemOpen
  \bibfield  {author} {\bibinfo {author} {\bibfnamefont {S.}~\bibnamefont {Weimann}}, \bibinfo {author} {\bibfnamefont {M.}~\bibnamefont {Kremer}}, \bibinfo {author} {\bibfnamefont {Y.}~\bibnamefont {Plotnik}}, \bibinfo {author} {\bibfnamefont {Y.}~\bibnamefont {Lumer}}, \bibinfo {author} {\bibfnamefont {S.}~\bibnamefont {Nolte}}, \bibinfo {author} {\bibfnamefont {K.~G.}\ \bibnamefont {Makris}}, \bibinfo {author} {\bibfnamefont {M.}~\bibnamefont {Segev}}, \bibinfo {author} {\bibfnamefont {M.~C.}\ \bibnamefont {Rechtsman}},\ and\ \bibinfo {author} {\bibfnamefont {A.}~\bibnamefont {Szameit}},\ }\bibfield  {title} {\bibinfo {title} {{Topologically protected bound states in photonic parity-time-symmetric crystals}},\ }\href {https://doi.org/10.1038/nmat4811} {\bibfield  {journal} {\bibinfo  {journal} {Nat. Mater.}\ }\textbf {\bibinfo {volume} {16}},\ \bibinfo {pages} {433} (\bibinfo {year} {2017})}\BibitemShut {NoStop}%
\bibitem [{\citenamefont {Xiao}\ \emph {et~al.}(2017)\citenamefont {Xiao}, \citenamefont {Zhan}, \citenamefont {Bian}, \citenamefont {Wang}, \citenamefont {Zhang}, \citenamefont {Wang}, \citenamefont {Li}, \citenamefont {Mochizuki}, \citenamefont {Kim}, \citenamefont {Kawakami}, \citenamefont {Yi}, \citenamefont {Obuse}, \citenamefont {Sanders},\ and\ \citenamefont {Xue}}]{Xiao-17}%
  \BibitemOpen
  \bibfield  {author} {\bibinfo {author} {\bibfnamefont {L.}~\bibnamefont {Xiao}}, \bibinfo {author} {\bibfnamefont {X.}~\bibnamefont {Zhan}}, \bibinfo {author} {\bibfnamefont {Z.~H.}\ \bibnamefont {Bian}}, \bibinfo {author} {\bibfnamefont {K.~K.}\ \bibnamefont {Wang}}, \bibinfo {author} {\bibfnamefont {X.}~\bibnamefont {Zhang}}, \bibinfo {author} {\bibfnamefont {X.~P.}\ \bibnamefont {Wang}}, \bibinfo {author} {\bibfnamefont {J.}~\bibnamefont {Li}}, \bibinfo {author} {\bibfnamefont {K.}~\bibnamefont {Mochizuki}}, \bibinfo {author} {\bibfnamefont {D.}~\bibnamefont {Kim}}, \bibinfo {author} {\bibfnamefont {N.}~\bibnamefont {Kawakami}}, \bibinfo {author} {\bibfnamefont {W.}~\bibnamefont {Yi}}, \bibinfo {author} {\bibfnamefont {H.}~\bibnamefont {Obuse}}, \bibinfo {author} {\bibfnamefont {B.~C.}\ \bibnamefont {Sanders}},\ and\ \bibinfo {author} {\bibfnamefont {P.}~\bibnamefont {Xue}},\ }\bibfield  {title} {\bibinfo {title} {{Observation of topological edge states in parity-time-symmetric quantum walks}},\ }\href
  {https://doi.org/10.1038/nphys4204} {\bibfield  {journal} {\bibinfo  {journal} {Nat. Phys.}\ }\textbf {\bibinfo {volume} {13}},\ \bibinfo {pages} {1117} (\bibinfo {year} {2017})}\BibitemShut {NoStop}%
\bibitem [{\citenamefont {St-Jean}\ \emph {et~al.}(2017)\citenamefont {St-Jean}, \citenamefont {Goblot}, \citenamefont {Galopin}, \citenamefont {Lema\^itre}, \citenamefont {Ozawa}, \citenamefont {Gratiet}, \citenamefont {Sagnes}, \citenamefont {Bloch},\ and\ \citenamefont {Amo}}]{St-Jean-17}%
  \BibitemOpen
  \bibfield  {author} {\bibinfo {author} {\bibfnamefont {P.}~\bibnamefont {St-Jean}}, \bibinfo {author} {\bibfnamefont {V.}~\bibnamefont {Goblot}}, \bibinfo {author} {\bibfnamefont {E.}~\bibnamefont {Galopin}}, \bibinfo {author} {\bibfnamefont {A.}~\bibnamefont {Lema\^itre}}, \bibinfo {author} {\bibfnamefont {T.}~\bibnamefont {Ozawa}}, \bibinfo {author} {\bibfnamefont {L.~L.}\ \bibnamefont {Gratiet}}, \bibinfo {author} {\bibfnamefont {I.}~\bibnamefont {Sagnes}}, \bibinfo {author} {\bibfnamefont {J.}~\bibnamefont {Bloch}},\ and\ \bibinfo {author} {\bibfnamefont {A.}~\bibnamefont {Amo}},\ }\bibfield  {title} {\bibinfo {title} {{Lasing in topological edge states of a one-dimensional lattice}},\ }\href {https://doi.org/10.1038/s41566-017-0006-2} {\bibfield  {journal} {\bibinfo  {journal} {Nat. Photon.}\ }\textbf {\bibinfo {volume} {11}},\ \bibinfo {pages} {651} (\bibinfo {year} {2017})}\BibitemShut {NoStop}%
\bibitem [{\citenamefont {Bahari}\ \emph {et~al.}(2017)\citenamefont {Bahari}, \citenamefont {Ndao}, \citenamefont {Vallini}, \citenamefont {Amili}, \citenamefont {Fainman},\ and\ \citenamefont {Kant\'e}}]{Bahari-17}%
  \BibitemOpen
  \bibfield  {author} {\bibinfo {author} {\bibfnamefont {B.}~\bibnamefont {Bahari}}, \bibinfo {author} {\bibfnamefont {A.}~\bibnamefont {Ndao}}, \bibinfo {author} {\bibfnamefont {F.}~\bibnamefont {Vallini}}, \bibinfo {author} {\bibfnamefont {A.~E.}\ \bibnamefont {Amili}}, \bibinfo {author} {\bibfnamefont {Y.}~\bibnamefont {Fainman}},\ and\ \bibinfo {author} {\bibfnamefont {B.}~\bibnamefont {Kant\'e}},\ }\bibfield  {title} {\bibinfo {title} {{Nonreciprocal lasing in topological cavities of arbitrary geometries}},\ }\href {https://doi.org/10.1126/science.aao4551} {\bibfield  {journal} {\bibinfo  {journal} {Science}\ }\textbf {\bibinfo {volume} {358}},\ \bibinfo {pages} {636} (\bibinfo {year} {2017})}\BibitemShut {NoStop}%
\bibitem [{\citenamefont {Harari}\ \emph {et~al.}(2018)\citenamefont {Harari}, \citenamefont {Bandres}, \citenamefont {Lumer}, \citenamefont {Rechtsman}, \citenamefont {Chong}, \citenamefont {Khajavikhan}, \citenamefont {Christodoulides},\ and\ \citenamefont {Segev}}]{Harari-18}%
  \BibitemOpen
  \bibfield  {author} {\bibinfo {author} {\bibfnamefont {G.}~\bibnamefont {Harari}}, \bibinfo {author} {\bibfnamefont {M.~A.}\ \bibnamefont {Bandres}}, \bibinfo {author} {\bibfnamefont {Y.}~\bibnamefont {Lumer}}, \bibinfo {author} {\bibfnamefont {M.~C.}\ \bibnamefont {Rechtsman}}, \bibinfo {author} {\bibfnamefont {Y.~D.}\ \bibnamefont {Chong}}, \bibinfo {author} {\bibfnamefont {M.}~\bibnamefont {Khajavikhan}}, \bibinfo {author} {\bibfnamefont {D.~N.}\ \bibnamefont {Christodoulides}},\ and\ \bibinfo {author} {\bibfnamefont {M.}~\bibnamefont {Segev}},\ }\bibfield  {title} {\bibinfo {title} {{Topological insulator laser: Theory}},\ }\href {https://doi.org/10.1126/science.aar4003} {\bibfield  {journal} {\bibinfo  {journal} {Science}\ }\textbf {\bibinfo {volume} {359}},\ \bibinfo {pages} {eaar4003} (\bibinfo {year} {2018})}\BibitemShut {NoStop}%
\bibitem [{\citenamefont {Bandres}\ \emph {et~al.}(2018)\citenamefont {Bandres}, \citenamefont {Wittek}, \citenamefont {Harari}, \citenamefont {Parto}, \citenamefont {Ren}, \citenamefont {Segev}, \citenamefont {Christodoulides},\ and\ \citenamefont {Khajavikhan}}]{Bandres-18}%
  \BibitemOpen
  \bibfield  {author} {\bibinfo {author} {\bibfnamefont {M.~A.}\ \bibnamefont {Bandres}}, \bibinfo {author} {\bibfnamefont {S.}~\bibnamefont {Wittek}}, \bibinfo {author} {\bibfnamefont {G.}~\bibnamefont {Harari}}, \bibinfo {author} {\bibfnamefont {M.}~\bibnamefont {Parto}}, \bibinfo {author} {\bibfnamefont {J.}~\bibnamefont {Ren}}, \bibinfo {author} {\bibfnamefont {M.}~\bibnamefont {Segev}}, \bibinfo {author} {\bibfnamefont {D.}~\bibnamefont {Christodoulides}},\ and\ \bibinfo {author} {\bibfnamefont {M.}~\bibnamefont {Khajavikhan}},\ }\bibfield  {title} {\bibinfo {title} {{Topological insulator laser: Experiments}},\ }\href {https://doi.org/10.1126/science.aar4005} {\bibfield  {journal} {\bibinfo  {journal} {{Science}}\ }\textbf {\bibinfo {volume} {359}},\ \bibinfo {pages} {eaar4005} (\bibinfo {year} {2018})}\BibitemShut {NoStop}%
\bibitem [{\citenamefont {Zhao}\ \emph {et~al.}(2019)\citenamefont {Zhao}, \citenamefont {Qiao}, \citenamefont {Wu}, \citenamefont {Midya}, \citenamefont {Longhi},\ and\ \citenamefont {Feng}}]{Zhao-19}%
  \BibitemOpen
  \bibfield  {author} {\bibinfo {author} {\bibfnamefont {H.}~\bibnamefont {Zhao}}, \bibinfo {author} {\bibfnamefont {X.}~\bibnamefont {Qiao}}, \bibinfo {author} {\bibfnamefont {T.}~\bibnamefont {Wu}}, \bibinfo {author} {\bibfnamefont {B.}~\bibnamefont {Midya}}, \bibinfo {author} {\bibfnamefont {S.}~\bibnamefont {Longhi}},\ and\ \bibinfo {author} {\bibfnamefont {L.}~\bibnamefont {Feng}},\ }\bibfield  {title} {\bibinfo {title} {{Non-Hermitian topological light steering}},\ }\href {https://doi.org/10.1126/science.aay1064} {\bibfield  {journal} {\bibinfo  {journal} {Science}\ }\textbf {\bibinfo {volume} {365}},\ \bibinfo {pages} {1163} (\bibinfo {year} {2019})}\BibitemShut {NoStop}%
\bibitem [{\citenamefont {Brandenbourger}\ \emph {et~al.}(2019)\citenamefont {Brandenbourger}, \citenamefont {Locsin}, \citenamefont {Lerner},\ and\ \citenamefont {Coulais}}]{Brandenbourger-19-skin-exp}%
  \BibitemOpen
  \bibfield  {author} {\bibinfo {author} {\bibfnamefont {M.}~\bibnamefont {Brandenbourger}}, \bibinfo {author} {\bibfnamefont {X.}~\bibnamefont {Locsin}}, \bibinfo {author} {\bibfnamefont {E.}~\bibnamefont {Lerner}},\ and\ \bibinfo {author} {\bibfnamefont {C.}~\bibnamefont {Coulais}},\ }\bibfield  {title} {\bibinfo {title} {{Non-reciprocal robotic metamaterials}},\ }\href {https://doi.org/10.1038/s41467-019-12599-3} {\bibfield  {journal} {\bibinfo  {journal} {Nat. Commun.}\ }\textbf {\bibinfo {volume} {10}},\ \bibinfo {pages} {4608} (\bibinfo {year} {2019})}\BibitemShut {NoStop}%
\bibitem [{\citenamefont {Ghatak}\ \emph {et~al.}(2020)\citenamefont {Ghatak}, \citenamefont {Brandenbourger}, \citenamefont {van Wezel},\ and\ \citenamefont {Coulais}}]{Ghatak-19-skin-exp}%
  \BibitemOpen
  \bibfield  {author} {\bibinfo {author} {\bibfnamefont {A.}~\bibnamefont {Ghatak}}, \bibinfo {author} {\bibfnamefont {M.}~\bibnamefont {Brandenbourger}}, \bibinfo {author} {\bibfnamefont {J.}~\bibnamefont {van Wezel}},\ and\ \bibinfo {author} {\bibfnamefont {C.}~\bibnamefont {Coulais}},\ }\bibfield  {title} {\bibinfo {title} {{Observation of non-Hermitian topology and its bulk-edge correspondence in an active mechanical metamaterial}},\ }\href {https://doi.org/10.1073/pnas.2010580117} {\bibfield  {journal} {\bibinfo  {journal} {Proc. Natl. Acad. Sci. USA}\ }\textbf {\bibinfo {volume} {117}},\ \bibinfo {pages} {29561} (\bibinfo {year} {2020})}\BibitemShut {NoStop}%
\bibitem [{\citenamefont {Helbig}\ \emph {et~al.}(2020)\citenamefont {Helbig}, \citenamefont {Hofmann}, \citenamefont {Imhof}, \citenamefont {Abdelghany}, \citenamefont {Kiessling}, \citenamefont {Molenkamp}, \citenamefont {Lee}, \citenamefont {Szameit}, \citenamefont {Greiter},\ and\ \citenamefont {Thomale}}]{Helbig-19-skin-exp}%
  \BibitemOpen
  \bibfield  {author} {\bibinfo {author} {\bibfnamefont {T.}~\bibnamefont {Helbig}}, \bibinfo {author} {\bibfnamefont {T.}~\bibnamefont {Hofmann}}, \bibinfo {author} {\bibfnamefont {S.}~\bibnamefont {Imhof}}, \bibinfo {author} {\bibfnamefont {M.}~\bibnamefont {Abdelghany}}, \bibinfo {author} {\bibfnamefont {T.}~\bibnamefont {Kiessling}}, \bibinfo {author} {\bibfnamefont {L.~W.}\ \bibnamefont {Molenkamp}}, \bibinfo {author} {\bibfnamefont {C.~H.}\ \bibnamefont {Lee}}, \bibinfo {author} {\bibfnamefont {A.}~\bibnamefont {Szameit}}, \bibinfo {author} {\bibfnamefont {M.}~\bibnamefont {Greiter}},\ and\ \bibinfo {author} {\bibfnamefont {R.}~\bibnamefont {Thomale}},\ }\bibfield  {title} {\bibinfo {title} {{Generalized bulk-boundary correspondence in non-Hermitian topolectrical circuits}},\ }\href {https://doi.org/10.1038/s41567-020-0922-9} {\bibfield  {journal} {\bibinfo  {journal} {Nat. Phys.}\ }\textbf {\bibinfo {volume} {16}},\ \bibinfo {pages} {747} (\bibinfo {year} {2020})}\BibitemShut {NoStop}%
\bibitem [{\citenamefont {Hofmann}\ \emph {et~al.}(2020)\citenamefont {Hofmann}, \citenamefont {Helbig}, \citenamefont {Schindler}, \citenamefont {Salgo}, \citenamefont {Brzezi\'nska}, \citenamefont {Greiter}, \citenamefont {Kiessling}, \citenamefont {Wolf}, \citenamefont {Vollhardt}, \citenamefont {Kaba\v{s}i}, \citenamefont {Lee}, \citenamefont {Bilu\v{s}i\'c}, \citenamefont {Thomale},\ and\ \citenamefont {Neupert}}]{Hofmann-19-skin-exp}%
  \BibitemOpen
  \bibfield  {author} {\bibinfo {author} {\bibfnamefont {T.}~\bibnamefont {Hofmann}}, \bibinfo {author} {\bibfnamefont {T.}~\bibnamefont {Helbig}}, \bibinfo {author} {\bibfnamefont {F.}~\bibnamefont {Schindler}}, \bibinfo {author} {\bibfnamefont {N.}~\bibnamefont {Salgo}}, \bibinfo {author} {\bibfnamefont {M.}~\bibnamefont {Brzezi\'nska}}, \bibinfo {author} {\bibfnamefont {M.}~\bibnamefont {Greiter}}, \bibinfo {author} {\bibfnamefont {T.}~\bibnamefont {Kiessling}}, \bibinfo {author} {\bibfnamefont {D.}~\bibnamefont {Wolf}}, \bibinfo {author} {\bibfnamefont {A.}~\bibnamefont {Vollhardt}}, \bibinfo {author} {\bibfnamefont {A.}~\bibnamefont {Kaba\v{s}i}}, \bibinfo {author} {\bibfnamefont {C.~H.}\ \bibnamefont {Lee}}, \bibinfo {author} {\bibfnamefont {A.}~\bibnamefont {Bilu\v{s}i\'c}}, \bibinfo {author} {\bibfnamefont {R.}~\bibnamefont {Thomale}},\ and\ \bibinfo {author} {\bibfnamefont {T.}~\bibnamefont {Neupert}},\ }\bibfield  {title} {\bibinfo {title} {{Reciprocal skin effect and its realization in a
  topolectrical circuit}},\ }\href {https://doi.org/10.1103/PhysRevResearch.2.023265} {\bibfield  {journal} {\bibinfo  {journal} {Phys. Rev. Research}\ }\textbf {\bibinfo {volume} {2}},\ \bibinfo {pages} {023265} (\bibinfo {year} {2020})}\BibitemShut {NoStop}%
\bibitem [{\citenamefont {Xiao}\ \emph {et~al.}(2020)\citenamefont {Xiao}, \citenamefont {Deng}, \citenamefont {Wang}, \citenamefont {Zhu}, \citenamefont {Wang}, \citenamefont {Yi},\ and\ \citenamefont {Xue}}]{Xiao-19-skin-exp}%
  \BibitemOpen
  \bibfield  {author} {\bibinfo {author} {\bibfnamefont {L.}~\bibnamefont {Xiao}}, \bibinfo {author} {\bibfnamefont {T.}~\bibnamefont {Deng}}, \bibinfo {author} {\bibfnamefont {K.}~\bibnamefont {Wang}}, \bibinfo {author} {\bibfnamefont {G.}~\bibnamefont {Zhu}}, \bibinfo {author} {\bibfnamefont {Z.}~\bibnamefont {Wang}}, \bibinfo {author} {\bibfnamefont {W.}~\bibnamefont {Yi}},\ and\ \bibinfo {author} {\bibfnamefont {P.}~\bibnamefont {Xue}},\ }\bibfield  {title} {\bibinfo {title} {{Non-Hermitian bulk-boundary correspondence in quantum dynamics}},\ }\href {https://doi.org/10.1038/s41567-020-0836-6} {\bibfield  {journal} {\bibinfo  {journal} {Nat. Phys.}\ }\textbf {\bibinfo {volume} {16}},\ \bibinfo {pages} {761} (\bibinfo {year} {2020})}\BibitemShut {NoStop}%
\bibitem [{\citenamefont {Weidemann}\ \emph {et~al.}(2020)\citenamefont {Weidemann}, \citenamefont {Kremer}, \citenamefont {Helbig}, \citenamefont {Hofmann}, \citenamefont {Stegmaier}, \citenamefont {Greiter}, \citenamefont {Thomale},\ and\ \citenamefont {Szameit}}]{Weidemann-20-skin-exp}%
  \BibitemOpen
  \bibfield  {author} {\bibinfo {author} {\bibfnamefont {S.}~\bibnamefont {Weidemann}}, \bibinfo {author} {\bibfnamefont {M.}~\bibnamefont {Kremer}}, \bibinfo {author} {\bibfnamefont {T.}~\bibnamefont {Helbig}}, \bibinfo {author} {\bibfnamefont {T.}~\bibnamefont {Hofmann}}, \bibinfo {author} {\bibfnamefont {A.}~\bibnamefont {Stegmaier}}, \bibinfo {author} {\bibfnamefont {M.}~\bibnamefont {Greiter}}, \bibinfo {author} {\bibfnamefont {R.}~\bibnamefont {Thomale}},\ and\ \bibinfo {author} {\bibfnamefont {A.}~\bibnamefont {Szameit}},\ }\bibfield  {title} {\bibinfo {title} {{Topological funneling of light}},\ }\href {https://doi.org/10.1126/science.aaz8727} {\bibfield  {journal} {\bibinfo  {journal} {Science}\ }\textbf {\bibinfo {volume} {368}},\ \bibinfo {pages} {311} (\bibinfo {year} {2020})}\BibitemShut {NoStop}%
\bibitem [{\citenamefont {Palacios}\ \emph {et~al.}(2020)\citenamefont {Palacios}, \citenamefont {Tchoumakov}, \citenamefont {Guix}, \citenamefont {S\'anchez},\ and\ \citenamefont {Grushin}}]{Palacios-21}%
  \BibitemOpen
  \bibfield  {author} {\bibinfo {author} {\bibfnamefont {L.~S.}\ \bibnamefont {Palacios}}, \bibinfo {author} {\bibfnamefont {S.}~\bibnamefont {Tchoumakov}}, \bibinfo {author} {\bibfnamefont {M.}~\bibnamefont {Guix}}, \bibinfo {author} {\bibfnamefont {I.~P.~S.}\ \bibnamefont {S\'anchez}},\ and\ \bibinfo {author} {\bibfnamefont {A.~G.}\ \bibnamefont {Grushin}},\ }\bibfield  {title} {\bibinfo {title} {{Guided accumulation of active particles by topological design of a second-order skin effect}},\ }\href {https://doi.org/10.1038/s41467-021-24948-2} {\bibfield  {journal} {\bibinfo  {journal} {Nat. Commun.}\ }\textbf {\bibinfo {volume} {12}},\ \bibinfo {pages} {4691} (\bibinfo {year} {2020})}\BibitemShut {NoStop}%
\bibitem [{\citenamefont {Zhao}\ \emph {et~al.}(2025)\citenamefont {Zhao}, \citenamefont {Wang}, \citenamefont {He}, \citenamefont {Poon}, \citenamefont {Pak}, \citenamefont {Liu}, \citenamefont {Ren}, \citenamefont {Liu},\ and\ \citenamefont {Jo}}]{Zhao-25}%
  \BibitemOpen
  \bibfield  {author} {\bibinfo {author} {\bibfnamefont {E.}~\bibnamefont {Zhao}}, \bibinfo {author} {\bibfnamefont {Z.}~\bibnamefont {Wang}}, \bibinfo {author} {\bibfnamefont {C.}~\bibnamefont {He}}, \bibinfo {author} {\bibfnamefont {T.~F.~J.}\ \bibnamefont {Poon}}, \bibinfo {author} {\bibfnamefont {K.~K.}\ \bibnamefont {Pak}}, \bibinfo {author} {\bibfnamefont {Y.-J.}\ \bibnamefont {Liu}}, \bibinfo {author} {\bibfnamefont {P.}~\bibnamefont {Ren}}, \bibinfo {author} {\bibfnamefont {X.-J.}\ \bibnamefont {Liu}},\ and\ \bibinfo {author} {\bibfnamefont {G.-B.}\ \bibnamefont {Jo}},\ }\bibfield  {title} {\bibinfo {title} {{Two-dimensional non-Hermitian skin effect in an ultracold Fermi gas}},\ }\href {https://doi.org/https://doi.org/10.1038/s41586-024-08347-3} {\bibfield  {journal} {\bibinfo  {journal} {Nature}\ }\textbf {\bibinfo {volume} {637}},\ \bibinfo {pages} {565} (\bibinfo {year} {2025})}\BibitemShut {NoStop}%
\bibitem [{\citenamefont {Shen}\ \emph {et~al.}(2025)\citenamefont {Shen}, \citenamefont {Chen}, \citenamefont {Yang},\ and\ \citenamefont {Lee}}]{Shen-25}%
  \BibitemOpen
  \bibfield  {author} {\bibinfo {author} {\bibfnamefont {R.}~\bibnamefont {Shen}}, \bibinfo {author} {\bibfnamefont {T.}~\bibnamefont {Chen}}, \bibinfo {author} {\bibfnamefont {B.}~\bibnamefont {Yang}},\ and\ \bibinfo {author} {\bibfnamefont {C.~H.}\ \bibnamefont {Lee}},\ }\bibfield  {title} {\bibinfo {title} {{Observation of the non-Hermitian skin effect and Fermi skin on a digital quantum computer}},\ }\href {https://doi.org/https://doi.org/10.1038/s41467-025-55953-4} {\bibfield  {journal} {\bibinfo  {journal} {Nat. Commun.}\ }\textbf {\bibinfo {volume} {16}},\ \bibinfo {pages} {1340} (\bibinfo {year} {2025})}\BibitemShut {NoStop}%
\bibitem [{\citenamefont {Konotop}\ \emph {et~al.}(2016)\citenamefont {Konotop}, \citenamefont {Yang},\ and\ \citenamefont {Zezyulin}}]{Konotop-review}%
  \BibitemOpen
  \bibfield  {author} {\bibinfo {author} {\bibfnamefont {V.~V.}\ \bibnamefont {Konotop}}, \bibinfo {author} {\bibfnamefont {J.}~\bibnamefont {Yang}},\ and\ \bibinfo {author} {\bibfnamefont {D.~A.}\ \bibnamefont {Zezyulin}},\ }\bibfield  {title} {\bibinfo {title} {{Nonlinear waves in $\mathcal{PT}$-symmetric systems}},\ }\href {https://doi.org/10.1103/RevModPhys.88.035002} {\bibfield  {journal} {\bibinfo  {journal} {Rev. Mod. Phys.}\ }\textbf {\bibinfo {volume} {88}},\ \bibinfo {pages} {035002} (\bibinfo {year} {2016})}\BibitemShut {NoStop}%
\bibitem [{\citenamefont {El-Ganainy}\ \emph {et~al.}(2018)\citenamefont {El-Ganainy}, \citenamefont {Makris}, \citenamefont {Khajavikhan}, \citenamefont {Musslimani}, \citenamefont {Rotter},\ and\ \citenamefont {Christodoulides}}]{Christodoulides-review}%
  \BibitemOpen
  \bibfield  {author} {\bibinfo {author} {\bibfnamefont {R.}~\bibnamefont {El-Ganainy}}, \bibinfo {author} {\bibfnamefont {K.~G.}\ \bibnamefont {Makris}}, \bibinfo {author} {\bibfnamefont {M.}~\bibnamefont {Khajavikhan}}, \bibinfo {author} {\bibfnamefont {Z.~H.}\ \bibnamefont {Musslimani}}, \bibinfo {author} {\bibfnamefont {S.}~\bibnamefont {Rotter}},\ and\ \bibinfo {author} {\bibfnamefont {D.~N.}\ \bibnamefont {Christodoulides}},\ }\bibfield  {title} {\bibinfo {title} {{Non-Hermitian physics and PT symmetry}},\ }\href {https://doi.org/10.1038/nphys4323} {\bibfield  {journal} {\bibinfo  {journal} {Nat. Phys.}\ }\textbf {\bibinfo {volume} {14}},\ \bibinfo {pages} {11} (\bibinfo {year} {2018})}\BibitemShut {NoStop}%
\bibitem [{\citenamefont {Yoshida}\ \emph {et~al.}(2020)\citenamefont {Yoshida}, \citenamefont {Mizoguchi},\ and\ \citenamefont {Hatsugai}}]{Yoshida-20}%
  \BibitemOpen
  \bibfield  {author} {\bibinfo {author} {\bibfnamefont {T.}~\bibnamefont {Yoshida}}, \bibinfo {author} {\bibfnamefont {T.}~\bibnamefont {Mizoguchi}},\ and\ \bibinfo {author} {\bibfnamefont {Y.}~\bibnamefont {Hatsugai}},\ }\bibfield  {title} {\bibinfo {title} {{Mirror skin effect and its electric circuit simulation}},\ }\href {https://doi.org/10.1103/PhysRevResearch.2.022062} {\bibfield  {journal} {\bibinfo  {journal} {Phys. Rev. Research}\ }\textbf {\bibinfo {volume} {2}},\ \bibinfo {pages} {022062} (\bibinfo {year} {2020})}\BibitemShut {NoStop}%
\bibitem [{\citenamefont {Okugawa}\ \emph {et~al.}(2020)\citenamefont {Okugawa}, \citenamefont {Takahashi},\ and\ \citenamefont {Yokomizo}}]{Okugawa-20}%
  \BibitemOpen
  \bibfield  {author} {\bibinfo {author} {\bibfnamefont {R.}~\bibnamefont {Okugawa}}, \bibinfo {author} {\bibfnamefont {R.}~\bibnamefont {Takahashi}},\ and\ \bibinfo {author} {\bibfnamefont {K.}~\bibnamefont {Yokomizo}},\ }\bibfield  {title} {\bibinfo {title} {{Second-order topological non-Hermitian skin effects}},\ }\href {https://doi.org/10.1103/PhysRevB.102.241202} {\bibfield  {journal} {\bibinfo  {journal} {Phys. Rev. B}\ }\textbf {\bibinfo {volume} {102}},\ \bibinfo {pages} {241202(R)} (\bibinfo {year} {2020})}\BibitemShut {NoStop}%
\bibitem [{\citenamefont {Kawabata}\ \emph {et~al.}(2020{\natexlab{b}})\citenamefont {Kawabata}, \citenamefont {Sato},\ and\ \citenamefont {Shiozaki}}]{KSS-20}%
  \BibitemOpen
  \bibfield  {author} {\bibinfo {author} {\bibfnamefont {K.}~\bibnamefont {Kawabata}}, \bibinfo {author} {\bibfnamefont {M.}~\bibnamefont {Sato}},\ and\ \bibinfo {author} {\bibfnamefont {K.}~\bibnamefont {Shiozaki}},\ }\bibfield  {title} {\bibinfo {title} {{Higher-order non-Hermitian skin effect}},\ }\href {https://doi.org/10.1103/PhysRevB.102.205118} {\bibfield  {journal} {\bibinfo  {journal} {Phys. Rev. B}\ }\textbf {\bibinfo {volume} {102}},\ \bibinfo {pages} {205118} (\bibinfo {year} {2020}{\natexlab{b}})}\BibitemShut {NoStop}%
\bibitem [{\citenamefont {Okugawa}\ \emph {et~al.}(2021)\citenamefont {Okugawa}, \citenamefont {Takahashi},\ and\ \citenamefont {Yokomizo}}]{Okugawa-21}%
  \BibitemOpen
  \bibfield  {author} {\bibinfo {author} {\bibfnamefont {R.}~\bibnamefont {Okugawa}}, \bibinfo {author} {\bibfnamefont {R.}~\bibnamefont {Takahashi}},\ and\ \bibinfo {author} {\bibfnamefont {K.}~\bibnamefont {Yokomizo}},\ }\bibfield  {title} {\bibinfo {title} {{Non-Hermitian band topology with generalized inversion symmetry}},\ }\href {https://doi.org/10.1103/PhysRevB.103.205205} {\bibfield  {journal} {\bibinfo  {journal} {Phys. Rev. B}\ }\textbf {\bibinfo {volume} {103}},\ \bibinfo {pages} {205205} (\bibinfo {year} {2021})}\BibitemShut {NoStop}%
\bibitem [{\citenamefont {Vecsei}\ \emph {et~al.}(2021)\citenamefont {Vecsei}, \citenamefont {Denner}, \citenamefont {Neupert},\ and\ \citenamefont {Schindler}}]{Vecsei-21}%
  \BibitemOpen
  \bibfield  {author} {\bibinfo {author} {\bibfnamefont {P.~M.}\ \bibnamefont {Vecsei}}, \bibinfo {author} {\bibfnamefont {M.~M.}\ \bibnamefont {Denner}}, \bibinfo {author} {\bibfnamefont {T.}~\bibnamefont {Neupert}},\ and\ \bibinfo {author} {\bibfnamefont {F.}~\bibnamefont {Schindler}},\ }\bibfield  {title} {\bibinfo {title} {{Symmetry indicators for inversion-symmetric non-Hermitian topological band structures}},\ }\href {https://doi.org/10.1103/PhysRevB.103.L201114} {\bibfield  {journal} {\bibinfo  {journal} {Phys. Rev. B}\ }\textbf {\bibinfo {volume} {103}},\ \bibinfo {pages} {L201114} (\bibinfo {year} {2021})}\BibitemShut {NoStop}%
\bibitem [{\citenamefont {Shiozaki}\ and\ \citenamefont {Ono}(2021)}]{Shiozaki-21}%
  \BibitemOpen
  \bibfield  {author} {\bibinfo {author} {\bibfnamefont {K.}~\bibnamefont {Shiozaki}}\ and\ \bibinfo {author} {\bibfnamefont {S.}~\bibnamefont {Ono}},\ }\bibfield  {title} {\bibinfo {title} {{Symmetry indicator in non-Hermitian systems}},\ }\href {https://doi.org/10.1103/PhysRevB.104.035424} {\bibfield  {journal} {\bibinfo  {journal} {Phys. Rev. B}\ }\textbf {\bibinfo {volume} {104}},\ \bibinfo {pages} {035424} (\bibinfo {year} {2021})}\BibitemShut {NoStop}%
\bibitem [{\citenamefont {Schindler}\ \emph {et~al.}(2023)\citenamefont {Schindler}, \citenamefont {Gu}, \citenamefont {Lian},\ and\ \citenamefont {Kawabata}}]{Schindler-23}%
  \BibitemOpen
  \bibfield  {author} {\bibinfo {author} {\bibfnamefont {F.}~\bibnamefont {Schindler}}, \bibinfo {author} {\bibfnamefont {K.}~\bibnamefont {Gu}}, \bibinfo {author} {\bibfnamefont {B.}~\bibnamefont {Lian}},\ and\ \bibinfo {author} {\bibfnamefont {K.}~\bibnamefont {Kawabata}},\ }\bibfield  {title} {\bibinfo {title} {{Hermitian Bulk -- Non-Hermitian Boundary Correspondence}},\ }\href {https://doi.org/10.1103/PRXQuantum.4.030315} {\bibfield  {journal} {\bibinfo  {journal} {PRX Quantum}\ }\textbf {\bibinfo {volume} {4}},\ \bibinfo {pages} {030315} (\bibinfo {year} {2023})}\BibitemShut {NoStop}%
\bibitem [{\citenamefont {Tanaka}\ \emph {et~al.}(2024)\citenamefont {Tanaka}, \citenamefont {Takahashi},\ and\ \citenamefont {Okugawa}}]{Tanaka-24NSG}%
  \BibitemOpen
  \bibfield  {author} {\bibinfo {author} {\bibfnamefont {Y.}~\bibnamefont {Tanaka}}, \bibinfo {author} {\bibfnamefont {R.}~\bibnamefont {Takahashi}},\ and\ \bibinfo {author} {\bibfnamefont {R.}~\bibnamefont {Okugawa}},\ }\bibfield  {title} {\bibinfo {title} {{Non-Hermitian skin effect enforced by nonsymmorphic symmetries}},\ }\href {https://doi.org/10.1103/PhysRevB.109.035131} {\bibfield  {journal} {\bibinfo  {journal} {Phys. Rev. B}\ }\textbf {\bibinfo {volume} {109}},\ \bibinfo {pages} {035131} (\bibinfo {year} {2024})}\BibitemShut {NoStop}%
\bibitem [{\citenamefont {Ishikawa}\ and\ \citenamefont {Yoshida}(2024)}]{Ishikawa-24}%
  \BibitemOpen
  \bibfield  {author} {\bibinfo {author} {\bibfnamefont {S.}~\bibnamefont {Ishikawa}}\ and\ \bibinfo {author} {\bibfnamefont {T.}~\bibnamefont {Yoshida}},\ }\bibfield  {title} {\bibinfo {title} {{Non-Hermitian ${\mathbb{Z}}_{4}$ skin effect protected by glide symmetry}},\ }\href {https://doi.org/10.1103/PhysRevB.110.115301} {\bibfield  {journal} {\bibinfo  {journal} {Phys. Rev. B}\ }\textbf {\bibinfo {volume} {110}},\ \bibinfo {pages} {115301} (\bibinfo {year} {2024})}\BibitemShut {NoStop}%
\bibitem [{\citenamefont {Wang}()}]{YifanWang-24}%
  \BibitemOpen
  \bibfield  {author} {\bibinfo {author} {\bibfnamefont {Y.}~\bibnamefont {Wang}},\ }\bibfield  {title} {\bibinfo {title} {{Classifying Order-Two Spatial Symmetries in Non-Hermitian Hamiltonians: Point-gapped AZ and AZ$^{\dag}$ Classes}},\ }\Eprint {https://arxiv.org/abs/2411.03410} {arXiv:2411.03410} \BibitemShut {NoStop}%
\bibitem [{\citenamefont {Tanaka}\ \emph {et~al.}()\citenamefont {Tanaka}, \citenamefont {Nakamura}, \citenamefont {Okugawa},\ and\ \citenamefont {Kawabata}}]{Tanaka-24}%
  \BibitemOpen
  \bibfield  {author} {\bibinfo {author} {\bibfnamefont {Y.}~\bibnamefont {Tanaka}}, \bibinfo {author} {\bibfnamefont {D.}~\bibnamefont {Nakamura}}, \bibinfo {author} {\bibfnamefont {R.}~\bibnamefont {Okugawa}},\ and\ \bibinfo {author} {\bibfnamefont {K.}~\bibnamefont {Kawabata}},\ }\bibfield  {title} {\bibinfo {title} {{Exceptional Second-Order Topological Insulators}},\ }\Eprint {https://arxiv.org/abs/2411.06898} {arXiv:2411.06898} \BibitemShut {NoStop}%
\bibitem [{\citenamefont {Wang}\ and\ \citenamefont {Benalcazar}()}]{WangBenalcazar-24}%
  \BibitemOpen
  \bibfield  {author} {\bibinfo {author} {\bibfnamefont {Y.}~\bibnamefont {Wang}}\ and\ \bibinfo {author} {\bibfnamefont {W.~A.}\ \bibnamefont {Benalcazar}},\ }\bibfield  {title} {\bibinfo {title} {{Higher-order Topological Knots and Nonreciprocal Dynamics in non-Hermitian lattices}},\ }\Eprint {https://arxiv.org/abs/2412.05809} {arXiv:2412.05809} \BibitemShut {NoStop}%
\bibitem [{sup()}]{supplement}%
  \BibitemOpen
  \href@noop {} {}\bibinfo {note} {{See the Supplemental Material for details on classification of nonsymmorphic non-Hermitian topology, nonsymmorphic topological phases in one dimension, $\mathbb{Z}_2$ and $\mathbb{Z}_4$ topological invariants in two dimensions, and systematic model construction in class AIII + $\mathcal{U}_{+}$.}}\BibitemShut {Stop}%
\bibitem [{\citenamefont {Su}\ \emph {et~al.}(1979)\citenamefont {Su}, \citenamefont {Schrieffer},\ and\ \citenamefont {Heeger}}]{SSH-79}%
  \BibitemOpen
  \bibfield  {author} {\bibinfo {author} {\bibfnamefont {W.~P.}\ \bibnamefont {Su}}, \bibinfo {author} {\bibfnamefont {J.~R.}\ \bibnamefont {Schrieffer}},\ and\ \bibinfo {author} {\bibfnamefont {A.~J.}\ \bibnamefont {Heeger}},\ }\bibfield  {title} {\bibinfo {title} {{Solitons in Polyacetylene}},\ }\href {https://doi.org/10.1103/PhysRevLett.42.1698} {\bibfield  {journal} {\bibinfo  {journal} {Phys. Rev. Lett.}\ }\textbf {\bibinfo {volume} {42}},\ \bibinfo {pages} {1698} (\bibinfo {year} {1979})}\BibitemShut {NoStop}%
\bibitem [{\citenamefont {Fu}\ and\ \citenamefont {Kane}(2006)}]{Fu-06}%
  \BibitemOpen
  \bibfield  {author} {\bibinfo {author} {\bibfnamefont {L.}~\bibnamefont {Fu}}\ and\ \bibinfo {author} {\bibfnamefont {C.~L.}\ \bibnamefont {Kane}},\ }\bibfield  {title} {\bibinfo {title} {{Time reversal polarization and a ${Z}_{2}$ adiabatic spin pump}},\ }\href {https://doi.org/10.1103/PhysRevB.74.195312} {\bibfield  {journal} {\bibinfo  {journal} {Phys. Rev. B}\ }\textbf {\bibinfo {volume} {74}},\ \bibinfo {pages} {195312} (\bibinfo {year} {2006})}\BibitemShut {NoStop}%
\end{thebibliography}%
\let\addcontentsline\oldaddcontentsline

\clearpage
\widetext

\setcounter{secnumdepth}{3}

\renewcommand{\theequation}{S\arabic{equation}}
\renewcommand{\thefigure}{S\arabic{figure}}
\renewcommand{\thetable}{S\arabic{table}}
\setcounter{equation}{0}
\setcounter{figure}{0}
\setcounter{table}{0}
\setcounter{section}{0}
\setcounter{tocdepth}{0}

\numberwithin{equation}{section} 

\begin{center}
{\bf \large Supplemental Material for \\ \smallskip 
``Nonsymmorphic Topological Phases of Non-Hermitian Systems"}
\end{center}


\section{Classification of nonsymmorphic non-Hermitian topology}
    \label{asec: classification}

We consider generic non-Hermitian Hamiltonians $H(\boldsymbol{k})$ respecting nonsymmorphic symmetry
\begin{equation}
    \mathcal{U}(\bm{k})H(\bm{k})\mathcal{V}^{-1}(\bm{k})=H(\sigma\bm{k}),
        \label{aeq: nsg}
\end{equation}
with $\bm{k}$-dependent unitary matrices $\mathcal{U}$ and $\mathcal{V}$ satisfying $\mathcal{U}(\sigma\bm{k})\mathcal{U}(\bm{k})=\mathcal{V}(\sigma\bm{k})\mathcal{V}(\bm{k})=e^{-ik_1}$,
or pseudo-nonsymmorphic symmetry
\begin{equation}
    \mathcal{U}(\bm{k})H^\dagger (\bm{k})\mathcal{V}^{-1}(\bm{k})= H (\sigma\bm{k})
        \label{aeq: nsg-pseudo}
\end{equation}
with $\bm{k}$-dependent unitary matrices $\mathcal{U}$ and $\mathcal{V}$ satisfying $\mathcal{U}(\sigma\bm{k}) \mathcal{V}(\bm{k}) = \mathcal{V}(\sigma\bm{k}) \mathcal{U}(\bm{k}) = e^{-ik_1}$.
Here, the transformed momentum is denoted by
\begin{align}
\sigma\bm{k} \coloneqq (k_1,k_2,\dots,k_{d-d_{\parallel}},-k_{d-d_{\parallel}+1},\dots,-k_{d}).
\end{align}
As also discussed in the main text, we introduce Hermitized Hamiltonians by
\begin{align}
    {\sf H}(\boldsymbol{k}) \coloneqq \begin{pmatrix}
        0 & H(\boldsymbol{k})\\
        H^{\dagger}(\boldsymbol{k}) & 0
    \end{pmatrix},
        \label{aeq: Hermitization}
\end{align}
where reference energy is chosen to be zero for simplicity.
By construction, ${\sf H}(\boldsymbol{k})$ respects chiral symmetry 
\begin{align}
    \Sigma {\sf H}(\boldsymbol{k}) \Sigma^{-1} = -{\sf H}(\boldsymbol{k}), \quad \Sigma^2=1, \quad \Sigma \coloneqq \begin{pmatrix}
        1 & 0 \\
        0 & -1
        \end{pmatrix}.
\end{align}
Point-gap topology of $H (\boldsymbol{k})$ is equivalent to Hermitian topology of ${\sf H}(\boldsymbol{k})$~\cite{Gong-18, KSUS-19}.
Accordingly, we associate nonsymmorphic-symmetric point-gap topology with  nonsymmorphic-symmetric Hermitized topology, which is classified on the basis of $K$-theory~\cite{Shiozaki-PRB-2016}.
This results in our classification tables of point-gap topology protected by nonsymmorphic or pseudo-nonsymmorphic symmetry in the Altland-Zirnbauer or Altland-Zirnbauer$^{\dag}$ classes without sublattice symmetry:
Tables~\ref{tab: complex AZ 0}, \ref{tab: real AZ 0}, and \ref{tab: real AZ dag 0} for $d_{\parallel} \equiv 0$ (mod $2$) and Tables~\ref{tab: complex AZ 1}, \ref{tab: real AZ 1}, and \ref{tab: real AZ dag 1} for $d_{\parallel} \equiv 1$ (mod $2$).

Now, we clarify the relationship between the two types of nonsymmorphic symmetries in Eqs.~\eqref{aeq: nsg} and~\eqref{aeq: nsg-pseudo} for non-Hermitian Hamiltonians $H(\boldsymbol{k})$ and nonsymmorphic symmetry for Hermitized Hamiltonians ${\sf H}(\boldsymbol{k})$ in Eq.~(\ref{aeq: Hermitization}):
\begin{align}\label{eq:ns_symmetry}
    {\sf U}(\boldsymbol{k}) {\sf H}(\boldsymbol{k}) {\sf U}^{\dagger}(\boldsymbol{k})={\sf H}(\sigma \boldsymbol{k}).
\end{align}
Here,
${\sf U}(\boldsymbol{k})$ is a unitary matrix satisfying 
\begin{align}\label{eq:u^2_exp-ikx}
    {\sf U}(\sigma \boldsymbol{k}){\sf U}(\boldsymbol{k}) = e^{-ik_1},
\end{align}
because of a half translation in the $x_1$ direction.
The symmetry operators satisfy the commutation or anticommutation relation~\cite{Shiozaki-21}
\begin{align}
    \Sigma{\sf U}(\boldsymbol{k}) = \epsilon_{\Sigma {\sf U}} {\sf U}(\boldsymbol{k})\Sigma, \quad \epsilon_{\Sigma {\sf U}}\in \{1,-1\},
\end{align}
depending on the sign of $\epsilon_{\Sigma {\sf U}}$, and the unitary matrix ${\sf U}(\boldsymbol{k})$ is given by
\begin{align}\label{eq:unitary}
    {\sf U}(\boldsymbol{k}) \coloneqq \begin{cases}
    \begin{pmatrix}
        \mathcal{U}(\boldsymbol{k}) & 0 \\
        0 & \mathcal{V}(\boldsymbol{k})
    \end{pmatrix} & (\epsilon_{\Sigma {\sf U}}=1), \\
    \\
    \begin{pmatrix}
         0 & \mathcal{V}(\boldsymbol{k}) \\
        \mathcal{U}(\boldsymbol{k}) & 0 
    \end{pmatrix} & (\epsilon_{\Sigma {\sf U}}=-1),
    \end{cases}
\end{align}
where $\mathcal{U}(\boldsymbol{k})$ and $\mathcal{V}(\boldsymbol{k})$ are unitary matrices periodic in the Brillouin zone. 
Substituting this equation into  Eq.~\eqref{eq:u^2_exp-ikx}, we obtain 
\begin{align}\label{eq:AandB_eq}
    \mathcal{U}(\sigma \boldsymbol{k}) \mathcal{U}(\boldsymbol{k}) = e^{-ik_1}, \quad \mathcal{V}(\sigma \boldsymbol{k}) \mathcal{V}(\boldsymbol{k}) = e^{-ik_1}, \quad (\epsilon_{\Sigma {\sf U}}=1), \\
    \mathcal{U}(\sigma \boldsymbol{k}) \mathcal{V}(\boldsymbol{k}) = e^{-ik_1}, \quad \mathcal{V}(\sigma \boldsymbol{k}) \mathcal{U}(\boldsymbol{k}) = e^{-ik_1}, \quad (\epsilon_{\Sigma {\sf U}}=-1).
\end{align}
In addition, by using the unitary matrices $\mathcal{U}(\boldsymbol{k})$ and $\mathcal{V}(\boldsymbol{k})$, Eq.~\eqref{eq:ns_symmetry} is rewritten as
\begin{align}\label{eq:AHcases}
    \mathcal{U}(\boldsymbol{k})H(\boldsymbol{k}) =
    \begin{cases}
    H(\sigma \boldsymbol{k})\mathcal{V}(\boldsymbol{k}) & (\epsilon_{\Sigma {\sf U}}=1), \\
    H^{\dagger}(\sigma \boldsymbol{k})\mathcal{V}(\boldsymbol{k}) & (\epsilon_{\Sigma {\sf U}}=-1).
    \end{cases}
\end{align}
For $\mathcal{U}(\boldsymbol{k})=\mathcal{V}(\boldsymbol{k})$, this equation is rewritten as 
\begin{gather}
    \mathcal{U}(\boldsymbol{k}) H(\boldsymbol{k}) \mathcal{U}^{-1}(\boldsymbol{k}) = H(\sigma \boldsymbol{k})\quad (\epsilon_{\Sigma {\sf U}}=1), \\
    \mathcal{U}(\boldsymbol{k}) H(\boldsymbol{k}) \mathcal{U}^{-1}(\boldsymbol{k}) = H^{\dagger}(\sigma \boldsymbol{k})\quad (\epsilon_{\Sigma {\sf U}}=-1), \label{eq:pseud-nss}\\
    \mathcal{U}(\sigma \boldsymbol{k}) \mathcal{U}(\boldsymbol{k}) = e^{-ik_x}.
\end{gather}
These equations represent nonsymmorphic symmetry and pseudo-nonsymmorphic symmetry for non-Hermitian Hamiltonians $H(\boldsymbol{k})$. 
More generally, $\mathcal{U}(\boldsymbol{k})$ and $\mathcal{V}(\boldsymbol{k})$ are not necessarily equal to each other, and nonsymmorphic symmetry of the Hermitized Hamiltonian leads to Eqs.~(\ref{aeq: nsg}) and (\ref{aeq: nsg-pseudo}).

\subsection{$d_{\parallel} \equiv 0$ (mod $2$)}

\begin{table}[H]
	\centering
	\caption{Classification of point-gap topology in the complex Altland-Zirnbauer symmetry classes protected by order-two nonsymmorphic unitary symmetry with $d_{\parallel} \equiv 0$ (mod $2$).
    ${\cal \M}$ and $\tilde{\cal \M}$ denote nonsymmorphic symmetry and pseudo-nonsymmorphic symmetry (or equivalently, nonsymmorphic symmetry$^{\dag}$), respectively.
    In class AIII, the subscript of ${\cal \M}_\pm$ or $\tilde{\cal \M}_\pm$ specifies the commutation ($+$) or anticommutation ($-$) relation with chiral symmetry.
    The topological indices highlighted in color indicate classification intrinsic to nonsymmorphic symmetry.}
	\label{tab: complex AZ 0}
     \begin{tabular}{cccccccccc} \hline \hline
    ~~Class~~ & ~~Hermitization~~ & ~~$d=1$~~ & ~~$d=2$~~ & ~~$d=3$~~ & ~~$d=4$~~ & ~~$d=5$~~ & ~~$d=6$~~ & ~~$d=7$~~ & ~~$d=8$~~ \\ \hline    
    A + ${\cal \M}$ & AIII + ${\sf \M}_+$ & $\mathbb{Z}$ & $0$ & $\mathbb{Z}$ & $0$ & $\mathbb{Z}$ & $0$ & $\mathbb{Z}$ & $0$ \\
    A + $\tilde{\cal \M}$ & AIII + ${\sf \M}_-$ & $0$ & \nsg{$\mathbb{Z}_2$} & $0$ & \nsg{$\mathbb{Z}_2$} & $0$ & \nsg{$\mathbb{Z}_2$} & $0$ & \nsg{$\mathbb{Z}_2$} \\ \hline
    AIII + ${\cal \M}_+$ & A + ${\mathsf \M}$ & $0$ & $\mathbb{Z}$ & $0$ & $\mathbb{Z}$ & $0$ & $\mathbb{Z}$ & $0$ & $\mathbb{Z}$ \\
    AIII + ${\cal \M}_-$ & A + $\bar{\mathsf \M}$ & \nsg{$\mathbb{Z}_2$} & $0$ & \nsg{$\mathbb{Z}_2$} & $0$ & \nsg{$\mathbb{Z}_2$} & $0$ & \nsg{$\mathbb{Z}_2$} & $0$ \\
    AIII + $\tilde{\cal \M}_+$ & A + $\bar{\mathsf \M}$ & \nsg{$\mathbb{Z}_2$} & $0$ & \nsg{$\mathbb{Z}_2$} & $0$ & \nsg{$\mathbb{Z}_2$} & $0$ & \nsg{$\mathbb{Z}_2$} & $0$ \\
    AIII + $\tilde{\cal \M}_-$ & A + ${\mathsf \M}$ & $0$ & $\mathbb{Z}$ & $0$ & $\mathbb{Z}$ & $0$ & $\mathbb{Z}$ & $0$ & $\mathbb{Z}$ \\ \hline \hline
  \end{tabular}
\end{table}

\clearpage
\begin{table}[H]
	\centering
	\caption{Classification of point-gap topology in the real Altland-Zirnbauer symmetry classes protected by order-two nonsymmorphic unitary symmetry with $d_{\parallel} \equiv 0$ (mod $2$).
    ${\cal \M}$ and $\tilde{\cal \M}$ denote nonsymmorphic symmetry and pseudo-nonsymmorphic symmetry (or equivalently, nonsymmorphic symmetry$^{\dag}$), respectively.
    The subscript of ${\cal \M}_\pm$ or $\tilde{\cal \M}_\pm$ specifies the commutation ($+$) or anticommutation ($-$) relation to time-reversal symmetry or particle-hole symmetry.
    For the symmetry classes involving both time-reversal symmetry and particle-hole symmetry (i.e., classes BDI, DIII, CII, and CI), the first subscript of ${\cal \M}_{\pm\pm}$ or $\tilde{\cal \M}_{\pm\pm}$ specifies the relation to time-reversal symmetry and the second one to particle-hole symmetry.
    The topological indices highlighted in color indicate classification intrinsic to nonsymmorphic symmetry.}
	\label{tab: real AZ 0}
     \begin{tabular}{cccccccccc} \hline \hline
    ~~Class~~ & ~~Hermitization~~ & ~~$d=1$~~ & ~~$d=2$~~ & ~~$d=3$~~ & ~~$d=4$~~ & ~~$d=5$~~ & ~~$d=6$~~ & ~~$d=7$~~ & ~~$d=8$~~ \\ \hline    
    AI + ${\cal \M}_+$ & ~~BDI + ${\sf \M}_{++}$~~ & ~~$\mathbb{Z} \oplus \nsg{\mathbb{Z}_2}$~~ & $0$ & $0$ & $0$ & $2\mathbb{Z}$ & $0$ & $\mathbb{Z}_2$ & ~~$\mathbb{Z}_2 \oplus \nsg{\mathbb{Z}_2}$~~ \\
    AI + ${\cal \M}_-$ & BDI + ${\sf \M}_{--}$ & ~~$\mathbb{Z} \oplus \nsg{\mathbb{Z}_2}$~~ & $0$ & $0$ & $0$ & $2\mathbb{Z}$ & $0$ & $\mathbb{Z}_2$ & ~~$\mathbb{Z}_2 \oplus \nsg{\mathbb{Z}_2}$~~ \\
    AI + $\tilde{\cal \M}_+$ & BDI + ${\sf \M}_{+-}$ & \nsg{$\mathbb{Z}_2$} & \nsg{$\mathbb{Z}_2$} & $0$ & $0$ & $0$ & \nsg{$\mathbb{Z}_2$} & $\mathbb{Z}_2$ & \nsg{$\mathbb{Z}_4$} \\
    AI + $\tilde{\cal \M}_-$ & BDI + ${\sf \M}_{-+}$ & \nsg{$\mathbb{Z}_2$} & \nsg{$\mathbb{Z}_2$} & $0$ & $0$ & $0$ & \nsg{$\mathbb{Z}_2$} & $\mathbb{Z}_2$ & \nsg{$\mathbb{Z}_4$} \\ \hline
    BDI + ${\cal \M}_{++}$ & D + ${\sf \M}_+$ & ~~$\mathbb{Z}_2 \oplus \nsg{\mathbb{Z}_2}$~~ & ~~$\mathbb{Z} \oplus \nsg{\mathbb{Z}_2}$~~ & $0$ & $0$ & $0$ & $2\mathbb{Z}$ & $0$ & $\mathbb{Z}_2$ \\
    BDI + ${\cal \M}_{+-}$ & D + $\bar{\sf \M}_+$ & \nsg{$\mathbb{Z}_4$} & \nsg{$\mathbb{Z}_2$} & \nsg{$\mathbb{Z}_2$} & $0$ & $0$ & $0$ & \nsg{$\mathbb{Z}_2$} & $\mathbb{Z}_2$ \\
    BDI + ${\cal \M}_{-+}$ & D + $\bar{\sf \M}_-$ & \nsg{$\mathbb{Z}_4$} & \nsg{$\mathbb{Z}_2$} & \nsg{$\mathbb{Z}_2$} & $0$ & $0$ & $0$ & \nsg{$\mathbb{Z}_2$} & $\mathbb{Z}_2$ \\
    BDI + ${\cal \M}_{--}$ & D + ${\sf \M}_-$ & ~~$\mathbb{Z}_2 \oplus \nsg{\mathbb{Z}_2}$~~ & ~~$\mathbb{Z} \oplus \nsg{\mathbb{Z}_2}$~~ & $0$ & $0$ & $0$ & $2\mathbb{Z}$ & $0$ & $\mathbb{Z}_2$ \\ 
    BDI + $\tilde{\cal \M}_{++}$ & D + $\bar{\sf \M}_+$ & \nsg{$\mathbb{Z}_4$} & \nsg{$\mathbb{Z}_2$} & \nsg{$\mathbb{Z}_2$} & $0$ & $0$ & $0$ & \nsg{$\mathbb{Z}_2$} & $\mathbb{Z}_2$ \\
    BDI + $\tilde{\cal \M}_{+-}$ & D + ${\sf \M}_-$ & ~~$\mathbb{Z}_2 \oplus \nsg{\mathbb{Z}_2}$~~ & ~~$\mathbb{Z} \oplus \nsg{\mathbb{Z}_2}$~~ & $0$ & $0$ & $0$ & $2\mathbb{Z}$ & $0$ & $\mathbb{Z}_2$ \\
    BDI + $\tilde{\cal \M}_{-+}$ & D + ${\sf \M}_+$ & ~~$\mathbb{Z}_2 \oplus \nsg{\mathbb{Z}_2}$~~ & ~~$\mathbb{Z} \oplus \nsg{\mathbb{Z}_2}$~~ & $0$ & $0$ & $0$ & $2\mathbb{Z}$ & $0$ & $\mathbb{Z}_2$ \\
    BDI + $\tilde{\cal \M}_{--}$ & D + $\bar{\sf \M}_-$ & \nsg{$\mathbb{Z}_4$} & \nsg{$\mathbb{Z}_2$} & \nsg{$\mathbb{Z}_2$} & $0$ & $0$ & $0$ & \nsg{$\mathbb{Z}_2$} & $\mathbb{Z}_2$ \\ \hline
    D + ${\cal \M}_+$ & ~~DIII + ${\mathsf \M}_{++}$~~ & $\mathbb{Z}_2$ & ~~$\mathbb{Z}_2 \oplus \nsg{\mathbb{Z}_2}$~~ & ~~$\mathbb{Z} \oplus \nsg{\mathbb{Z}_2}$~~ & $0$ & $0$ & $0$ & $2\mathbb{Z}$ & $0$ \\
    D + ${\cal \M}_-$ & DIII + ${\mathsf \M}_{--}$ & $\mathbb{Z}_2$ & ~~$\mathbb{Z}_2 \oplus \nsg{\mathbb{Z}_2}$~~ & ~~$\mathbb{Z} \oplus \nsg{\mathbb{Z}_2}$~~ & $0$ & $0$ & $0$ & $2\mathbb{Z}$ & $0$ \\
    D + $\tilde{\cal \M}_+$ & DIII + ${\mathsf \M}_{-+}$ & $\mathbb{Z}_2$ & \nsg{$\mathbb{Z}_4$} & \nsg{$\mathbb{Z}_2$} & \nsg{$\mathbb{Z}_2$} & $0$ & $0$ & $0$ & \nsg{$\mathbb{Z}_2$} \\
    D + $\tilde{\cal \M}_-$ & DIII + ${\mathsf \M}_{+-}$ & $\mathbb{Z}_2$ & \nsg{$\mathbb{Z}_4$} & \nsg{$\mathbb{Z}_2$} & \nsg{$\mathbb{Z}_2$} & $0$ & $0$ & $0$ & \nsg{$\mathbb{Z}_2$} \\ \hline
    ~~DIII + ${\cal \M}_{++}$~~ & AII + ${\sf \M}_{+}$ & $0$ & $\mathbb{Z}_2$ & ~~$\mathbb{Z}_2 \oplus \nsg{\mathbb{Z}_2}$~~ & ~~$\mathbb{Z} \oplus \nsg{\mathbb{Z}_2}$~~ & $0$ & $0$ & $0$ & $2\mathbb{Z}$ \\
    DIII + ${\cal \M}_{+-}$ & AII + $\bar{\sf \M}_+$ & \nsg{$\mathbb{Z}_2$} & $\mathbb{Z}_2$ & \nsg{$\mathbb{Z}_4$} & \nsg{$\mathbb{Z}_2$} & \nsg{$\mathbb{Z}_2$} & $0$ & $0$ & $0$ \\
    DIII + ${\cal \M}_{-+}$ & AII + $\bar{\sf \M}_-$ & \nsg{$\mathbb{Z}_2$} & $\mathbb{Z}_2$ & \nsg{$\mathbb{Z}_4$} & \nsg{$\mathbb{Z}_2$} & \nsg{$\mathbb{Z}_2$} & $0$ & $0$ & $0$ \\
    DIII + ${\cal \M}_{--}$ & AII + ${\sf \M}_{-}$ & $0$ & $\mathbb{Z}_2$ & ~~$\mathbb{Z}_2 \oplus \nsg{\mathbb{Z}_2}$~~ & ~~$\mathbb{Z} \oplus \nsg{\mathbb{Z}_2}$~~ & $0$ & $0$ & $0$ & $2\mathbb{Z}$ \\ 
    DIII + $\tilde{\cal \M}_{++}$ & AII + $\bar{\sf \M}_-$ & \nsg{$\mathbb{Z}_2$} & $\mathbb{Z}_2$ & \nsg{$\mathbb{Z}_4$} & \nsg{$\mathbb{Z}_2$} & \nsg{$\mathbb{Z}_2$} & $0$ & $0$ & $0$ \\
    DIII + $\tilde{\cal \M}_{+-}$ & AII + ${\sf \M}_{+}$ & $0$ & $\mathbb{Z}_2$ & ~~$\mathbb{Z}_2 \oplus \nsg{\mathbb{Z}_2}$~~ & ~~$\mathbb{Z} \oplus \nsg{\mathbb{Z}_2}$~~ & $0$ & $0$ & $0$ & $2\mathbb{Z}$ \\
    DIII + $\tilde{\cal \M}_{-+}$ & AII + ${\sf \M}_{-}$ & $0$ & $\mathbb{Z}_2$ & ~~$\mathbb{Z}_2 \oplus \nsg{\mathbb{Z}_2}$~~ & ~~$\mathbb{Z} \oplus \nsg{\mathbb{Z}_2}$~~ & $0$ & $0$ & $0$ & $2\mathbb{Z}$ \\
    DIII + $\tilde{\cal \M}_{--}$ & AII + $\bar{\sf \M}_+$ & \nsg{$\mathbb{Z}_2$} & $\mathbb{Z}_2$ & \nsg{$\mathbb{Z}_4$} & \nsg{$\mathbb{Z}_2$} & \nsg{$\mathbb{Z}_2$} & $0$ & $0$ & $0$ \\ \hline
    AII + ${\cal \M}_+$ & CII + ${\sf \M}_{++}$ & ~~$2\mathbb{Z}$~~ & $0$ & $\mathbb{Z}_2$ & ~~$\mathbb{Z}_2 \oplus \nsg{\mathbb{Z}_2}$~~ & ~~$\mathbb{Z} \oplus \nsg{\mathbb{Z}_2}$~~ & $0$ & $0$ & $0$ \\
    AII + ${\cal \M}_-$ & CII + ${\sf \M}_{--}$ & ~~$2\mathbb{Z}$~~ & $0$ & $\mathbb{Z}_2$ & ~~$\mathbb{Z}_2 \oplus \nsg{\mathbb{Z}_2}$~~ & ~~$\mathbb{Z} \oplus \nsg{\mathbb{Z}_2}$~~ & $0$ & $0$ & $0$ \\
    AII + $\tilde{\cal \M}_+$ & CII + ${\sf \M}_{+-}$ & $0$ & \nsg{$\mathbb{Z}_2$} & $\mathbb{Z}_2$ & \nsg{$\mathbb{Z}_4$} & \nsg{$\mathbb{Z}_2$} & \nsg{$\mathbb{Z}_2$} & $0$ & $0$ \\
    AII + $\tilde{\cal \M}_-$ & CII + ${\sf \M}_{-+}$ & $0$ & \nsg{$\mathbb{Z}_2$} & $\mathbb{Z}_2$ & \nsg{$\mathbb{Z}_4$} & \nsg{$\mathbb{Z}_2$} & \nsg{$\mathbb{Z}_2$} & $0$ & $0$ \\ \hline
    CII + ${\cal \M}_{++}$ & C + ${\sf \M}_+$ & $0$ & ~~$2\mathbb{Z}$~~ & $0$ & $\mathbb{Z}_2$ & ~~$\mathbb{Z}_2 \oplus \nsg{\mathbb{Z}_2}$~~ & ~~$\mathbb{Z} \oplus \nsg{\mathbb{Z}_2}$~~ & $0$ & $0$ \\
    CII + ${\cal \M}_{+-}$ & C + $\bar{\sf \M}_+$ & $0$ & $0$ & \nsg{$\mathbb{Z}_2$} & $\mathbb{Z}_2$ & \nsg{$\mathbb{Z}_4$} & \nsg{$\mathbb{Z}_2$} & \nsg{$\mathbb{Z}_2$} & $0$ \\
    CII + ${\cal \M}_{-+}$ & C + $\bar{\sf \M}_-$ & $0$ & $0$ & \nsg{$\mathbb{Z}_2$} & $\mathbb{Z}_2$ & \nsg{$\mathbb{Z}_4$} & \nsg{$\mathbb{Z}_2$} & \nsg{$\mathbb{Z}_2$} & $0$  \\
    CII + ${\cal \M}_{--}$ & C + ${\sf \M}_-$ & $0$ & ~~$2\mathbb{Z}$~~ & $0$ & $\mathbb{Z}_2$ & ~~$\mathbb{Z}_2 \oplus \nsg{\mathbb{Z}_2}$~~ & ~~$\mathbb{Z} \oplus \nsg{\mathbb{Z}_2}$~~ & $0$ & $0$ \\ 
    CII + $\tilde{\cal \M}_{++}$ & C + $\bar{\sf \M}_+$ & $0$ & $0$ & \nsg{$\mathbb{Z}_2$} & $\mathbb{Z}_2$ & \nsg{$\mathbb{Z}_4$} & \nsg{$\mathbb{Z}_2$} & \nsg{$\mathbb{Z}_2$} & $0$ \\
    CII + $\tilde{\cal \M}_{+-}$ & C + ${\sf \M}_-$ & $0$ & ~~$2\mathbb{Z}$~~ & $0$ & $\mathbb{Z}_2$ & ~~$\mathbb{Z}_2 \oplus \nsg{\mathbb{Z}_2}$~~ & ~~$\mathbb{Z} \oplus \nsg{\mathbb{Z}_2}$~~ & $0$ & $0$ \\
    CII + $\tilde{\cal \M}_{-+}$ & C + ${\sf \M}_+$ & $0$ & ~~$2\mathbb{Z}$~~ & $0$ & $\mathbb{Z}_2$ & ~~$\mathbb{Z}_2 \oplus \nsg{\mathbb{Z}_2}$~~ & ~~$\mathbb{Z} \oplus \nsg{\mathbb{Z}_2}$~~ & $0$ & $0$ \\
    CII + $\tilde{\cal \M}_{--}$ & C + $\bar{\sf \M}_-$ & $0$ & $0$ & \nsg{$\mathbb{Z}_2$} & $\mathbb{Z}_2$ & \nsg{$\mathbb{Z}_4$} & \nsg{$\mathbb{Z}_2$} & \nsg{$\mathbb{Z}_2$} & $0$ \\ \hline
    C + ${\cal \M}_+$ & CI + ${\sf \M}_{++}$ & $0$ & $0$ & $2\mathbb{Z}$ & $0$ & $\mathbb{Z}_2$ & ~~$\mathbb{Z}_2 \oplus \nsg{\mathbb{Z}_2}$~~ & ~~$\mathbb{Z} \oplus \nsg{\mathbb{Z}_2}$~~ & $0$ \\
    C + ${\cal \M}_-$ & CI + ${\sf \M}_{--}$ & $0$ & $0$ & $2\mathbb{Z}$ & $0$ & $\mathbb{Z}_2$ & ~~$\mathbb{Z}_2 \oplus \nsg{\mathbb{Z}_2}$~~ & ~~$\mathbb{Z} \oplus \nsg{\mathbb{Z}_2}$~~ & $0$ \\
    C + $\tilde{\cal \M}_+$ & CI + ${\sf \M}_{-+}$ & $0$ & $0$ & $0$ & \nsg{$\mathbb{Z}_2$} & $\mathbb{Z}_2$ & \nsg{$\mathbb{Z}_4$} & \nsg{$\mathbb{Z}_2$} & \nsg{$\mathbb{Z}_2$} \\
    C + $\tilde{\cal \M}_-$ & CI + ${\sf \M}_{+-}$ & $0$ & $0$ & $0$ & \nsg{$\mathbb{Z}_2$} & $\mathbb{Z}_2$ & \nsg{$\mathbb{Z}_4$} & \nsg{$\mathbb{Z}_2$} & \nsg{$\mathbb{Z}_2$} \\ \hline
    CI + ${\cal \M}_{++}$ & AI + ${\sf \M}_+$ & $0$ & $0$ & $0$ & $2\mathbb{Z}$ & $0$ & $\mathbb{Z}_2$ & ~~$\mathbb{Z}_2 \oplus \nsg{\mathbb{Z}_2}$~~ & ~~$\mathbb{Z} \oplus \nsg{\mathbb{Z}_2}$~~ \\
    CI + ${\cal \M}_{+-}$ & AI + $\bar{\sf \M}_+$ & \nsg{$\mathbb{Z}_2$} & $0$ & $0$ & $0$ & \nsg{$\mathbb{Z}_2$} & $\mathbb{Z}_2$ & \nsg{$\mathbb{Z}_4$} & \nsg{$\mathbb{Z}_2$} \\
    CI + ${\cal \M}_{-+}$ & AI + $\bar{\sf \M}_-$ & \nsg{$\mathbb{Z}_2$} & $0$ & $0$ & $0$ & \nsg{$\mathbb{Z}_2$} & $\mathbb{Z}_2$ & \nsg{$\mathbb{Z}_4$} & \nsg{$\mathbb{Z}_2$} \\
    CI + ${\cal \M}_{--}$ & AI + ${\sf \M}_-$ & $0$ & $0$ & $0$ & $2\mathbb{Z}$ & $0$ & $\mathbb{Z}_2$ & ~~$\mathbb{Z}_2 \oplus \nsg{\mathbb{Z}_2}$~~ & ~~$\mathbb{Z} \oplus \nsg{\mathbb{Z}_2}$~~ \\ 
    CI + $\tilde{\cal \M}_{++}$ & AI + $\bar{\sf \M}_-$ & \nsg{$\mathbb{Z}_2$} & $0$ & $0$ & $0$ & \nsg{$\mathbb{Z}_2$} & $\mathbb{Z}_2$ & \nsg{$\mathbb{Z}_4$} & \nsg{$\mathbb{Z}_2$} \\
    CI + $\tilde{\cal \M}_{+-}$ & AI + ${\sf \M}_+$ & $0$ & $0$ & $0$ & $2\mathbb{Z}$ & $0$ & $\mathbb{Z}_2$ & ~~$\mathbb{Z}_2 \oplus \nsg{\mathbb{Z}_2}$~~ & ~~$\mathbb{Z} \oplus \nsg{\mathbb{Z}_2}$~~ \\
    CI + $\tilde{\cal \M}_{-+}$ & AI + ${\sf \M}_-$ & $0$ & $0$ & $0$ & $2\mathbb{Z}$ & $0$ & $\mathbb{Z}_2$ & ~~$\mathbb{Z}_2 \oplus \nsg{\mathbb{Z}_2}$~~ & ~~$\mathbb{Z} \oplus \nsg{\mathbb{Z}_2}$~~ \\
    CI + $\tilde{\cal \M}_{--}$ & AI + $\bar{\sf \M}_+$ & \nsg{$\mathbb{Z}_2$} & $0$ & $0$ & $0$ & \nsg{$\mathbb{Z}_2$} & $\mathbb{Z}_2$ & \nsg{$\mathbb{Z}_4$} & \nsg{$\mathbb{Z}_2$} \\ \hline \hline
  \end{tabular}
\end{table}

\clearpage
\begin{table}[H]
	\centering
	\caption{Classification of point-gap topology in the real Altland-Zirnbauer$^{\dag}$ symmetry classes protected by order-two nonsymmorphic unitary symmetry with $d_{\parallel} \equiv 0$ (mod $2$).
    ${\cal \M}$ and $\tilde{\cal \M}$ denote nonsymmorphic symmetry and pseudo-nonsymmorphic symmetry (or equivalently, nonsymmorphic symmetry$^{\dag}$), respectively.
    The subscript of ${\cal \M}_\pm$ or $\tilde{\cal \M}_\pm$ specifies the commutation ($+$) or anticommutation ($-$) relation to time-reversal symmetry$^{\dag}$ or particle-hole symmetry$^{\dag}$.
    For the symmetry classes involving both time-reversal symmetry$^{\dag}$ and particle-hole symmetry$^{\dag}$ (i.e., classes BDI$^{\dag}$, DIII$^{\dag}$, CII$^{\dag}$, and CI$^{\dag}$), the first subscript of ${\cal \M}_{\pm\pm}$ or $\tilde{\cal \M}_{\pm\pm}$ specifies the relation to time-reversal symmetry$^{\dag}$ and the second one to particle-hole symmetry$^{\dag}$.
    The topological indices highlighted in color indicate classification intrinsic to nonsymmorphic symmetry.}
	\label{tab: real AZ dag 0}
     \begin{tabular}{cccccccccc} \hline \hline
    ~~Class~~ & ~~Hermitization~~ & ~~$d=1$~~ & ~~$d=2$~~ & ~~$d=3$~~ & ~~$d=4$~~ & ~~$d=5$~~ & ~~$d=6$~~ & ~~$d=7$~~ & ~~$d=8$~~ \\ \hline    
    AI$^{\dag}$ + ${\cal \M}_+$ & CI + ${\sf \M}_{++}$ & $0$ & $0$ & $2\mathbb{Z}$ & $0$ & $\mathbb{Z}_2$ & ~~$\mathbb{Z}_2 \oplus \nsg{\mathbb{Z}_2}$~~ & ~~$\mathbb{Z} \oplus \nsg{\mathbb{Z}_2}$~~ & $0$ \\
    AI$^{\dag}$ + ${\cal \M}_-$ & CI + ${\sf \M}_{--}$ & $0$ & $0$ & $2\mathbb{Z}$ & $0$ & $\mathbb{Z}_2$ & ~~$\mathbb{Z}_2 \oplus \nsg{\mathbb{Z}_2}$~~ & ~~$\mathbb{Z} \oplus \nsg{\mathbb{Z}_2}$~~ & $0$ \\
    AI$^{\dag}$ + $\tilde{\cal \M}_+$ & CI + ${\sf \M}_{+-}$ & $0$ & $0$ & $0$ & \nsg{$\mathbb{Z}_2$} & $\mathbb{Z}_2$ & \nsg{$\mathbb{Z}_4$} & \nsg{$\mathbb{Z}_2$} & \nsg{$\mathbb{Z}_2$} \\
    AI$^{\dag}$ + $\tilde{\cal \M}_-$ & CI + ${\sf \M}_{-+}$ & $0$ & $0$ & $0$ & \nsg{$\mathbb{Z}_2$} & $\mathbb{Z}_2$ & \nsg{$\mathbb{Z}_4$} & \nsg{$\mathbb{Z}_2$} & \nsg{$\mathbb{Z}_2$} \\ \hline
    BDI$^{\dag}$ + ${\cal \M}_{++}$ & AI + ${\sf \M}_+$ & $0$ & $0$ & $0$ & $2\mathbb{Z}$ & $0$ & $\mathbb{Z}_2$ & ~~$\mathbb{Z}_2 \oplus \nsg{\mathbb{Z}_2}$~~ & ~~$\mathbb{Z} \oplus \nsg{\mathbb{Z}_2}$~~ \\
    BDI$^{\dag}$ + ${\cal \M}_{+-}$ & AI + $\bar{\sf \M}_-$ & \nsg{$\mathbb{Z}_2$} & $0$ & $0$ & $0$ & \nsg{$\mathbb{Z}_2$} & $\mathbb{Z}_2$ & \nsg{$\mathbb{Z}_4$} & \nsg{$\mathbb{Z}_2$} \\
    BDI$^{\dag}$ + ${\cal \M}_{-+}$ & AI + $\bar{\sf \M}_+$ & \nsg{$\mathbb{Z}_2$} & $0$ & $0$ & $0$ & \nsg{$\mathbb{Z}_2$} & $\mathbb{Z}_2$ & \nsg{$\mathbb{Z}_4$} & \nsg{$\mathbb{Z}_2$} \\
    BDI$^{\dag}$ + ${\cal \M}_{--}$ & AI + ${\sf \M}_-$ & $0$ & $0$ & $0$ & $2\mathbb{Z}$ & $0$ & $\mathbb{Z}_2$ & ~~$\mathbb{Z}_2 \oplus \nsg{\mathbb{Z}_2}$~~ & ~~$\mathbb{Z} \oplus \nsg{\mathbb{Z}_2}$~~ \\ 
    BDI$^{\dag}$ + $\tilde{\cal \M}_{++}$ & AI + $\bar{\sf \M}_+$ & \nsg{$\mathbb{Z}_2$} & $0$ & $0$ & $0$ & \nsg{$\mathbb{Z}_2$} & $\mathbb{Z}_2$ & \nsg{$\mathbb{Z}_4$} & \nsg{$\mathbb{Z}_2$} \\
    BDI$^{\dag}$ + $\tilde{\cal \M}_{+-}$ & AI + ${\sf \M}_+$ & $0$ & $0$ & $0$ & $2\mathbb{Z}$ & $0$ & $\mathbb{Z}_2$ & ~~$\mathbb{Z}_2 \oplus \nsg{\mathbb{Z}_2}$~~ & ~~$\mathbb{Z} \oplus \nsg{\mathbb{Z}_2}$~~ \\
    BDI$^{\dag}$ + $\tilde{\cal \M}_{-+}$ & AI + ${\sf \M}_-$ & $0$ & $0$ & $0$ & $2\mathbb{Z}$ & $0$ & $\mathbb{Z}_2$ & ~~$\mathbb{Z}_2 \oplus \nsg{\mathbb{Z}_2}$~~ & ~~$\mathbb{Z} \oplus \nsg{\mathbb{Z}_2}$~~ \\
    BDI$^{\dag}$ + $\tilde{\cal \M}_{--}$ & AI + $\bar{\sf \M}_-$ & \nsg{$\mathbb{Z}_2$} & $0$ & $0$ & $0$ & \nsg{$\mathbb{Z}_2$} & $\mathbb{Z}_2$ & \nsg{$\mathbb{Z}_4$} & \nsg{$\mathbb{Z}_2$} \\ \hline
    D$^{\dag}$ + ${\cal \M}_+$ & BDI + ${\sf \M}_{++}$ & ~~$\mathbb{Z} \oplus \nsg{\mathbb{Z}_2}$~~ & $0$ & $0$ & $0$ & $2\mathbb{Z}$ & $0$ & $\mathbb{Z}_2$ & ~~$\mathbb{Z}_2 \oplus \nsg{\mathbb{Z}_2}$~~ \\
    D$^{\dag}$ + ${\cal \M}_-$ & BDI + ${\sf \M}_{--}$ & ~~$\mathbb{Z} \oplus \nsg{\mathbb{Z}_2}$~~ & $0$ & $0$ & $0$ & $2\mathbb{Z}$ & $0$ & $\mathbb{Z}_2$ & ~~$\mathbb{Z}_2 \oplus \nsg{\mathbb{Z}_2}$~~ \\
    D$^{\dag}$ + $\tilde{\cal \M}_+$ & BDI + ${\sf \M}_{-+}$ & \nsg{$\mathbb{Z}_2$} & \nsg{$\mathbb{Z}_2$} & $0$ & $0$ & $0$ & \nsg{$\mathbb{Z}_2$} & $\mathbb{Z}_2$ & \nsg{$\mathbb{Z}_4$} \\
    D$^{\dag}$ + $\tilde{\cal \M}_-$ & BDI + ${\sf \M}_{+-}$ & \nsg{$\mathbb{Z}_2$} & \nsg{$\mathbb{Z}_2$} & $0$ & $0$ & $0$ & \nsg{$\mathbb{Z}_2$} & $\mathbb{Z}_2$ & \nsg{$\mathbb{Z}_4$} \\ \hline
    ~~DIII$^{\dag}$ + ${\cal \M}_{++}$~~ & D + ${\sf \M}_+$ & ~~$\mathbb{Z}_2 \oplus \nsg{\mathbb{Z}_2}$~~ & ~~$\mathbb{Z} \oplus \nsg{\mathbb{Z}_2}$~~ & $0$ & $0$ & $0$ & $2\mathbb{Z}$ & $0$ & $\mathbb{Z}_2$ \\
    DIII$^{\dag}$ + ${\cal \M}_{+-}$ & D + $\bar{\sf \M}_-$ & \nsg{$\mathbb{Z}_4$} & \nsg{$\mathbb{Z}_2$} & \nsg{$\mathbb{Z}_2$} & $0$ & $0$ & $0$ & \nsg{$\mathbb{Z}_2$} & $\mathbb{Z}_2$ \\
    DIII$^{\dag}$ + ${\cal \M}_{-+}$ & D + $\bar{\sf \M}_+$ & \nsg{$\mathbb{Z}_4$} & \nsg{$\mathbb{Z}_2$} & \nsg{$\mathbb{Z}_2$} & $0$ & $0$ & $0$ & \nsg{$\mathbb{Z}_2$} & $\mathbb{Z}_2$ \\
    DIII$^{\dag}$ + ${\cal \M}_{--}$ & D + ${\sf \M}_-$ & ~~$\mathbb{Z}_2 \oplus \nsg{\mathbb{Z}_2}$~~ & ~~$\mathbb{Z} \oplus \nsg{\mathbb{Z}_2}$~~ & $0$ & $0$ & $0$ & $2\mathbb{Z}$ & $0$ & $\mathbb{Z}_2$ \\ 
    DIII$^{\dag}$ + $\tilde{\cal \M}_{++}$ & D + $\bar{\sf \M}_-$ & \nsg{$\mathbb{Z}_4$} & \nsg{$\mathbb{Z}_2$} & \nsg{$\mathbb{Z}_2$} & $0$ & $0$ & $0$ & \nsg{$\mathbb{Z}_2$} & $\mathbb{Z}_2$ \\
    DIII$^{\dag}$ + $\tilde{\cal \M}_{+-}$ & D + ${\sf \M}_-$ & ~~$\mathbb{Z}_2 \oplus \nsg{\mathbb{Z}_2}$~~ & ~~$\mathbb{Z} \oplus \nsg{\mathbb{Z}_2}$~~ & $0$ & $0$ & $0$ & $2\mathbb{Z}$ & $0$ & $\mathbb{Z}_2$ \\
    DIII$^{\dag}$ + $\tilde{\cal \M}_{-+}$ & D + ${\sf \M}_+$ & ~~$\mathbb{Z}_2 \oplus \nsg{\mathbb{Z}_2}$~~ & ~~$\mathbb{Z} \oplus \nsg{\mathbb{Z}_2}$~~ & $0$ & $0$ & $0$ & $2\mathbb{Z}$ & $0$ & $\mathbb{Z}_2$ \\
    DIII$^{\dag}$ + $\tilde{\cal \M}_{--}$ & D + $\bar{\sf \M}_+$ & \nsg{$\mathbb{Z}_4$} & \nsg{$\mathbb{Z}_2$} & \nsg{$\mathbb{Z}_2$} & $0$ & $0$ & $0$ & \nsg{$\mathbb{Z}_2$} & $\mathbb{Z}_2$ \\ \hline
    AII$^{\dag}$ + ${\cal \M}_+$ & ~~DIII + ${\sf \M}_{++}$~~ & $\mathbb{Z}_2$ & ~~$\mathbb{Z}_2 \oplus \nsg{\mathbb{Z}_2}$~~ & ~~$\mathbb{Z} \oplus \nsg{\mathbb{Z}_2}$~~ & $0$ & $0$ & $0$ & $2\mathbb{Z}$ & $0$ \\
    AII$^{\dag}$ + ${\cal \M}_-$ & DIII + ${\sf \M}_{--}$ & $\mathbb{Z}_2$ & ~~$\mathbb{Z}_2 \oplus \nsg{\mathbb{Z}_2}$~~ & ~~$\mathbb{Z} \oplus \nsg{\mathbb{Z}_2}$~~ & $0$ & $0$ & $0$ & $2\mathbb{Z}$ & $0$ \\
    AII$^{\dag}$ + $\tilde{\cal \M}_+$ & DIII + ${\sf \M}_{+-}$ & $\mathbb{Z}_2$ & \nsg{$\mathbb{Z}_4$} & \nsg{$\mathbb{Z}_2$} & \nsg{$\mathbb{Z}_2$} & $0$ & $0$ & $0$ & \nsg{$\mathbb{Z}_2$} \\
    AII$^{\dag}$ + $\tilde{\cal \M}_-$ & DIII + ${\sf \M}_{-+}$ & $\mathbb{Z}_2$ & \nsg{$\mathbb{Z}_4$} & \nsg{$\mathbb{Z}_2$} & \nsg{$\mathbb{Z}_2$} & $0$ & $0$ & $0$ & \nsg{$\mathbb{Z}_2$} \\ \hline
    CII$^{\dag}$ + ${\cal \M}_{++}$ & AII + ${\sf \M}_+$ & $0$ & $\mathbb{Z}_2$ & ~~$\mathbb{Z}_2 \oplus \nsg{\mathbb{Z}_2}$~~ & ~~$\mathbb{Z} \oplus \nsg{\mathbb{Z}_2}$~~ & $0$ & $0$ & $0$ & $2\mathbb{Z}$ \\
    CII$^{\dag}$ + ${\cal \M}_{+-}$ & AII + $\bar{\sf \M}_-$ & \nsg{$\mathbb{Z}_2$} & $\mathbb{Z}_2$ & \nsg{$\mathbb{Z}_4$} & \nsg{$\mathbb{Z}_2$} & \nsg{$\mathbb{Z}_2$} & $0$ & $0$ & $0$ \\
    CII$^{\dag}$ + ${\cal \M}_{-+}$ & AII + $\bar{\sf \M}_+$ & \nsg{$\mathbb{Z}_2$} & $\mathbb{Z}_2$ & \nsg{$\mathbb{Z}_4$} & \nsg{$\mathbb{Z}_2$} & \nsg{$\mathbb{Z}_2$} & $0$ & $0$ & $0$ \\
    CII$^{\dag}$ + ${\cal \M}_{--}$ & AII + ${\sf \M}_-$ & $0$ & $\mathbb{Z}_2$ & ~~$\mathbb{Z}_2 \oplus \nsg{\mathbb{Z}_2}$~~ & ~~$\mathbb{Z} \oplus \nsg{\mathbb{Z}_2}$~~ & $0$ & $0$ & $0$ & $2\mathbb{Z}$ \\ 
    CII$^{\dag}$ + $\tilde{\cal \M}_{++}$ & AII + $\bar{\sf \M}_+$ & \nsg{$\mathbb{Z}_2$} & $\mathbb{Z}_2$ & \nsg{$\mathbb{Z}_4$} & \nsg{$\mathbb{Z}_2$} & \nsg{$\mathbb{Z}_2$} & $0$ & $0$ & $0$ \\
    CII$^{\dag}$ + $\tilde{\cal \M}_{+-}$ & AII + ${\sf \M}_+$ & $0$ & $\mathbb{Z}_2$ & ~~$\mathbb{Z}_2 \oplus \nsg{\mathbb{Z}_2}$~~ & ~~$\mathbb{Z} \oplus \nsg{\mathbb{Z}_2}$~~ & $0$ & $0$ & $0$ & $2\mathbb{Z}$ \\
    CII$^{\dag}$ + $\tilde{\cal \M}_{-+}$ & AII + ${\sf \M}_-$ & $0$ & $\mathbb{Z}_2$ & ~~$\mathbb{Z}_2 \oplus \nsg{\mathbb{Z}_2}$~~ & ~~$\mathbb{Z} \oplus \nsg{\mathbb{Z}_2}$~~ & $0$ & $0$ & $0$ & $2\mathbb{Z}$ \\
    CII$^{\dag}$ + $\tilde{\cal \M}_{--}$ & AII + $\bar{\sf \M}_-$ & \nsg{$\mathbb{Z}_2$} & $\mathbb{Z}_2$ & \nsg{$\mathbb{Z}_4$} & \nsg{$\mathbb{Z}_2$} & \nsg{$\mathbb{Z}_2$} & $0$ & $0$ & $0$ \\ \hline
    C$^{\dag}$ + ${\cal \M}_+$ & CII + ${\sf \M}_{++}$ & ~~$2\mathbb{Z}$~~ & $0$ & $\mathbb{Z}_2$ & ~~$\mathbb{Z}_2 \oplus \nsg{\mathbb{Z}_2}$~~ & ~~$\mathbb{Z} \oplus \nsg{\mathbb{Z}_2}$~~ & $0$ & $0$ & $0$ \\
    C$^{\dag}$ + ${\cal \M}_-$ & CII + ${\sf \M}_{--}$ & ~~$2\mathbb{Z}$~~ & $0$ & $\mathbb{Z}_2$ & ~~$\mathbb{Z}_2 \oplus \nsg{\mathbb{Z}_2}$~~ & ~~$\mathbb{Z} \oplus \nsg{\mathbb{Z}_2}$~~ & $0$ & $0$ & $0$ \\
    C$^{\dag}$ + $\tilde{\cal \M}_+$ & CII + ${\sf \M}_{-+}$ & $0$ & \nsg{$\mathbb{Z}_2$} & $\mathbb{Z}_2$ & \nsg{$\mathbb{Z}_4$} & \nsg{$\mathbb{Z}_2$} & \nsg{$\mathbb{Z}_2$} & $0$ & $0$ \\
    C$^{\dag}$ + $\tilde{\cal \M}_-$ & CII + ${\sf \M}_{+-}$ & $0$ & \nsg{$\mathbb{Z}_2$} & $\mathbb{Z}_2$ & \nsg{$\mathbb{Z}_4$} & \nsg{$\mathbb{Z}_2$} & \nsg{$\mathbb{Z}_2$} & $0$ & $0$ \\ \hline
    CI$^{\dag}$ + ${\cal \M}_{++}$ & C + ${\sf \M}_+$ & $0$ & ~~$2\mathbb{Z}$~~ & $0$ & $\mathbb{Z}_2$ & ~~$\mathbb{Z}_2 \oplus \nsg{\mathbb{Z}_2}$~~ & ~~$\mathbb{Z} \oplus \nsg{\mathbb{Z}_2}$~~ & $0$ & $0$ \\
    CI$^{\dag}$ + ${\cal \M}_{+-}$ & C + $\bar{\sf \M}_-$ & $0$ & $0$ & \nsg{$\mathbb{Z}_2$} & $\mathbb{Z}_2$ & \nsg{$\mathbb{Z}_4$} & \nsg{$\mathbb{Z}_2$} & \nsg{$\mathbb{Z}_2$} & $0$ \\
    CI$^{\dag}$ + ${\cal \M}_{-+}$ & C + $\bar{\sf \M}_+$ & $0$ & $0$ & \nsg{$\mathbb{Z}_2$} & $\mathbb{Z}_2$ & \nsg{$\mathbb{Z}_4$} & \nsg{$\mathbb{Z}_2$} & \nsg{$\mathbb{Z}_2$} & $0$ \\
    CI$^{\dag}$ + ${\cal \M}_{--}$ & C + ${\sf \M}_-$ & $0$ & ~~$2\mathbb{Z}$~~ & $0$ & $\mathbb{Z}_2$ & ~~$\mathbb{Z}_2 \oplus \nsg{\mathbb{Z}_2}$~~ & ~~$\mathbb{Z} \oplus \nsg{\mathbb{Z}_2}$~~ & $0$ & $0$ \\ 
    CI$^{\dag}$ + $\tilde{\cal \M}_{++}$ & C + $\bar{\sf \M}_-$ & $0$ & $0$ & \nsg{$\mathbb{Z}_2$} & $\mathbb{Z}_2$ & \nsg{$\mathbb{Z}_4$} & \nsg{$\mathbb{Z}_2$} & \nsg{$\mathbb{Z}_2$} & $0$ \\
    CI$^{\dag}$ + $\tilde{\cal \M}_{+-}$ & C + ${\sf \M}_-$ & $0$ & ~~$2\mathbb{Z}$~~ & $0$ & $\mathbb{Z}_2$ & ~~$\mathbb{Z}_2 \oplus \nsg{\mathbb{Z}_2}$~~ & ~~$\mathbb{Z} \oplus \nsg{\mathbb{Z}_2}$~~ & $0$ & $0$ \\
    CI$^{\dag}$ + $\tilde{\cal \M}_{-+}$ & C + ${\sf \M}_+$ & $0$ & ~~$2\mathbb{Z}$~~ & $0$ & $\mathbb{Z}_2$ & ~~$\mathbb{Z}_2 \oplus \nsg{\mathbb{Z}_2}$~~ & ~~$\mathbb{Z} \oplus \nsg{\mathbb{Z}_2}$~~ & $0$ & $0$ \\
    CI$^{\dag}$ + $\tilde{\cal \M}_{--}$ & C + $\bar{\sf \M}_+$ & $0$ & $0$ & \nsg{$\mathbb{Z}_2$} & $\mathbb{Z}_2$ & \nsg{$\mathbb{Z}_4$} & \nsg{$\mathbb{Z}_2$} & \nsg{$\mathbb{Z}_2$} & $0$ \\ \hline \hline
  \end{tabular}
\end{table}

\subsection{$d_{\parallel} \equiv 1$ (mod $2$)}

\begin{table}[H]
	\centering
	\caption{Classification of point-gap topology in the complex Altland-Zirnbauer symmetry classes protected by order-two nonsymmorphic unitary symmetry with $d_{\parallel} \equiv 1$ (mod $2$).
    ${\cal \M}$ and $\tilde{\cal \M}$ denote nonsymmorphic symmetry and pseudo-nonsymmorphic symmetry (or equivalently, nonsymmorphic symmetry$^{\dag}$), respectively.
    In class AIII, the subscript of ${\cal \M}_\pm$ or $\tilde{\cal \M}_\pm$ specifies the commutation ($+$) or anticommutation ($-$) relation with chiral symmetry.
    The topological indices highlighted in color indicate classification intrinsic to nonsymmorphic symmetry.}
	\label{tab: complex AZ 1}
     \begin{tabular}{cccccccccc} \hline \hline
    ~~Class~~ & ~~Hermitization~~ & ~~$d=1$~~ & ~~$d=2$~~ & ~~$d=3$~~ & ~~$d=4$~~ & ~~$d=5$~~ & ~~$d=6$~~ & ~~$d=7$~~ & ~~$d=8$~~ \\ \hline    
    A + ${\cal \M}$ & AIII + ${\sf \M}_+$ & $0$ & \nsg{$\mathbb{Z}_2$} & $0$ & \nsg{$\mathbb{Z}_2$} & $0$ & \nsg{$\mathbb{Z}_2$} & $0$ & \nsg{$\mathbb{Z}_2$} \\
    A + $\tilde{\cal \M}$ & AIII + ${\sf \M}_-$ & $\mathbb{Z}$ & $0$ & $\mathbb{Z}$ & $0$ & $\mathbb{Z}$ & $0$ & $\mathbb{Z}$ & $0$ \\ \hline
    AIII + ${\cal \M}_+$ & A + ${\mathsf \M}$ & \nsg{$\mathbb{Z}_2$} & $0$ & \nsg{$\mathbb{Z}_2$} & $0$ & \nsg{$\mathbb{Z}_2$} & $0$ & \nsg{$\mathbb{Z}_2$} & $0$ \\
    AIII + ${\cal \M}_-$ & A + $\bar{\mathsf \M}$ & $0$ & $\mathbb{Z}$ & $0$ & $\mathbb{Z}$ & $0$ & $\mathbb{Z}$ & $0$ & $\mathbb{Z}$ \\
    AIII + $\tilde{\cal \M}_+$ & A + $\bar{\mathsf \M}$ & $0$ & $\mathbb{Z}$ & $0$ & $\mathbb{Z}$ & $0$ & $\mathbb{Z}$ & $0$ & $\mathbb{Z}$ \\
    AIII + $\tilde{\cal \M}_-$ & A + ${\mathsf \M}$ & \nsg{$\mathbb{Z}_2$} & $0$ & \nsg{$\mathbb{Z}_2$} & $0$ & \nsg{$\mathbb{Z}_2$} & $0$ & \nsg{$\mathbb{Z}_2$} & $0$ \\ \hline \hline
  \end{tabular}
\end{table}

\clearpage
\begin{table}[H]
	\centering
	\caption{Classification of point-gap topology in the real Altland-Zirnbauer symmetry classes protected by order-two nonsymmorphic unitary symmetry with $d_{\parallel} \equiv 1$ (mod $2$).
    ${\cal \M}$ and $\tilde{\cal \M}$ denote nonsymmorphic symmetry and pseudo-nonsymmorphic symmetry (or equivalently, nonsymmorphic symmetry$^{\dag}$), respectively.
    The subscript of ${\cal \M}_\pm$ or $\tilde{\cal \M}_\pm$ specifies the commutation ($+$) or anticommutation ($-$) relation to time-reversal symmetry or particle-hole symmetry.
    For the symmetry classes involving both time-reversal symmetry and particle-hole symmetry (i.e., classes BDI, DIII, CII, and CI), the first subscript of ${\cal \M}_{\pm\pm}$ or $\tilde{\cal \M}_{\pm\pm}$ specifies the relation to time-reversal symmetry and the second one to particle-hole symmetry.
    The topological indices highlighted in color indicate classification intrinsic to nonsymmorphic symmetry.}
	\label{tab: real AZ 1}
     \begin{tabular}{cccccccccc} \hline \hline
    ~~Class~~ & ~~Hermitization~~ & ~~$d=1$~~ & ~~$d=2$~~ & ~~$d=3$~~ & ~~$d=4$~~ & ~~$d=5$~~ & ~~$d=6$~~ & ~~$d=7$~~ & ~~$d=8$~~ \\ \hline    
    AI + ${\cal \M}_+$ & ~~BDI + ${\sf \M}_{++}$~~ & \nsg{$\mathbb{Z}_2$} & \nsg{$\mathbb{Z}_2$} & $0$ & $0$ & $0$ & \nsg{$\mathbb{Z}_2$} & $\mathbb{Z}_2$ & \nsg{$\mathbb{Z}_4$} \\
    AI + ${\cal \M}_-$ & BDI + ${\sf \M}_{--}$ & \nsg{$\mathbb{Z}_2$} & \nsg{$\mathbb{Z}_2$} & $0$ & $0$ & $0$ & \nsg{$\mathbb{Z}_2$} & $\mathbb{Z}_2$ & \nsg{$\mathbb{Z}_4$} \\
    AI + $\tilde{\cal \M}_+$ & BDI + ${\sf \M}_{+-}$ & ~~$\mathbb{Z} \oplus \nsg{\mathbb{Z}_2}$~~ & $0$ & $0$ & $0$ & $2\mathbb{Z}$ & $0$ & $\mathbb{Z}_2$ & $\mathbb{Z}_2 \oplus \nsg{\mathbb{Z}_2}$ \\
    AI + $\tilde{\cal \M}_-$ & BDI + ${\sf \M}_{-+}$ & ~~$\mathbb{Z} \oplus \nsg{\mathbb{Z}_2}$~~ & $0$ & $0$ & $0$ & $2\mathbb{Z}$ & $0$ & $\mathbb{Z}_2$ & $\mathbb{Z}_2 \oplus \nsg{\mathbb{Z}_2}$ \\ \hline
    BDI + ${\cal \M}_{++}$ & D + ${\sf \M}_+$ & \nsg{$\mathbb{Z}_4$} & \nsg{$\mathbb{Z}_2$} & \nsg{$\mathbb{Z}_2$} & $0$ & $0$ & $0$ & \nsg{$\mathbb{Z}_2$} & $\mathbb{Z}_2$ \\
    BDI + ${\cal \M}_{+-}$ & D + $\bar{\sf \M}_+$ & ~~$\mathbb{Z}_2 \oplus \nsg{\mathbb{Z}_2}$~~ & ~~$\mathbb{Z} \oplus \nsg{\mathbb{Z}_2}$~~ & $0$ & $0$ & $0$ & $2\mathbb{Z}$ & $0$ & $\mathbb{Z}_2$ \\
    BDI + ${\cal \M}_{-+}$ & D + $\bar{\sf \M}_-$ & ~~$\mathbb{Z}_2 \oplus \nsg{\mathbb{Z}_2}$~~ & ~~$\mathbb{Z} \oplus \nsg{\mathbb{Z}_2}$~~ & $0$ & $0$ & $0$ & $2\mathbb{Z}$ & $0$ & $\mathbb{Z}_2$ \\
    BDI + ${\cal \M}_{--}$ & D + ${\sf \M}_-$ & \nsg{$\mathbb{Z}_4$} & \nsg{$\mathbb{Z}_2$} & \nsg{$\mathbb{Z}_2$} & $0$ & $0$ & $0$ & \nsg{$\mathbb{Z}_2$} & $\mathbb{Z}_2$ \\ 
    BDI + $\tilde{\cal \M}_{++}$ & D + $\bar{\sf \M}_+$ & ~~$\mathbb{Z}_2 \oplus \nsg{\mathbb{Z}_2}$~~ & ~~$\mathbb{Z} \oplus \nsg{\mathbb{Z}_2}$~~ & $0$ & $0$ & $0$ & $2\mathbb{Z}$ & $0$ & $\mathbb{Z}_2$ \\
    BDI + $\tilde{\cal \M}_{+-}$ & D + ${\sf \M}_-$ & \nsg{$\mathbb{Z}_4$} & \nsg{$\mathbb{Z}_2$} & \nsg{$\mathbb{Z}_2$} & $0$ & $0$ & $0$ & \nsg{$\mathbb{Z}_2$} & $\mathbb{Z}_2$ \\
    BDI + $\tilde{\cal \M}_{-+}$ & D + ${\sf \M}_+$ & \nsg{$\mathbb{Z}_4$} & \nsg{$\mathbb{Z}_2$} & \nsg{$\mathbb{Z}_2$} & $0$ & $0$ & $0$ & \nsg{$\mathbb{Z}_2$} & $\mathbb{Z}_2$ \\
    BDI + $\tilde{\cal \M}_{--}$ & D + $\bar{\sf \M}_-$ & ~~$\mathbb{Z}_2 \oplus \nsg{\mathbb{Z}_2}$~~ & ~~$\mathbb{Z} \oplus \nsg{\mathbb{Z}_2}$~~ & $0$ & $0$ & $0$ & $2\mathbb{Z}$ & $0$ & $\mathbb{Z}_2$ \\ \hline
    D + ${\cal \M}_+$ & ~~DIII + ${\mathsf \M}_{++}$~~ & $\mathbb{Z}_2$ & \nsg{$\mathbb{Z}_4$} & \nsg{$\mathbb{Z}_2$} & \nsg{$\mathbb{Z}_2$} & $0$ & $0$ & $0$ & \nsg{$\mathbb{Z}_2$} \\
    D + ${\cal \M}_-$ & DIII + ${\mathsf \M}_{--}$ & $\mathbb{Z}_2$ & \nsg{$\mathbb{Z}_4$} & \nsg{$\mathbb{Z}_2$} & \nsg{$\mathbb{Z}_2$} & $0$ & $0$ & $0$ & \nsg{$\mathbb{Z}_2$} \\
    D + $\tilde{\cal \M}_+$ & DIII + ${\mathsf \M}_{-+}$ & $\mathbb{Z}_2$ & ~~$\mathbb{Z}_2 \oplus \nsg{\mathbb{Z}_2}$~~ & ~~$\mathbb{Z} \oplus \nsg{\mathbb{Z}_2}$~~ & $0$ & $0$ & $0$ & $2\mathbb{Z}$ & $0$ \\
    D + $\tilde{\cal \M}_-$ & DIII + ${\mathsf \M}_{+-}$ & $\mathbb{Z}_2$ & ~~$\mathbb{Z}_2 \oplus \nsg{\mathbb{Z}_2}$~~ & ~~$\mathbb{Z} \oplus \nsg{\mathbb{Z}_2}$~~ & $0$ & $0$ & $0$ & $2\mathbb{Z}$ & $0$ \\ \hline
    ~~DIII + ${\cal \M}_{++}$~~ & AII + ${\sf \M}_{+}$ & \nsg{$\mathbb{Z}_2$} & $\mathbb{Z}_2$ & \nsg{$\mathbb{Z}_4$} & \nsg{$\mathbb{Z}_2$} & \nsg{$\mathbb{Z}_2$} & $0$ & $0$ & $0$ \\
    DIII + ${\cal \M}_{+-}$ & AII + $\bar{\sf \M}_+$ & $0$ & $\mathbb{Z}_2$ & ~~$\mathbb{Z}_2 \oplus \nsg{\mathbb{Z}_2}$~~ & ~~$\mathbb{Z} \oplus \nsg{\mathbb{Z}_2}$~~ & $0$ & $0$ & $0$ & $2\mathbb{Z}$ \\
    DIII + ${\cal \M}_{-+}$ & AII + $\bar{\sf \M}_-$ & $0$ & $\mathbb{Z}_2$ & ~~$\mathbb{Z}_2 \oplus \nsg{\mathbb{Z}_2}$~~ & ~~$\mathbb{Z} \oplus \nsg{\mathbb{Z}_2}$~~ & $0$ & $0$ & $0$ & $2\mathbb{Z}$ \\
    DIII + ${\cal \M}_{--}$ & AII + ${\sf \M}_{-}$ & \nsg{$\mathbb{Z}_2$} & $\mathbb{Z}_2$ & \nsg{$\mathbb{Z}_4$} & \nsg{$\mathbb{Z}_2$} & \nsg{$\mathbb{Z}_2$} & $0$ & $0$ & $0$ \\ 
    DIII + $\tilde{\cal \M}_{++}$ & AII + $\bar{\sf \M}_-$ & $0$ & $\mathbb{Z}_2$ & ~~$\mathbb{Z}_2 \oplus \nsg{\mathbb{Z}_2}$~~ & ~~$\mathbb{Z} \oplus \nsg{\mathbb{Z}_2}$~~ & $0$ & $0$ & $0$ & $2\mathbb{Z}$ \\
    DIII + $\tilde{\cal \M}_{+-}$ & AII + ${\sf \M}_{+}$ & \nsg{$\mathbb{Z}_2$} & $\mathbb{Z}_2$ & \nsg{$\mathbb{Z}_4$} & \nsg{$\mathbb{Z}_2$} & \nsg{$\mathbb{Z}_2$} & $0$ & $0$ & $0$ \\
    DIII + $\tilde{\cal \M}_{-+}$ & AII + ${\sf \M}_{-}$ & \nsg{$\mathbb{Z}_2$} & $\mathbb{Z}_2$ & \nsg{$\mathbb{Z}_4$} & \nsg{$\mathbb{Z}_2$} & \nsg{$\mathbb{Z}_2$} & $0$ & $0$ & $0$ \\
    DIII + $\tilde{\cal \M}_{--}$ & AII + $\bar{\sf \M}_+$ & $0$ & $\mathbb{Z}_2$ & ~~$\mathbb{Z}_2 \oplus \nsg{\mathbb{Z}_2}$~~ & ~~$\mathbb{Z} \oplus \nsg{\mathbb{Z}_2}$~~ & $0$ & $0$ & $0$ & $2\mathbb{Z}$ \\ \hline
    AII + ${\cal \M}_+$ & CII + ${\sf \M}_{++}$ & $0$ & \nsg{$\mathbb{Z}_2$} & $\mathbb{Z}_2$ & \nsg{$\mathbb{Z}_4$} & \nsg{$\mathbb{Z}_2$} & \nsg{$\mathbb{Z}_2$} & $0$ & $0$ \\
    AII + ${\cal \M}_-$ & CII + ${\sf \M}_{--}$ & $0$ & \nsg{$\mathbb{Z}_2$} & $\mathbb{Z}_2$ & \nsg{$\mathbb{Z}_4$} & \nsg{$\mathbb{Z}_2$} & \nsg{$\mathbb{Z}_2$} & $0$ & $0$ \\
    AII + $\tilde{\cal \M}_+$ & CII + ${\sf \M}_{+-}$ & $2\mathbb{Z}$ & $0$ & $\mathbb{Z}_2$ & ~~$\mathbb{Z}_2 \oplus \nsg{\mathbb{Z}_2}$~~ & ~~$\mathbb{Z} \oplus \nsg{\mathbb{Z}_2}$~~ & $0$ & $0$ & $0$ \\
    AII + $\tilde{\cal \M}_-$ & CII + ${\sf \M}_{-+}$ & $2\mathbb{Z}$ & $0$ & $\mathbb{Z}_2$ & ~~$\mathbb{Z}_2 \oplus \nsg{\mathbb{Z}_2}$~~ & ~~$\mathbb{Z} \oplus \nsg{\mathbb{Z}_2}$~~ & $0$ & $0$ & $0$ \\ \hline
    CII + ${\cal \M}_{++}$ & C + ${\sf \M}_+$ & $0$ & $0$ & \nsg{$\mathbb{Z}_2$} & $\mathbb{Z}_2$ & \nsg{$\mathbb{Z}_4$} & \nsg{$\mathbb{Z}_2$} & \nsg{$\mathbb{Z}_2$} & $0$ \\
    CII + ${\cal \M}_{+-}$ & C + $\bar{\sf \M}_+$ & $0$ & $2\mathbb{Z}$ & $0$ & $\mathbb{Z}_2$ & ~~$\mathbb{Z}_2 \oplus \nsg{\mathbb{Z}_2}$~~ & ~~$\mathbb{Z} \oplus \nsg{\mathbb{Z}_2}$~~ & $0$ & $0$ \\
    CII + ${\cal \M}_{-+}$ & C + $\bar{\sf \M}_-$ & $0$ & $2\mathbb{Z}$ & $0$ & $\mathbb{Z}_2$ & ~~$\mathbb{Z}_2 \oplus \nsg{\mathbb{Z}_2}$~~ & ~~$\mathbb{Z} \oplus \nsg{\mathbb{Z}_2}$~~ & $0$ & $0$ \\
    CII + ${\cal \M}_{--}$ & C + ${\sf \M}_-$ & $0$ & $0$ & \nsg{$\mathbb{Z}_2$} & $\mathbb{Z}_2$ & \nsg{$\mathbb{Z}_4$} & \nsg{$\mathbb{Z}_2$} & \nsg{$\mathbb{Z}_2$} & $0$ \\ 
    CII + $\tilde{\cal \M}_{++}$ & C + $\bar{\sf \M}_+$ & $0$ & $2\mathbb{Z}$ & $0$ & $\mathbb{Z}_2$ & ~~$\mathbb{Z}_2 \oplus \nsg{\mathbb{Z}_2}$~~ & ~~$\mathbb{Z} \oplus \nsg{\mathbb{Z}_2}$~~ & $0$ & $0$ \\
    CII + $\tilde{\cal \M}_{+-}$ & C + ${\sf \M}_-$ & $0$ & $0$ & \nsg{$\mathbb{Z}_2$} & $\mathbb{Z}_2$ & \nsg{$\mathbb{Z}_4$} & \nsg{$\mathbb{Z}_2$} & \nsg{$\mathbb{Z}_2$} & $0$ \\
    CII + $\tilde{\cal \M}_{-+}$ & C + ${\sf \M}_+$ & $0$ & $0$ & \nsg{$\mathbb{Z}_2$} & $\mathbb{Z}_2$ & \nsg{$\mathbb{Z}_4$} & \nsg{$\mathbb{Z}_2$} & \nsg{$\mathbb{Z}_2$} & $0$ \\
    CII + $\tilde{\cal \M}_{--}$ & C + $\bar{\sf \M}_-$ & $0$ & $2\mathbb{Z}$ & $0$ & $\mathbb{Z}_2$ & ~~$\mathbb{Z}_2 \oplus \nsg{\mathbb{Z}_2}$~~ & ~~$\mathbb{Z} \oplus \nsg{\mathbb{Z}_2}$~~ & $0$ & $0$ \\ \hline
    C + ${\cal \M}_+$ & CI + ${\sf \M}_{++}$ & $0$ & $0$ & $0$ & \nsg{$\mathbb{Z}_2$} & $\mathbb{Z}_2$ & \nsg{$\mathbb{Z}_4$} & \nsg{$\mathbb{Z}_2$} & \nsg{$\mathbb{Z}_2$} \\
    C + ${\cal \M}_-$ & CI + ${\sf \M}_{--}$ & $0$ & $0$ & $0$ & \nsg{$\mathbb{Z}_2$} & $\mathbb{Z}_2$ & \nsg{$\mathbb{Z}_4$} & \nsg{$\mathbb{Z}_2$} & \nsg{$\mathbb{Z}_2$}  \\
    C + $\tilde{\cal \M}_+$ & CI + ${\sf \M}_{-+}$ & $0$ & $0$ & $2\mathbb{Z}$ & $0$ & $\mathbb{Z}_2$ & ~~$\mathbb{Z}_2 \oplus \nsg{\mathbb{Z}_2}$~~ & ~~$\mathbb{Z} \oplus \nsg{\mathbb{Z}_2}$~~ & $0$ \\
    C + $\tilde{\cal \M}_-$ & CI + ${\sf \M}_{+-}$ & $0$ & $0$ & $2\mathbb{Z}$ & $0$ & $\mathbb{Z}_2$ & ~~$\mathbb{Z}_2 \oplus \nsg{\mathbb{Z}_2}$~~ & ~~$\mathbb{Z} \oplus \nsg{\mathbb{Z}_2}$~~ & $0$ \\ \hline
    CI + ${\cal \M}_{++}$ & AI + ${\sf \M}_+$ & \nsg{$\mathbb{Z}_2$} & $0$ & $0$ & $0$ & \nsg{$\mathbb{Z}_2$} & $\mathbb{Z}_2$ & \nsg{$\mathbb{Z}_4$} & \nsg{$\mathbb{Z}_2$} \\
    CI + ${\cal \M}_{+-}$ & AI + $\bar{\sf \M}_+$ & $0$ & $0$ & $0$ & $2\mathbb{Z}$ & $0$ & $\mathbb{Z}_2$ & ~~$\mathbb{Z}_2 \oplus \nsg{\mathbb{Z}_2}$~~ & ~~$\mathbb{Z} \oplus \nsg{\mathbb{Z}_2}$~~ \\
    CI + ${\cal \M}_{-+}$ & AI + $\bar{\sf \M}_-$ & $0$ & $0$ & $0$ & $2\mathbb{Z}$ & $0$ & $\mathbb{Z}_2$ & ~~$\mathbb{Z}_2 \oplus \nsg{\mathbb{Z}_2}$~~ & ~~$\mathbb{Z} \oplus \nsg{\mathbb{Z}_2}$~~ \\
    CI + ${\cal \M}_{--}$ & AI + ${\sf \M}_-$ & \nsg{$\mathbb{Z}_2$} & $0$ & $0$ & $0$ & \nsg{$\mathbb{Z}_2$} & $\mathbb{Z}_2$ & \nsg{$\mathbb{Z}_4$} & \nsg{$\mathbb{Z}_2$} \\ 
    CI + $\tilde{\cal \M}_{++}$ & AI + $\bar{\sf \M}_-$ & $0$ & $0$ & $0$ & $2\mathbb{Z}$ & $0$ & $\mathbb{Z}_2$ & ~~$\mathbb{Z}_2 \oplus \nsg{\mathbb{Z}_2}$~~ & ~~$\mathbb{Z} \oplus \nsg{\mathbb{Z}_2}$~~ \\
    CI + $\tilde{\cal \M}_{+-}$ & AI + ${\sf \M}_+$ & \nsg{$\mathbb{Z}_2$} & $0$ & $0$ & $0$ & \nsg{$\mathbb{Z}_2$} & $\mathbb{Z}_2$ & \nsg{$\mathbb{Z}_4$} & \nsg{$\mathbb{Z}_2$} \\
    CI + $\tilde{\cal \M}_{-+}$ & AI + ${\sf \M}_-$ & \nsg{$\mathbb{Z}_2$} & $0$ & $0$ & $0$ & \nsg{$\mathbb{Z}_2$} & $\mathbb{Z}_2$ & \nsg{$\mathbb{Z}_4$} & \nsg{$\mathbb{Z}_2$} \\
    CI + $\tilde{\cal \M}_{--}$ & AI + $\bar{\sf \M}_+$ & $0$ & $0$ & $0$ & $2\mathbb{Z}$ & $0$ & $\mathbb{Z}_2$ & ~~$\mathbb{Z}_2 \oplus \nsg{\mathbb{Z}_2}$~~ & ~~$\mathbb{Z} \oplus \nsg{\mathbb{Z}_2}$~~ \\ \hline \hline
  \end{tabular}
\end{table}

\clearpage
\begin{table}[H]
	\centering
	\caption{Classification of point-gap topology in the real Altland-Zirnbauer$^{\dag}$ symmetry classes protected by order-two nonsymmorphic unitary symmetry with $d_{\parallel} \equiv 1$ (mod $2$).
    ${\cal \M}$ and $\tilde{\cal \M}$ denote nonsymmorphic symmetry and pseudo-nonsymmorphic symmetry (or equivalently, nonsymmorphic symmetry$^{\dag}$), respectively.
    The subscript of ${\cal \M}_\pm$ or $\tilde{\cal \M}_\pm$ specifies the commutation ($+$) or anticommutation ($-$) relation to time-reversal symmetry$^{\dag}$ or particle-hole symmetry$^{\dag}$.
    For the symmetry classes involving both time-reversal symmetry$^{\dag}$ and particle-hole symmetry$^{\dag}$ (i.e., classes BDI$^{\dag}$, DIII$^{\dag}$, CII$^{\dag}$, and CI$^{\dag}$), the first subscript of ${\cal \M}_{\pm\pm}$ or $\tilde{\cal \M}_{\pm\pm}$ specifies the relation to time-reversal symmetry$^{\dag}$ and the second one to particle-hole symmetry$^{\dag}$.
    The topological indices highlighted in color indicate classification intrinsic to nonsymmorphic symmetry.}
	\label{tab: real AZ dag 1}
     \begin{tabular}{cccccccccc} \hline \hline
    ~~Class~~ & ~~Hermitization~~ & ~~$d=1$~~ & ~~$d=2$~~ & ~~$d=3$~~ & ~~$d=4$~~ & ~~$d=5$~~ & ~~$d=6$~~ & ~~$d=7$~~ & ~~$d=8$~~ \\ \hline    
    AI$^{\dag}$ + ${\cal \M}_+$ & CI + ${\sf \M}_{++}$ & $0$ & $0$ & $0$ & \nsg{$\mathbb{Z}_2$} & $\mathbb{Z}_2$ & \nsg{$\mathbb{Z}_4$} & \nsg{$\mathbb{Z}_2$} & \nsg{$\mathbb{Z}_2$}  \\
    AI$^{\dag}$ + ${\cal \M}_-$ & CI + ${\sf \M}_{--}$ & $0$ & $0$ & $0$ & \nsg{$\mathbb{Z}_2$} & $\mathbb{Z}_2$ & \nsg{$\mathbb{Z}_4$} & \nsg{$\mathbb{Z}_2$} & \nsg{$\mathbb{Z}_2$}  \\
    AI$^{\dag}$ + $\tilde{\cal \M}_+$ & CI + ${\sf \M}_{+-}$ & $0$ & $0$ & $2\mathbb{Z}$ & $0$ & $\mathbb{Z}_2$ & ~~$\mathbb{Z}_2 \oplus \nsg{\mathbb{Z}_2}$~~ & ~~$\mathbb{Z} \oplus \nsg{\mathbb{Z}_2}$~~ & $0$ \\
    AI$^{\dag}$ + $\tilde{\cal \M}_-$ & CI + ${\sf \M}_{-+}$ & $0$ & $0$ & $2\mathbb{Z}$ & $0$ & $\mathbb{Z}_2$ & ~~$\mathbb{Z}_2 \oplus \nsg{\mathbb{Z}_2}$~~ & ~~$\mathbb{Z} \oplus \nsg{\mathbb{Z}_2}$~~ & $0$ \\ \hline
    BDI$^{\dag}$ + ${\cal \M}_{++}$ & AI + ${\sf \M}_+$ & \nsg{$\mathbb{Z}_2$} & $0$ & $0$ & $0$ & \nsg{$\mathbb{Z}_2$} & $\mathbb{Z}_2$ & \nsg{$\mathbb{Z}_4$} & \nsg{$\mathbb{Z}_2$} \\
    BDI$^{\dag}$ + ${\cal \M}_{+-}$ & AI + $\bar{\sf \M}_-$ & $0$ & $0$ & $0$ & $2\mathbb{Z}$ & $0$ & $\mathbb{Z}_2$ & ~~$\mathbb{Z}_2 \oplus \nsg{\mathbb{Z}_2}$~~ & ~~$\mathbb{Z} \oplus \nsg{\mathbb{Z}_2}$~~ \\
    BDI$^{\dag}$ + ${\cal \M}_{-+}$ & AI + $\bar{\sf \M}_+$ & $0$ & $0$ & $0$ & $2\mathbb{Z}$ & $0$ & $\mathbb{Z}_2$ & ~~$\mathbb{Z}_2 \oplus \nsg{\mathbb{Z}_2}$~~ & ~~$\mathbb{Z} \oplus \nsg{\mathbb{Z}_2}$~~ \\
    BDI$^{\dag}$ + ${\cal \M}_{--}$ & AI + ${\sf \M}_-$ & \nsg{$\mathbb{Z}_2$} & $0$ & $0$ & $0$ & \nsg{$\mathbb{Z}_2$} & $\mathbb{Z}_2$ & \nsg{$\mathbb{Z}_4$} & \nsg{$\mathbb{Z}_2$} \\ 
    BDI$^{\dag}$ + $\tilde{\cal \M}_{++}$ & AI + $\bar{\sf \M}_+$ & $0$ & $0$ & $0$ & $2\mathbb{Z}$ & $0$ & $\mathbb{Z}_2$ & ~~$\mathbb{Z}_2 \oplus \nsg{\mathbb{Z}_2}$~~ & ~~$\mathbb{Z} \oplus \nsg{\mathbb{Z}_2}$~~ \\
    BDI$^{\dag}$ + $\tilde{\cal \M}_{+-}$ & AI + ${\sf \M}_+$ & \nsg{$\mathbb{Z}_2$} & $0$ & $0$ & $0$ & \nsg{$\mathbb{Z}_2$} & $\mathbb{Z}_2$ & \nsg{$\mathbb{Z}_4$} & \nsg{$\mathbb{Z}_2$} \\
    BDI$^{\dag}$ + $\tilde{\cal \M}_{-+}$ & AI + ${\sf \M}_-$ & \nsg{$\mathbb{Z}_2$} & $0$ & $0$ & $0$ & \nsg{$\mathbb{Z}_2$} & $\mathbb{Z}_2$ & \nsg{$\mathbb{Z}_4$} & \nsg{$\mathbb{Z}_2$} \\
    BDI$^{\dag}$ + $\tilde{\cal \M}_{--}$ & AI + $\bar{\sf \M}_-$ & $0$ & $0$ & $0$ & $2\mathbb{Z}$ & $0$ & $\mathbb{Z}_2$ & ~~$\mathbb{Z}_2 \oplus \nsg{\mathbb{Z}_2}$~~ & ~~$\mathbb{Z} \oplus \nsg{\mathbb{Z}_2}$~~ \\ \hline
    D$^{\dag}$ + ${\cal \M}_+$ & BDI + ${\sf \M}_{++}$ & \nsg{$\mathbb{Z}_2$} & \nsg{$\mathbb{Z}_2$} & $0$ & $0$ & $0$ & \nsg{$\mathbb{Z}_2$} & $\mathbb{Z}_2$ & \nsg{$\mathbb{Z}_4$} \\
    D$^{\dag}$ + ${\cal \M}_-$ & BDI + ${\sf \M}_{--}$ & \nsg{$\mathbb{Z}_2$} & \nsg{$\mathbb{Z}_2$} & $0$ & $0$ & $0$ & \nsg{$\mathbb{Z}_2$} & $\mathbb{Z}_2$ & \nsg{$\mathbb{Z}_4$} \\
    D$^{\dag}$ + $\tilde{\cal \M}_+$ & BDI + ${\sf \M}_{-+}$ & ~~$\mathbb{Z} \oplus \nsg{\mathbb{Z}_2}$~~ & $0$ & $0$ & $0$ & $2\mathbb{Z}$ & $0$ & $\mathbb{Z}_2$ & $\mathbb{Z}_2 \oplus \nsg{\mathbb{Z}_2}$ \\
    D$^{\dag}$ + $\tilde{\cal \M}_-$ & BDI + ${\sf \M}_{+-}$ & ~~$\mathbb{Z} \oplus \nsg{\mathbb{Z}_2}$~~ & $0$ & $0$ & $0$ & $2\mathbb{Z}$ & $0$ & $\mathbb{Z}_2$ & $\mathbb{Z}_2 \oplus \nsg{\mathbb{Z}_2}$ \\ \hline
    ~~DIII$^{\dag}$ + ${\cal \M}_{++}$~~ & D + ${\sf \M}_+$ & \nsg{$\mathbb{Z}_4$} & \nsg{$\mathbb{Z}_2$} & \nsg{$\mathbb{Z}_2$} & $0$ & $0$ & $0$ & \nsg{$\mathbb{Z}_2$} & $\mathbb{Z}_2$ \\
    DIII$^{\dag}$ + ${\cal \M}_{+-}$ & D + $\bar{\sf \M}_-$ & ~~$\mathbb{Z}_2 \oplus \nsg{\mathbb{Z}_2}$~~ & ~~$\mathbb{Z} \oplus \nsg{\mathbb{Z}_2}$~~ & $0$ & $0$ & $0$ & $2\mathbb{Z}$ & $0$ & $\mathbb{Z}_2$ \\
    DIII$^{\dag}$ + ${\cal \M}_{-+}$ & D + $\bar{\sf \M}_+$ & ~~$\mathbb{Z}_2 \oplus \nsg{\mathbb{Z}_2}$~~ & ~~$\mathbb{Z} \oplus \nsg{\mathbb{Z}_2}$~~ & $0$ & $0$ & $0$ & $2\mathbb{Z}$ & $0$ & $\mathbb{Z}_2$ \\
    DIII$^{\dag}$ + ${\cal \M}_{--}$ & D + ${\sf \M}_-$ & \nsg{$\mathbb{Z}_4$} & \nsg{$\mathbb{Z}_2$} & \nsg{$\mathbb{Z}_2$} & $0$ & $0$ & $0$ & \nsg{$\mathbb{Z}_2$} & $\mathbb{Z}_2$ \\ 
    DIII$^{\dag}$ + $\tilde{\cal \M}_{++}$ & D + $\bar{\sf \M}_-$ & ~~$\mathbb{Z}_2 \oplus \nsg{\mathbb{Z}_2}$~~ & ~~$\mathbb{Z} \oplus \nsg{\mathbb{Z}_2}$~~ & $0$ & $0$ & $0$ & $2\mathbb{Z}$ & $0$ & $\mathbb{Z}_2$ \\
    DIII$^{\dag}$ + $\tilde{\cal \M}_{+-}$ & D + ${\sf \M}_-$ & \nsg{$\mathbb{Z}_4$} & \nsg{$\mathbb{Z}_2$} & \nsg{$\mathbb{Z}_2$} & $0$ & $0$ & $0$ & \nsg{$\mathbb{Z}_2$} & $\mathbb{Z}_2$ \\
    DIII$^{\dag}$ + $\tilde{\cal \M}_{-+}$ & D + ${\sf \M}_+$ & \nsg{$\mathbb{Z}_4$} & \nsg{$\mathbb{Z}_2$} & \nsg{$\mathbb{Z}_2$} & $0$ & $0$ & $0$ & \nsg{$\mathbb{Z}_2$} & $\mathbb{Z}_2$ \\
    DIII$^{\dag}$ + $\tilde{\cal \M}_{--}$ & D + $\bar{\sf \M}_+$ & ~~$\mathbb{Z}_2 \oplus \nsg{\mathbb{Z}_2}$~~ & ~~$\mathbb{Z} \oplus \nsg{\mathbb{Z}_2}$~~ & $0$ & $0$ & $0$ & $2\mathbb{Z}$ & $0$ & $\mathbb{Z}_2$ \\ \hline
    AII$^{\dag}$ + ${\cal \M}_+$ & ~~DIII + ${\sf \M}_{++}$~~ & $\mathbb{Z}_2$ & \nsg{$\mathbb{Z}_4$} & \nsg{$\mathbb{Z}_2$} & \nsg{$\mathbb{Z}_2$} & $0$ & $0$ & $0$ & \nsg{$\mathbb{Z}_2$} \\
    AII$^{\dag}$ + ${\cal \M}_-$ & DIII + ${\sf \M}_{--}$ & $\mathbb{Z}_2$ & \nsg{$\mathbb{Z}_4$} & \nsg{$\mathbb{Z}_2$} & \nsg{$\mathbb{Z}_2$} & $0$ & $0$ & $0$ & \nsg{$\mathbb{Z}_2$} \\
    AII$^{\dag}$ + $\tilde{\cal \M}_+$ & DIII + ${\sf \M}_{+-}$ & $\mathbb{Z}_2$ & ~~$\mathbb{Z}_2 \oplus \nsg{\mathbb{Z}_2}$~~ & ~~$\mathbb{Z} \oplus \nsg{\mathbb{Z}_2}$~~ & $0$ & $0$ & $0$ & $2\mathbb{Z}$ & $0$ \\
    AII$^{\dag}$ + $\tilde{\cal \M}_-$ & DIII + ${\sf \M}_{-+}$ & $\mathbb{Z}_2$ & ~~$\mathbb{Z}_2 \oplus \nsg{\mathbb{Z}_2}$~~ & ~~$\mathbb{Z} \oplus \nsg{\mathbb{Z}_2}$~~ & $0$ & $0$ & $0$ & $2\mathbb{Z}$ & $0$ \\ \hline
    CII$^{\dag}$ + ${\cal \M}_{++}$ & AII + ${\sf \M}_+$ & \nsg{$\mathbb{Z}_2$} & $\mathbb{Z}_2$ & \nsg{$\mathbb{Z}_4$} & \nsg{$\mathbb{Z}_2$} & \nsg{$\mathbb{Z}_2$} & $0$ & $0$ & $0$ \\
    CII$^{\dag}$ + ${\cal \M}_{+-}$ & AII + $\bar{\sf \M}_-$ & $0$ & $\mathbb{Z}_2$ & ~~$\mathbb{Z}_2 \oplus \nsg{\mathbb{Z}_2}$~~ & ~~$\mathbb{Z} \oplus \nsg{\mathbb{Z}_2}$~~ & $0$ & $0$ & $0$ & $2\mathbb{Z}$ \\
    CII$^{\dag}$ + ${\cal \M}_{-+}$ & AII + $\bar{\sf \M}_+$ & $0$ & $\mathbb{Z}_2$ & ~~$\mathbb{Z}_2 \oplus \nsg{\mathbb{Z}_2}$~~ & ~~$\mathbb{Z} \oplus \nsg{\mathbb{Z}_2}$~~ & $0$ & $0$ & $0$ & $2\mathbb{Z}$ \\
    CII$^{\dag}$ + ${\cal \M}_{--}$ & AII + ${\sf \M}_-$ & \nsg{$\mathbb{Z}_2$} & $\mathbb{Z}_2$ & \nsg{$\mathbb{Z}_4$} & \nsg{$\mathbb{Z}_2$} & \nsg{$\mathbb{Z}_2$} & $0$ & $0$ & $0$ \\ 
    CII$^{\dag}$ + $\tilde{\cal \M}_{++}$ & AII + $\bar{\sf \M}_+$ & $0$ & $\mathbb{Z}_2$ & ~~$\mathbb{Z}_2 \oplus \nsg{\mathbb{Z}_2}$~~ & ~~$\mathbb{Z} \oplus \nsg{\mathbb{Z}_2}$~~ & $0$ & $0$ & $0$ & $2\mathbb{Z}$ \\
    CII$^{\dag}$ + $\tilde{\cal \M}_{+-}$ & AII + ${\sf \M}_+$ & \nsg{$\mathbb{Z}_2$} & $\mathbb{Z}_2$ & \nsg{$\mathbb{Z}_4$} & \nsg{$\mathbb{Z}_2$} & \nsg{$\mathbb{Z}_2$} & $0$ & $0$ & $0$ \\
    CII$^{\dag}$ + $\tilde{\cal \M}_{-+}$ & AII + ${\sf \M}_-$ & \nsg{$\mathbb{Z}_2$} & $\mathbb{Z}_2$ & \nsg{$\mathbb{Z}_4$} & \nsg{$\mathbb{Z}_2$} & \nsg{$\mathbb{Z}_2$} & $0$ & $0$ & $0$ \\
    CII$^{\dag}$ + $\tilde{\cal \M}_{--}$ & AII + $\bar{\sf \M}_-$ & $0$ & $\mathbb{Z}_2$ & ~~$\mathbb{Z}_2 \oplus \nsg{\mathbb{Z}_2}$~~ & ~~$\mathbb{Z} \oplus \nsg{\mathbb{Z}_2}$~~ & $0$ & $0$ & $0$ & $2\mathbb{Z}$ \\ \hline
    C$^{\dag}$ + ${\cal \M}_+$ & CII + ${\sf \M}_{++}$ & $0$ & \nsg{$\mathbb{Z}_2$} & $\mathbb{Z}_2$ & \nsg{$\mathbb{Z}_4$} & \nsg{$\mathbb{Z}_2$} & \nsg{$\mathbb{Z}_2$} & $0$ & $0$ \\
    C$^{\dag}$ + ${\cal \M}_-$ & CII + ${\sf \M}_{--}$ & $0$ & \nsg{$\mathbb{Z}_2$} & $\mathbb{Z}_2$ & \nsg{$\mathbb{Z}_4$} & \nsg{$\mathbb{Z}_2$} & \nsg{$\mathbb{Z}_2$} & $0$ & $0$ \\
    C$^{\dag}$ + $\tilde{\cal \M}_+$ & CII + ${\sf \M}_{-+}$ & $2\mathbb{Z}$ & $0$ & $\mathbb{Z}_2$ & ~~$\mathbb{Z}_2 \oplus \nsg{\mathbb{Z}_2}$~~ & ~~$\mathbb{Z} \oplus \nsg{\mathbb{Z}_2}$~~ & $0$ & $0$ & $0$ \\
    C$^{\dag}$ + $\tilde{\cal \M}_-$ & CII + ${\sf \M}_{+-}$ & $2\mathbb{Z}$ & $0$ & $\mathbb{Z}_2$ & ~~$\mathbb{Z}_2 \oplus \nsg{\mathbb{Z}_2}$~~ & ~~$\mathbb{Z} \oplus \nsg{\mathbb{Z}_2}$~~ & $0$ & $0$ & $0$ \\ \hline
    CI$^{\dag}$ + ${\cal \M}_{++}$ & C + ${\sf \M}_+$ & $0$ & $0$ & \nsg{$\mathbb{Z}_2$} & $\mathbb{Z}_2$ & \nsg{$\mathbb{Z}_4$} & \nsg{$\mathbb{Z}_2$} & \nsg{$\mathbb{Z}_2$} & $0$ \\
    CI$^{\dag}$ + ${\cal \M}_{+-}$ & C + $\bar{\sf \M}_-$ & $0$ & $2\mathbb{Z}$ & $0$ & $\mathbb{Z}_2$ & ~~$\mathbb{Z}_2 \oplus \nsg{\mathbb{Z}_2}$~~ & ~~$\mathbb{Z} \oplus \nsg{\mathbb{Z}_2}$~~ & $0$ & $0$ \\
    CI$^{\dag}$ + ${\cal \M}_{-+}$ & C + $\bar{\sf \M}_+$ & $0$ & $2\mathbb{Z}$ & $0$ & $\mathbb{Z}_2$ & ~~$\mathbb{Z}_2 \oplus \nsg{\mathbb{Z}_2}$~~ & ~~$\mathbb{Z} \oplus \nsg{\mathbb{Z}_2}$~~ & $0$ & $0$ \\
    CI$^{\dag}$ + ${\cal \M}_{--}$ & C + ${\sf \M}_-$ & $0$ & $0$ & \nsg{$\mathbb{Z}_2$} & $\mathbb{Z}_2$ & \nsg{$\mathbb{Z}_4$} & \nsg{$\mathbb{Z}_2$} & \nsg{$\mathbb{Z}_2$} & $0$ \\ 
    CI$^{\dag}$ + $\tilde{\cal \M}_{++}$ & C + $\bar{\sf \M}_-$ & $0$ & $2\mathbb{Z}$ & $0$ & $\mathbb{Z}_2$ & ~~$\mathbb{Z}_2 \oplus \nsg{\mathbb{Z}_2}$~~ & ~~$\mathbb{Z} \oplus \nsg{\mathbb{Z}_2}$~~ & $0$ & $0$ \\
    CI$^{\dag}$ + $\tilde{\cal \M}_{+-}$ & C + ${\sf \M}_-$ & $0$ & $0$ & \nsg{$\mathbb{Z}_2$} & $\mathbb{Z}_2$ & \nsg{$\mathbb{Z}_4$} & \nsg{$\mathbb{Z}_2$} & \nsg{$\mathbb{Z}_2$} & $0$ \\
    CI$^{\dag}$ + $\tilde{\cal \M}_{-+}$ & C + ${\sf \M}_+$ & $0$ & $0$ & \nsg{$\mathbb{Z}_2$} & $\mathbb{Z}_2$ & \nsg{$\mathbb{Z}_4$} & \nsg{$\mathbb{Z}_2$} & \nsg{$\mathbb{Z}_2$} & $0$ \\
    CI$^{\dag}$ + $\tilde{\cal \M}_{--}$ & C + $\bar{\sf \M}_+$ & $0$ & $2\mathbb{Z}$ & $0$ & $\mathbb{Z}_2$ & ~~$\mathbb{Z}_2 \oplus \nsg{\mathbb{Z}_2}$~~ & ~~$\mathbb{Z} \oplus \nsg{\mathbb{Z}_2}$~~ & $0$ & $0$ \\ \hline \hline
  \end{tabular}
\end{table}

\section{Generic $2\times2$ Hamiltonians with pseudo-nonsymmorphic symmetry}

We provide general forms of $2\times2$ non-Hermitian Hamiltonians respecting pseudo-nonsymmorphic symmetry with $d_{\parallel}=0$:
\begin{align}
    \mathcal{U}_{\alpha}(k_x) H_{\alpha}^{\dagger}(\boldsymbol{k}) \mathcal{V}_{\alpha}^{-1}(k_x) = H_{\alpha}(\boldsymbol{k}),\quad \mathcal{U}_{\alpha}(k_x)\mathcal{V}_{\alpha}(k_x)=e^{-ik_x}\quad(\alpha={\rm I, II}),
\end{align}
where $\mathcal{U}_{\alpha}(k_x)$ and $\mathcal{V}_{\alpha}(k_x)$ are given as 
\begin{align}
    \mathcal{U}_{\rm I}(k_x)=\sigma_0,\quad
    \mathcal{V}_{\rm I}(k_x)=e^{-ik_x}\sigma_0
    ;\quad
    \mathcal{U}_{\rm II}(k_x)=\mathcal{V}_{\rm II}(k_x)= 
    \begin{pmatrix}
        0 & e^{-ik_x} \\
        1 & 0
    \end{pmatrix}.
\end{align}
For each symmetry representation, $H_{\alpha}(\bm{k})$ takes the form of 
\begin{align}
    H_{\rm I}(\bm{k})=
    \begin{pmatrix}
        g(\bm{k})e^{ik_x/2} & f(\bm{k}) \\
        f^{*}(\bm{k})e^{ik_x} & h(\bm{k})e^{ik_x/2}
    \end{pmatrix},\quad
    H_{\rm II}(\bm{k})=
    \begin{pmatrix}
        f(\bm{k}) & g(\bm{k})e^{-ik_x/2} \\
        h(\bm{k})e^{ik_x/2} & f^{*}(\bm{k})
    \end{pmatrix},
\end{align}
where $f$ is a complex function, and $g$ and $h$ are real functions.
Owing to the $2\pi$-periodicity of the Hamiltonians, $f$ is $2\pi$-periodic for arbitrary momentum directions, while $g$ and $h$ are $2\pi$-antiperiodic for the $k_x$ direction but $2\pi$-periodic for the other directions.

Notably, $\mathcal{U}_{\rm I}$ and $\mathcal{U}_{\rm II}$ ($\mathcal{V}_{\rm I}$ and $\mathcal{V}_{\rm II}$) are not unitary equivalent since $\mathcal{U}_{\rm I}$ and $\mathcal{V}_{\rm I}$ are proportional to the $2\times2$ identity matrix $\sigma_0$, while their corresponding unitary matrices for the Hermitized Hamiltonians [see Eqs.~(\ref{aeq: Hermitization}) and (\ref{eq:unitary})] are:

\begin{align}
    {\sf U}_{\rm I}\left(k_x\right)
    =
    {\sf S}\,
    {\sf U}_{\rm II}\left(k_x\right)\,
    {\sf S}^{\dagger},
\end{align}
with
\begin{align}
    {\sf U}_{\alpha={\rm I,II}}\,\left(k_x\right)
    \coloneqq
    \begin{pmatrix}
        0 & \mathcal{U}_{\alpha}\left(k_x\right)\\
        \mathcal{V}_{\alpha}\left(k_x\right)
        & 0
    \end{pmatrix},\quad
    {\sf S} 
    \coloneqq
    \begin{pmatrix}
        \sigma_0 & 0 \\
        0 & \mathcal{U}_{\rm II}\,\left(k_x\right)
    \end{pmatrix}.
\end{align}
This can be relevant to the origin of the different boundary physics discussed in the main text, meriting further investigation.

\section{Nonsymmorphic topological phases in one dimension}

\subsection{Class A + $\bar{\sf U}$ ($d_{\parallel} = 0$; Hermitian)}

Following Ref.~\cite{Shiozaki-PRB-2015}, we investigate the $\mathbb{Z}_2$ topological phase of one-dimensional Hermitian systems in class A + $\bar{\sf U}$ with $d_{\parallel} = 0$.
The condition of nonsymmorphic symmetry is expressed as
\begin{equation}
    \bar{\sf U} \left( k_x \right) {\sf H} \left( k_x \right) \bar{\sf U}^{-1} \left( k_x \right) = - {\sf H} \left( k_x \right), \quad \bar{\sf U}^2 \left( k_x \right) = e^{-i k_x}.
\end{equation}
The $2\pi$ periodicity of $\bar{\sf U} \left( k_x \right)$ constrains its form to
\begin{equation}
    \bar{\sf U} \left( k_x \right) = \begin{pmatrix}
        0 & e^{-i k_x} \\
        1 & 0
    \end{pmatrix}.
\end{equation}
Accordingly, the Hamiltonian ${\sf H} \left( k_x \right)$ is given as
\begin{equation}
    {\sf H} \left( k_x \right) = \begin{pmatrix}
        X \left( k_x \right) & -i Y \left( k_x \right) e^{-i k_x/2} \\
        i Y \left( k_x \right) e^{i k_x/2} & -X \left( k_x \right)
    \end{pmatrix}
        \label{aeq: 1D}
\end{equation}
with $N \times N$ Hermitian matrices $X \left( k_x \right)$ and $Y \left( k_x \right)$.
The $2\pi$ periodicity of ${\sf H} \left( k_x \right)$ imposes the constraints
\begin{equation}
    X \left( k_x + 2\pi \right) = X \left( k_x \right), \quad Y \left( k_x + 2\pi \right) = - Y \left( k_x \right).
\end{equation}
Introducing an $N \times N$ non-Hermitian matrix $Z \left( k_x \right)$ by 
\begin{equation}
    Z \left( k_x \right) \coloneqq X \left( k_x \right) + i Y \left( k_x \right),
\end{equation}
we have
\begin{equation}
    Z \left( k_x + 2\pi \right) = Z^{\dag} \left( k_x \right).
\end{equation}
Owing to $\det \mathsf{H} \left( k_x \right) = \left| \det Z \left( k_x \right) \right|^2$, the existence of a spectral gap around $E=0$ in $\mathsf{H} \left( k_x \right)$ is equivalent to the opening of a point gap around $E=0$ in $Z \left( k_x \right)$.
We define the real and imaginary parts of $\det Z \left( k_x \right)$ as
\begin{equation}
    x \left( k_x \right) \coloneqq \mathrm{Re} \left[ \det Z \left( k_x \right) \right], \quad y \left( k_x \right) \coloneqq \mathrm{Im} \left[ \det Z \left( k_x \right) \right],
\end{equation}
satisfying
\begin{equation}
    x \left( k_x + 2\pi \right) = x \left( k_x \right), \quad y \left( k_x + 2\pi \right) = - y \left( k_x \right).
\end{equation}
The $\mathbb{Z}_2$ topological invariant is determined by the parity of the number of times the trajectory $\left( x \left( k_x \right), y \left( k_x \right) \right)$ crosses the negative $x$ axis as $k_x$ varies from $0$ to $2\pi$.
Simple representations of trivial and nontrivial phases are given by $\left( x, y \right) = \left( \pm 1, 0 \right)$ ($N=1$).

\subsection{Class AIII + ${\cal U}_{-}$ ($d_{\parallel} = 0$; non-Hermitian)}

According to Table~\ref{tab: complex AZ 0}, one-dimensional non-Hermitian systems in class AIII + $\mathcal{U}_{-}$ with $d_{\parallel} = 0$ support the $\mathbb{Z}_2$-classified point-gap topological phase.
Owing to the anticommutation relation between nonsymmorphic symmetry and chiral symmetry, the chiral symmetry operator can be chosen as
\begin{equation}
    \Gamma \coloneqq \begin{pmatrix}
        1 & 0 \\
        0 & -1
    \end{pmatrix}.
\end{equation}
Then, from the Hermitian Hamiltonian ${\sf H} \left( k_x \right)$ in Eq.~(\ref{aeq: 1D}), the non-Hermitian Hamiltonian $\tilde{H} \left( k_x \right)$ can be introduced as
\begin{equation}
    H \left( k_x \right) \coloneqq - i {\sf H} \left( k_x \right) \Gamma = \begin{pmatrix}
        -i X & Y e^{-i k_x/2} \\
        Y e^{i k_x/2} & -i X
    \end{pmatrix},
\end{equation}
respecting the symmetry conditions,
\begin{equation}
    \bar{\sf U} \left( k_x \right) H \left( k_x \right) \bar{\sf U}^{-1} \left( k_x \right) = H \left( k_x \right), \quad 
    \Gamma H^{\dag} \left( k_x \right) \Gamma^{-1} = - H \left( k_x \right).
\end{equation}
Analogous to the corresponding Hermitian systems ${\sf H} \left( k_x \right)$, the point-gapped non-Hermitian systems $H \left( k_x \right)$ realize the $\mathbb{Z}_2$ topological phase.
Similarly, non-Hermitian systems in class AIII + $\tilde{\cal U}_{+}$ ($d_{\parallel} = 0$) host the $\mathbb{Z}_2$ point-gap topological phase.

\section{$\mathbb{Z}_2$ topological invariant and Fermi-surface-like formula in 2D Class A + $\tilde{\mathcal{U}}$}

By introducing
\begin{equation}
    \tilde{H}(\bm{k}) \coloneqq \mathcal{V}(\bm{k}) \, H(\bm{k}) \, e^{i k_x/2},
\end{equation}
the symmetry constraint in Eq.~(\ref{aeq: nsg-pseudo}) for $d_{\parallel}=0$ becomes
\begin{equation}
    \tilde{H}^\dagger(\bm{k}) = \tilde{H}(\bm{k}), \quad 
    \tilde{H}(k_x + 2\pi,\,k_y) = -\tilde{H}(k_x,\,k_y).
        \label{eq:apbc_tilde_h}
\end{equation}
The Berry phase of the occupied bands along the loop in the $k_y$ direction at fixed $k_x$, denoted by $e^{i\gamma(k_x)}$, satisfies \cite{Shiozaki-PRB-2015}
\begin{equation}
e^{i\gamma(k_x + 2\pi)} = e^{-i\gamma(k_x)}.
\end{equation}
If the trajectory of $e^{i\gamma(k_x)}$ crosses the positive real axis an odd (even) number of times as $k_x$ goes from $0$ to $2\pi$, then $\tilde{H}(\bm{k})$ is topologically nontrivial (trivial), which defines the $\mathbb{Z}_2$ invariant $\nu$. An illustrative example is given by the $2\times 2$ Hamiltonian
\begin{equation}
\tilde{H}(\bm{k}) 
=
\begin{pmatrix}
  0 & e^{i( k_x/2 + n k_y )} \\
  e^{-i( k_x/2 + n k_y )} & 0
\end{pmatrix},
\quad
n \in \mathbb{Z},
\end{equation}
whose $\mathbb{Z}_2$ invariant is 
\begin{equation}
\nu \equiv n \quad \left(\bmod~2 \right).
\end{equation}

Next, we discuss a Fermi-surface-like formula for this $\mathbb{Z}_2$ index $\nu$ in generic $2\times 2$ models of the form
\begin{equation}
\tilde{H}(\bm{k}) 
= 
c(\bm{k})\, \sigma_0
\;+\;
\begin{pmatrix}
    \varepsilon(\bm{k}) & \Delta(\bm{k}) \\
    \Delta(\bm{k})^*    & -\,\varepsilon(\bm{k})
\end{pmatrix}.
\label{eq:model_2by2}
\end{equation}
Because of the antiperiodicity in $k_x$ in Eq.~\eqref{eq:apbc_tilde_h}, the real-valued function $\varepsilon(\bm{k})$ must cross zero an odd number of times as $k_x$ increases by $2\pi$ for each fixed $k_y$. Analogously to superconducting systems, we refer to the loops on which $\varepsilon(\bm{k})=0$ as \textit{Fermi loops}.

To evaluate the $\mathbb{Z}_2$ invariant, we consider the one-parameter family of Hamiltonians,
\begin{equation}
\tilde{H}(\bm{k};\,a)
=
c(\bm{k}) \sigma_0
\;+\;
\begin{pmatrix}
    \varepsilon(\bm{k}) & a\,\Delta(\bm{k}) \\
    a\,\Delta(\bm{k})^* & -\,\varepsilon(\bm{k})
\end{pmatrix}
\quad
\left( a > 0 \right),
\end{equation}
and take the limit $a \to 0^+$. In this limit, away from the Fermi loops [i.e., $\varepsilon(\bm{k}) \neq 0$], the sign of $\varepsilon(\bm{k})$ determines that the occupied state is either $(1,0)^T$ or $(0,1)^T$, both of which yield zero Berry phase. 
Only on the Fermi loops [i.e., $\varepsilon(\bm{k}_{\mathrm{FL}})=0$], a nontrivial Berry phase can arise. 
Specifically, at such points, the occupied eigenstate becomes $\left(\frac{\Delta(\bm{k}_{\mathrm{FL}})}{|\Delta(\bm{k}_{\mathrm{FL}})|},-1\right)^T$, whose Berry phase along each Fermi loop is given by
\begin{equation}
e^{i\gamma_{\mathrm{FL}}}
=
\exp \left[
  -\frac{i}{2} \oint_{\varepsilon(\bm{k}_{\mathrm{FL}})=0}
  d\,\log \,\Delta(\bm{k})
\right]
=
(-1)^{N_{\mathrm{FL}}},
\end{equation}
where 
\begin{equation}
N_{\mathrm{FL}}
\coloneqq
\frac{1}{2\pi}
\oint_{\varepsilon(\bm{k}_{\mathrm{FL}})=0}
d\,\log \,\Delta(\bm{k})
\;\;\in\;\mathbb{Z}
\end{equation}
is the winding number of the phase of $\Delta(\bm{k})$ around the Fermi loop.

In the presence of multiple Fermi loops, each loop $\mathrm{FL}_\alpha$ carries its own integer winding number $N_{\mathrm{FL}_\alpha}$, and the total $\mathbb{Z}_2$ invariant is given by
\begin{equation}
\nu
\equiv
\sum_{\alpha} N_{\mathrm{FL}_\alpha} \quad \left(\bmod~2 \right).
\end{equation}
Hence, $\nu$ counts the parity of the sum of these winding numbers over all Fermi loops.

\section{Parity of $\mathbb{Z}_4$ topological invariant in 2D class AII$^{\dag}$ + $\tilde{\cal \M}_{-}$}\label{asec:Z4=1_is_Z2=1}

We show that the $\mathbb{Z}_4$ topological invariant $\nu \equiv 1,3$ ($\mathrm{mod}~4$) yields the conventional $\mathbb{Z}_2$ skin effect~\cite{OKSS-20} in two-dimensional non-Hermitian systems for class AII$^{\dag}$ + $\tilde{\cal \M}_{-}$ ($d_{\parallel}=0$). 
We consider the Hermitized Hamiltonian ${\sf H}(\boldsymbol{k})$ in class DIII + ${\sf U}_{{-+}}$ corresponding to the non-Hermitian Hamiltonian $H(\boldsymbol{k})$ in class AII$^\dagger$ + $\tilde{\mathcal{U}}_{{-}}$ (see Table~\ref{tab: real AZ dag 0}). 
In class DIII, the Hermitized Hamiltonian ${\sf H}(\boldsymbol{k})$ respects time-reversal symmetry, particle-hole symmetry, and chiral symmetry:
\begin{align}
    &{\sf T}{\sf H}^{*}(\boldsymbol{k}){\sf T}^{-1} = {\sf H}(-\boldsymbol{k}),  \quad {\sf T}{\sf T}^{*}=-1, \label{eq:DIII_TRS} \\
    &{\sf C}{\sf H}^{T}(\boldsymbol{k}){\sf C}^{-1} = -{\sf H}(-\boldsymbol{k}), \quad {\sf C}{\sf C}^{*}=1, \label{eq:DIII_PHS} \\
    &{\sf \Gamma}{\sf H}(\boldsymbol{k}){\sf \Gamma}^{-1}=-{\sf H}(\boldsymbol{k}), \quad  {\sf \Gamma}^2=1. \label{eq:DIII_CS}
\end{align}
Furthermore, the Hermitized Hamiltonian ${\sf H(\boldsymbol{k})}$ respects nonsymmorphic symmetry ($d_{\parallel}=0$)
\begin{align}\label{eq:Hermitized_NSG}
    {\sf U}(k_x) {\sf H}(\boldsymbol{k}){\sf U}^{-1}(k_x) = {\sf H}
    (\boldsymbol{k}), \quad {\sf U}^2(k_x)=e^{-ik_x},
\end{align}
with the symmetry eigenvalues of nonsymmorphic symmetry, $g_{\pm}(k_x)=\pm e^{ik_x/2}$.

\subsection{Time-reversal polarization}
 The Hermitized Hamiltonian with $2N$ bands possesses $N$ Kramers pairs $\{ \ket{u_n^{\rm I}(\boldsymbol{k})}, \ket{u_n^{\rm II} (\boldsymbol{k})}\}$ with $n=1,2,\cdots N$.
We define the polarization associated with one of the Kramers pairs by
\begin{align}
    P^{s} \coloneqq \frac{i}{2\pi}\int^{\pi}_{-\pi}dk~\mathrm{tr} \mathcal{A}^{s}(k) \quad (s=\mathrm{I}, \mathrm{II}),
\end{align}
where $\mathcal{A}^{s}(k)$ is the Berry connection of occupied states with $s=\mathrm{I}$ or $\mathrm{II}$, and the trace is taken over all the occupied bands. We introduce time-reversal polarization~\cite{Fu-06} defined by
\begin{align}
    P_{\mathsf{T}} \coloneqq P^{\mathrm{I}}-P^{\mathrm{II}}.
\end{align}
The sum of the time-reversal polarizations associated with occupied states and unoccupied states is zero: 
\begin{align}
    P_{\mathsf{T}} + \overline{P_{\mathsf{T}}} \equiv 0\quad (\mathrm{mod}\ 2), \label{aeq:PT+barPT=0}
\end{align}
where $\overline{P_{\mathsf{T}}}$ is the time-reversal polarization associated with unoccupied states. 
In addition, in the presence of particle-hole symmetry, the time-reversal polarizations $P_{\mathsf{T}}$ and $\overline{P_{\mathsf{T}}}$ satisfy 
\begin{align}
    P_{\mathsf{T}} \equiv \overline{P_{\mathsf{T}}} \quad (\mathrm{mod}\ 2). \label{aeq:PT=barPT}
\end{align}
From Eqs.~\eqref{aeq:PT+barPT=0} and \eqref{aeq:PT=barPT}, we obtain
\begin{align}
    P_{\mathsf{T}} \equiv \overline{P_{\mathsf{T}}} \in \{ 0, 1 \} \quad (\mathrm{mod}\ 2). \label{aeq:QTRP}
\end{align}
This quantized time-reversal polarization is a one-dimensional $\mathbb{Z}_2$ topological invariant in class DIII.

\subsection{$\mathbb{Z}_2$ topological invariant in 2D class DIII}

Using the time-reversal polarization in Eq.~\eqref{aeq:QTRP}, we define a two-dimensional $\mathbb{Z}_2$ topological invariant in class DIII 
\begin{align}
   \nu'\coloneqq P_{\mathsf{T}}(\Gamma Y)-P_{\mathsf{T}}(XM) \in \{0,1 \} \quad (\mathrm{mod}\ 2), \label{aeq:2DZ2topoinv_DIII}
\end{align}
where $P_{\mathsf{T}}(\Gamma Y)$ and $P_{\mathsf{T}}(XM)$ are the time-reversal polarization along the $k_y$ direction between the high-symmetry points in the Brillouin zone, and the high-symmetry points are defined as $\Gamma=(0,0)$, $X=(\pi,0)$, $Y=(0,\pi)$, and $M=(\pi, \pi)$.
This topological invariant $\nu'$ detects the helical edge states along the $x$ direction protected by internal symmetries in class DIII. 

The symmetry eigenvalues $g_{\pm}(k_x)$ take the values $\pm i$ at $k_x=\pi$, and the symmetry operators satisfy $\mathsf{T}\mathsf{U}^{*}(k_x)=\mathsf{U}(-k_x)\mathsf{T}$.
Consequently, the time-reversal operator maps the $g_{\pm}(\pi)$ sector into itself. Thus, we can define the time-reversal polarization in each $g_{\pm}(\pi)$ sector at $k_x=\pi$.
Conversely, since the symmetry operators satisfy $\mathsf{C}\mathsf{U}^{*}(k_x)=\mathsf{U}(-k_x)\mathsf{C}$, the particle-hole operator exchanges the $g_{+}(\pi)$ and $g_{-}(\pi)$ sectors, yielding
\begin{align}
    P^{(+)}_{\mathsf{T}}(XM) \equiv \overline{P^{(-)}_{\mathsf{T}}}(XM)\quad ({\rm mod}\ 2), \label{aeq:PTp=barPTm}
\end{align}
where $P^{(\pm)}_{\mathsf{T}}(XM)$ $\left( \overline{P^{(\pm)}_{\mathsf{T}}}(XM)\right)$ is the time-reversal polarization associated with the $g_{\pm}(k_x)$ sectors for occupied (unoccupied) states along the $XM$ line.
In addition, the sum of the time-reversal polarizations $P^{(\pm)}_{\mathsf{T}}(XM)$ and $ \overline{P^{(\pm)}_{\mathsf{T}}}(XM)$ in each $g_{\pm}(k_x)$ sector is zero:
\begin{align}
        P^{(\pm)}_{\mathsf{T}}(XM) + \overline{P^{(\pm)}_{\mathsf{T}}}(XM) \equiv 0 \quad ({\rm mod}\ 2). \label{aeq:PTp+barPTm=0}
\end{align}
From Eqs.~\eqref{aeq:PTp=barPTm} and \eqref{aeq:PTp+barPTm=0}, we get
\begin{align}
    P_{\mathsf{T}}(XM)  \equiv P^{(+)}_{\mathsf{T}}(XM) + {P^{(-)}_{\mathsf{T}}}(XM)  \equiv P^{(+)}_{\mathsf{T}}(XM) - \overline{P^{(-)}_{\mathsf{T}}}(XM) 
     \equiv 0 \quad ({\rm mod}\ 2),
\end{align}
which shows that the one-dimensional $\mathbb{Z}_2$ topological invariant along the $XM$ line is trivial. Thus, in the presence of nonsymmorphic symmetry, the $\mathbb{Z}_2$ topological invariant $\nu'$ in  Eq.~\eqref{aeq:2DZ2topoinv_DIII} reduces to
\begin{align}
    \nu' \equiv P_{\mathsf{T}}(\Gamma Y) \quad ({\rm mod}\ 2). \label{aeq:nup=PT}
\end{align}

On the other hand, the symmetry eigenvalues $g_{\pm}(k_x)$ are $\pm 1$ at $k_x=0$, and the symmetry operators satisfy $\mathsf{T}\mathsf{U}^{*}(k_x)=-\mathsf{U}(-k_x)\mathsf{T}$. 
Therefore, the time-reversal operator exchanges the $g_{+}(0)$ and $g_{-}(0)$ sectors, leading to \begin{gather}
    P^{(+)}_{k_x=0} \equiv  -P^{(-)}_{k_x=0} \quad (\mathrm{mod}\ 2),
    \end{gather}
where $P^{(\pm)}_{k_x=0}$ is the polarization along the $y$ direction
\begin{align}
        P^{(\pm)}_{k_x=0} \coloneqq \frac{i}{2\pi} \int^{\pi}_{-\pi} dk_y {\rm tr}\mathcal{A}_{\pm}(0,k_y) \quad ({\mathrm{mod}}\ 2), 
\end{align}
in the $g_{\pm}$ sector at $k_x=0$. 
Thus, the time-reversal polarization reads
    \begin{gather}
    P_{\mathsf{T}}(\Gamma Y) \equiv 2P^{(+)}_{k_x=0}\quad (\mathrm{mod}\ 2). \label{aeq:polariz_+=polariz_-}
\end{gather}
Using Eqs.~\eqref{aeq:nup=PT} and \eqref{aeq:polariz_+=polariz_-}, we obtain
\begin{align}
    \nu' \equiv 2P^{(+)}_{k_x=0} \equiv \frac{i}{\pi} \int^{\pi}_{-\pi} dk_y {\rm tr}\mathcal{A}_+(0,k_y) \quad ({\rm mod}\ 2). \label{aeq:nup_mod2} 
\end{align}

\subsection{$\mathbb{Z}_4$ topological invariant in 2D class DIII + ${\sf U}_{-+}$}

The $\mathbb{Z}_4$ topological invariant is defined by~\cite{Shiozaki-PRB-2016}
\begin{align}
    \nu \coloneqq \frac{2i}{\pi}\int^{\pi}_{-\pi}dk_y {\rm tr}\mathcal{A}^{\mathrm{I}}_{+}(\pi,k_y)-\frac{i}{\pi}\int^{\pi}_{0}dk_x \int^{\pi}_{-\pi}dk_y{\rm tr}\mathcal{F}_{+}(k_x,k_y) \quad ({\rm mod}\ 4),
\end{align}
where $\mathcal{A}^{\mathrm{I,II}}_{+}(\pi,k_y)$ is the Berry connection of occupied states $\ket{u_n^{(+)\rm I,II}(\boldsymbol{k})}$ in the $g_{+}(k_x)$ sector, and $\mathcal{F}_{+}$ is the Berry curvature in the $g_{+}(k_x)$ sector. 
From Stokes' theorem, the second term of $\nu$ reads
\begin{align}
    -\frac{i}{\pi}\int^{\pi}_{0}dk_x\int^{\pi}_{-\pi}dk_y {\rm tr}\mathcal{F}_{+}(k_x,k_y) &\equiv -\frac{i}{\pi}\int^{\pi}_{-\pi}dk_y{\rm tr}\mathcal{A}_{+}(\pi,k_y)+\frac{i}{\pi}\int^{\pi}_{-\pi}dk_y {\rm tr}\mathcal{A}_+(0,k_y)\quad ({\rm mod}\ 2)\nonumber \\
    &\equiv - \frac{2i}{\pi}\int^{\pi}_{-\pi}dk_y {\rm tr}\mathcal{A}^{\rm I}_{+}(\pi,k_y)+\frac{i}{\pi}\int^{\pi}_{-\pi} dk_y {\rm tr}\mathcal{A}_{+}(0,k_y)\quad ({\rm mod}\ 2),
\end{align}
where we used $\mathrm{tr}\mathcal{A}_{+}(\pi,k_y)=\mathrm{tr}\mathcal{A}^{\mathrm{I}}_{+}(\pi,k_y)+\mathrm{tr}\mathcal{A}^{\mathrm{II}}_{+}(\pi,k_y)$ and $\mathrm{tr}\mathcal{A}^{\mathrm{I}}_{+}(\pi,k_y)=\mathrm{tr}\mathcal{A}^{\mathrm{II}}_{+}(\pi,-k_y)$.
Thus, $\nu$ reads
\begin{align}
    \nu= \frac{i}{\pi} \int^{\pi}_{-\pi} dk_y {\rm tr}\mathcal{A}_+(0,k_y) \quad ({\rm mod}\ 2). \label{aeq:nu_mod2}
\end{align}
From Eqs.~\eqref{aeq:nup_mod2} and \eqref{aeq:nu_mod2}, it follows that 
\begin{align}
    \nu \equiv \nu' \quad (\mathrm{mod}\ 2),
\end{align}
showing that the parity of the $\mathbb{Z}_4$ topological invariant $\nu$ reduces to the $\mathbb{Z}_2$ topological invariant $\nu'$.
Since $\nu'=1$ corresponds to the $\mathbb{Z}_2$ topological phase protected by internal symmetries, the non-Hermitian Hamiltonian corresponding to the Hermitized Hamiltonian with $\nu=1,3$ exhibits a point-gap topological phase protected by internal symmetry in class AII$^\dagger$, which is the conventional $\mathbb{Z}_2$ skin effect \cite{OKSS-20}.

\section{Hermitized Hamiltonian from 2D class AII$^{\dag}$ + $\tilde{\cal \M}_{-}$}\label{asec:Hermitized_Z4=2}

\subsection{Hermitized continuum model}

When a given non-Hermitian Hamiltonian $H(\boldsymbol{k})$ belongs to class AII$^\dagger$ + $\tilde{\mathcal{U}}_{-}$, the Hermitized Hamiltonian ${\sf H}(\boldsymbol{k})$ belongs to class DIII + ${\sf U}_{-+}$ (see Table~\ref{tab: real AZ dag 0}). 
In class DIII, the Hermitized Hamiltonian ${\sf H}(\boldsymbol{k})$ respects time-reversal symmetry in Eq.~\eqref{eq:DIII_TRS}, particle-hole symmetry in Eq.~\eqref{eq:DIII_PHS}, and chiral symmetry in Eq.~\eqref{eq:DIII_CS} with
\begin{align}
    {\sf T}= i\tau_x \sigma_y, \quad {\sf C} = \tau_y \sigma_y, \quad 
    {\sf \Gamma} = \tau_z.
\end{align}
Furthermore, the Hamiltonian ${\sf H(\boldsymbol{k})}$ respects nonsymmorphic symmetry ($d_{\parallel}=0$) in Eq.~\eqref{eq:Hermitized_NSG}
with 
\begin{align}\label{eq:NSG_rep_DIII}
{\sf U}(k_x)\coloneqq \begin{pmatrix}
        0 & \mathcal{U}(k_x)  \\
        \mathcal{V}(k_x) & 0  \\
    \end{pmatrix}_{\tau}, \quad 
    \mathcal{U}(k_x)\coloneqq\begin{pmatrix}
         i & 0 \\
         0 & -ie^{-ik_x} \\
    \end{pmatrix}_{\sigma}, \quad
    \mathcal{V}(k_x) \coloneqq
    \begin{pmatrix}
        -ie^{-ik_x} & 0  \\
        0 & i 
    \end{pmatrix}_{\sigma}.
\end{align}
Because of ${\sf U}^2(k_x)=e^{-ik_x}$, $\mathcal{U}(k_x)$ and $\mathcal{V}(k_x)$ satisfy
\begin{align}
    \mathcal{U}(k_x) =  e^{-ik_x}\mathcal{V}^{-1}(k_x).
\end{align} 
These symmetry representations obey the following commutation and anticommutation relations: 
\begin{align}
    {\sf T}{\sf U}^{*}(k_x) =-{\sf U}(-k_x){\sf T}, \quad {\sf C}{\sf U}^{*}(k_x) = {\sf U}(-k_x){\sf C}.
\end{align}
Therefore, the Hermitized Hamiltonian ${\sf H}(\boldsymbol{k})$ belongs to class DIII + ${\sf U}_{-+}$.

From the non-Hermitian Hamiltonian
\begin{equation}\label{aeq:AIIdagger_surface}
    H_{{\rm AII}^{\dagger}}
    (k_x) = \begin{pmatrix}
     ie^{-ik_x/2} g(k_x) & f(k_x) \\
     f^{*}(k_x) & ie^{ik_x/2} g(-k_x)
    \end{pmatrix},
\end{equation}
we obtain the following Hermitized Hamiltonian
\begin{align}\label{eq:surface_DIII}
    {\sf H}_{\rm DIII}(k_x)\coloneqq \begin{pmatrix}
        0 & H_{{\rm AII}^{\dagger}}(k_x) \\
        H_{{\rm AII}^{\dagger}}^\dagger(k_x) & 0 
    \end{pmatrix}_{\tau}.
\end{align}
When we choose $g(k_x)=v_1 \sin (k_x/2)$ and $f(k_x)=v_2 \sin k_x+iv_3 \sin k_x$
($v_1, v_2, v_3 \in \mathbb{R}$) in Eq.~(\ref{aeq:AIIdagger_surface}), the Hermitized Hamiltonian is written as
\begin{align}
    {\sf H}_{\rm DIII}(k_x) = v_1\sin^2(k_x/2) \tau_x + \frac{v_1}{2}\sin k_x \tau_y \sigma_z + v_2\sin k_x \tau_x \sigma_x+ v_3\sin k_x \tau_x \sigma_y.
\end{align}
This Hamiltonian exhibits gapless boundary states. In the simultaneous presence of the internal symmetries and pseudo-nonsymmorphic symmetry, no mass terms open the gap, showing the stability of the gapless boundary states.
We confirm that this Hamiltonian ${\sf H}_{\rm DIII}(k_x)$ is characterized by the $\mathbb{Z}_4$ topological invariant $\nu \equiv 2$ ($\mathrm{mod}~4$).
For that purpose, we consider its doubled Hamiltonian, 
\begin{align}
    {\sf {H}}_{d}(k_x)\coloneqq 
    {\sf H}_{\rm DIII}(k_x) \oplus {\sf H}_{\rm DIII}(k_x),
\end{align}
which can be gapped by adding symmetry-preserving mass terms, $m_1\mu_y \tau_y \sigma_x$ and $m_2\mu_y \tau_y \sigma_y$ ($m_1,m_2\in \mathbb{R}$).
Here, the symmetry operators are chosen as
\begin{align}
    {\sf {U}}_{d}(k_x)\coloneqq \mu_x \otimes {\sf U}(k_x),\quad {\sf {T}}_{d}\coloneqq \mu_0 \otimes {\sf T}, \quad {\sf {\Gamma}}_{d}\coloneqq \mu_0 \otimes {\sf \Gamma},
\end{align}
with the $2\times2$ identity matrix $\mu_0$ and the Pauli matrix $\mu_x$. 
Thus, this doubled Hamiltonian ${\sf {H}}_{d}(k_x)$ hosts the trivial topological invariant $\nu \equiv 0$ ($\mathrm{mod}~4$), further implying $\nu \equiv 2$ for ${\sf {H}}_{\rm DIII}(k_x)$.

\subsection{Hermitized lattice model}

We discuss a two-dimensional lattice model of Hermitian $\mathbb{Z}_4$ topological nonsymmorphic crystalline insulator in class DIII + ${\sf U}_{-+}$.
To begin with, we introduce a $\mathbb{Z}_2$ topological nonsymmorphic crystalline insulator in class D~\cite{Shiozaki-PRB-2016},
\begin{align}
    {\sf H}_{e_+ - e_-}(\boldsymbol{k})\coloneqq (-t_y\cos k_y-m) \sigma_z +g(k_x)\left[ 
    \cos (k_x/2) \tau_x +\sin(k_x/2) \tau_y \right] \sigma_x
    + \Delta \sin k_y \sigma_y,
\end{align}
with $g(-k_x)=-g(k_x)$ and $g(k_x+2\pi)=-g(k_x)$.
Particle-hole symmetry is given by
\begin{align}
       {\sf C}_{\rm D}{\sf H}^{T}_{e_+ - e_-}(\boldsymbol{k}){\sf C}_{\rm D}^{-1} = -{\sf H}_{ e_+ - e_-}(-\boldsymbol{k}), \quad {\sf C}_{\rm D}{\sf C}_{\rm D}^{*}=+1,\quad {\sf C}_{\rm D} =\sigma_x.
\end{align}
In addition, nonsymmorphic symmetry ($d_{\parallel}=0$) reads
\begin{align}\label{eq:nonsym_classD}
    {\sf U}_{\rm D}(k_x) {\sf H}_{e_+ - e_-}(\boldsymbol{k}){\sf U}_{\rm D}^{-1}(k_x) = {\sf H}_{e_+ - e_-}(\boldsymbol{k}), \quad {\sf U}_{\rm D}^2(k_x)=e^{-ik_x}, \quad 
    {\sf U}_{\rm D}(k_x)\coloneqq \begin{pmatrix}
        0 & e^{-ik_x} \\
        1 & 0
    \end{pmatrix}_\tau \otimes \sigma_0.
\end{align}

Doubling the model ${\sf H}_{e_+ - e_-}(\boldsymbol{k})$ in such a way that the doubled Hamiltonian respects time-reversal symmetry, particle-hole symmetry, and nonsymmorphic symmetry, we obtain a topological nonsymmorphic crystalline insulator with $\nu \equiv 2$ ($\mathrm{mod}~4$) in class DIII.
Adding a degree of freedom $\mu_i$ ($i=x,y,z$), we choose symmetry representations for class DIII, 
\begin{align}
    {\sf C}_{\rm DIII}=\mu_x  \tau_0\sigma_x,\quad {\sf T}_{\rm DIII} = \mu_y  \tau_z \sigma_z, \quad {\sf \Gamma}_{\rm DIII}=\mu_z \tau_z \sigma_y,
\end{align}
where ${\sf C}_{\rm DIII}$ and ${\sf T}_{\rm DIII}$ satisfy ${\sf C}_{\rm DIII}{\sf C}^{*}_{\rm DIII} = +1$ and ${\sf T}_{\rm DIII}{\sf T}^{*}_{\rm DIII} = -1$.
Further doubling the model ${\sf H}_{e_+ - e_-}(\boldsymbol{k})$ under these symmetries, we get the eight-band Hamiltonian:
\begin{align}\label{eq:DIII}
    {\sf H}_{\rm DIII}(\boldsymbol{k})\coloneqq &(-t_y\cos k_y-m) \mu_0 \tau_0 \sigma_z +g(k_x) \mu_x \left[ 
    \cos (k_x/2) \tau_x +\sin(k_x/2) \tau_y \right] \sigma_x
    + \Delta \sin k_y \mu_x \tau_0 \sigma_y \nonumber \\
    &+v\sin k_y \mu_z \tau_0 \sigma_z + t'\mu_x \left[ \frac{1+\cos k_x}{2}\tau_x + \frac{\sin k_x}{2}\tau_y \right]\sigma_z,
\end{align}
where the fourth and fifth terms are newly added symmetry-preserving terms. 
This Hamiltonian ${\sf H}_{\rm DIII}(\boldsymbol{k})$ respects time-reversal symmetry, particle-hole symmetry, and nonsymmorphic symmetry: 
\begin{gather}
        {\sf T}_{\rm DIII}{\sf H}^{*}_{\rm DIII}(\boldsymbol{k}){\sf T}_{\rm DIII}^{-1} = {\sf H}_{\rm DIII}(-\boldsymbol{k}), \quad {\sf T}_{\rm DIII}{\sf T}_{\rm DIII}^{*}=-1, \\ 
            {\sf C}_{\rm DIII}{\sf H}^{T}_{\rm DIII}(\boldsymbol{k}){\sf C}_{\rm DIII}^{-1} = -{\sf H}_{\rm DIII}(-\boldsymbol{k}), \quad {\sf C}_{\rm DIII}{\sf C}_{\rm DIII}^{*}=1, \\
            {\sf U}_{\rm DIII}(k_x) {\sf H}_{\rm DIII}(\boldsymbol{k}){\sf U}_{\rm DIII}^{-1}(k_x) = {\sf H}_{\rm DIII}(\boldsymbol{k}), \quad {\sf U}_{\rm DIII}^2(k_x)=e^{-ik_x}, \quad 
    {\sf U}_{\rm DIII}(k_x)\coloneqq \mu_0\otimes \begin{pmatrix}
        0 & e^{-ik_x} \\
        1 & 0
    \end{pmatrix}_\tau \otimes \sigma_0. \label{eq:dIII_nonsymm}
\end{gather}

\begin{figure}
\begin{center}
\includegraphics[width=1.\columnwidth]{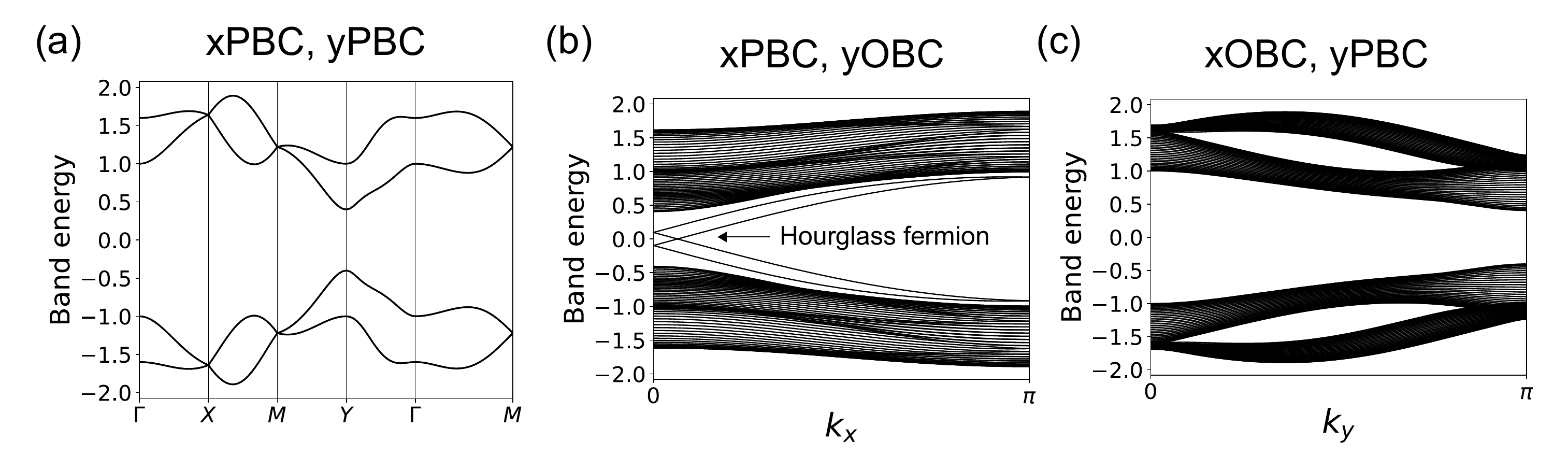}
\caption{Energy spectra of the Hamiltonian in Eq.~(\ref{eq:DIII}) under the (a)~periodic boundary conditions (PBC) in both $x$ and $y$ directions, (b)~PBC in the $x$ direction and open boundary conditions (OBC) in the $y$ direction, and (c)~OBC in the $x$ direction and PBC in the $y$ direction  [$g(k_x)=c \sin(k_x/2)$, $m=0.3$,  $t=\Delta=c=1$, $v=0.4$, $t'=0.3$, $L_i=30$ ($i=x,y$)].}\label{fig:hourglass}
\end{center}
\end{figure}

Figure~\ref{fig:hourglass}\,(a) shows the bulk band structure of the Hamiltonian ${\sf H}_{\rm DIII}(\boldsymbol{k})$ in Eq.~(\ref{eq:DIII}).
Under the open boundary conditions in the $y$ direction, the gapless boundary states appear [Fig.~\ref{fig:hourglass}\,(b)].
These boundary states have hourglass-like dispersions~\cite{Shiozaki-PRB-2016, Wang-16}, characteristic of the nonsymmorphic topological phases with $\nu \equiv 2$ ($\mathrm{mod}~4$). 
On the other hand, the gapless boundary states do not appear under the open boundary conditions in the $x$ direction [Fig.~\ref{fig:hourglass}\,(c)] because nonsymmorphic symmetry is broken owing to the boundary conditions. 

To transform ${\sf U}_{\rm DIII}(k_x)$ into a similar representation to Eq.~(\ref{eq:NSG_rep_DIII}), we consider the following unitary transformation:
\begin{align}\label{eq:DIII_2}
    {\sf H}'_{\rm DIII}(\boldsymbol{k})\coloneqq &V^{\dagger}(\boldsymbol{k}){\sf H}_{\rm DIII}(\boldsymbol{k})V(\boldsymbol{k})\nonumber \\
    =&(t_y\cos k_y+m) \mu_x \tau_x \sigma_x +g(k_x)  \left[ 
    \cos (k_x/2) \mu_y \tau_x \sigma_0 +\sin(k_x/2) \mu_x \tau_x \sigma_z \right] 
    + \Delta \sin k_y \mu_x \tau_0 \sigma_y \nonumber \\
    &+v\sin k_y \mu_x \tau_y \sigma_y - t'/2 \left[  (1+\cos k_x)\mu_y\tau_y \sigma_0 + \sin k_x\mu_x\tau_y \sigma_z \right],
\end{align}
with 
\begin{equation}
\begin{split}
V(\boldsymbol{k})\coloneqq&\begin{pmatrix}
    v_1, & v_2, &v_3, & v_4, & v_5, & v_6, & v_7, & v_8
    \end{pmatrix}, \\
   v_1^{T} \coloneqq&e^{-i{k_x}/2}/2\begin{pmatrix}
    1 - i, & 1 + i, & 0, & 0,& 0,& 0,& 0,& 0
     \end{pmatrix}, \\
    v_2^{T} \coloneqq&e^{ik_x/2}/2\begin{pmatrix}
        0, & 0, & 0, & 0, & 0, & 0, & -1+i,& -1 - i    
     \end{pmatrix}, \\
     v_3^{T} \coloneqq&e^{-ik_x/2}/2\begin{pmatrix}
        0,& 0,& 0,& 0,& 1 + i,& 1 - i,& 0,& 0
     \end{pmatrix}, \\
    v_4^{T} \coloneqq&e^{ik_x/2}/2\begin{pmatrix}
    0, & 0, & 1 + i, & 1 - i, & 0, & 0, & 0, & 0
     \end{pmatrix}, \\
     v_5^{T} \coloneqq&e^{ik_x/2}/2\begin{pmatrix}
    0, &0,& -1 - i, & 1 - i, & 0,& 0,& 0,& 0 
    \end{pmatrix}, \\
     v_6^{T} \coloneqq&e^{-ik_x/2}/2\begin{pmatrix}
    0,& 0,& 0,& 0,& -1-i,& 1 - i,& 0,& 0,
    \end{pmatrix}, \\ 
     v_7^{T} \coloneqq&e^{ik_x/2}/2\begin{pmatrix}
     0,& 0,& 0,& 0,& 0,& 0,& 1-i ,& -1 -i,
    \end{pmatrix}, \\
     v_8^{T} \coloneqq&e^{-ik_x/2}/2\begin{pmatrix}
    -1+i,& 1 + i,& 0,& 0,& 0,& 0,& 0,& 0
    \end{pmatrix}. 
\end{split}
\end{equation}
We calculate the tensor product $A\otimes B \otimes C$ in order of $A\otimes (B \otimes C)$, and the tensor product is defined by $A \otimes B =(A_{ij}B)_{ij}$.
Nonsymmorphic symmetry is also transformed into the following form via this unitary transformation:
\begin{gather}
    {\sf U}'_{\rm DIII}(k_x)\coloneqq V^{\dagger}(\boldsymbol{k}){\sf U}_{\rm DIII}(k_x)V(\boldsymbol{k})
    =\begin{pmatrix}
        0 & \mathcal{U}(\boldsymbol{k}) \\
        \mathcal{V}(\boldsymbol{k}) & 0
    \end{pmatrix}_{\mu}, \\
    \mathcal{U}(\boldsymbol{k})= \tau_0 \otimes \begin{pmatrix}
        -i & 0 \\
         0 & ie^{-ik_x}
    \end{pmatrix}_{\sigma}, \quad \mathcal{V}(\boldsymbol{k})= \tau_0 \otimes \begin{pmatrix}
        ie^{-ik_x} & 0 \\
         0 & -i
    \end{pmatrix}_{\sigma}.
\end{gather}
The Hamiltonian ${\sf H}'_{\rm DIII}(\boldsymbol{k})$ respects time-reversal symmetry, particle-hole symmetry, and chiral symmetry with
\begin{align}
    {\sf T}'= i\mu_x \tau_z \sigma_y, \quad {\sf C}'= \mu_y \tau_z \sigma_y, \quad {\sf \Gamma}'= \mu_z.
\end{align}

The off-diagonal part of the Hermitian Hamiltonian ${\sf H}_{\rm DIII}(\boldsymbol{k})$ in Eq.~(\ref{eq:DIII}) is a non-Hermitian matrix
\begin{align}
    {H}'_{{\rm AII}^{\dagger}}(\boldsymbol{k})\coloneqq &(t_y\cos k_y+m)  \tau_x \sigma_x +g(k_x)  \left[ 
    -i\cos (k_x/2) \tau_x \sigma_0 +\sin(k_x/2) \tau_x \sigma_z \right] 
    + \Delta \sin k_y \tau_0 \sigma_y \nonumber \\
    &+v\sin k_y \tau_y \sigma_y - (t'/2) \left[ -i(1+\cos k_x)\tau_y \sigma_0 + \sin k_x \tau_y \sigma_z \right],
\end{align}
which can be considered as
the non-Hermitian Hamiltonian exhibiting the point-gap topological phase with $\nu \equiv 2$ ($\mathrm{mod}~4$), as discussed in the main text.
Since this non-Hermitian Hamiltonian exhibits the $\mathbb{Z}_4$ topological phase even for $v=t'=0$, we set $v=t'=0$ for simplicity. 
In addition, we add a symmetry-preserving term $t_x\cos k_x\sigma_x \tau_x$ to the non-Hermitian Hamiltonian. 
In this manner, we obtain the model in the main text. 

\section{Systematic model construction in class AIII + $\mathcal{U}_{+}$}

We provide a general method to construct non-Hermitian models in class AIII + $\mathcal{U}_{+}$ from Hermitian models in class A + $\mathsf{\M}$. 
We first apply the Hermitization mapping in Eq.~(\ref{aeq: Hermitization}) to non-Hermitian point-gapped Hamiltonians $H \left( {\bm k} \right)$ in class AIII + $\mathcal{U}_{+}$,
\begin{align}
    &{\mathcal{U} \left( k_1 \right)}\,{H} \left( {\bm k} \right) \,{\mathcal{V}^{-1} \left( k_1 \right)} 
    = {H} \left( \sigma{\bm k} \right)
    ,\quad \left[{\mathcal{U} \left( k_1 \right)}\right]^2 = \left[{\mathcal{V} \left( k_1 \right)}\right]^2 = e^{-ik_1},\\
    &{\Gamma}\,{H}^{\dagger} \left( {\bm k} \right) \,{\Gamma}^{-1} 
    = - {H} \left( {\bm k} \right)
    ,\quad \Gamma^2 = 1,\\
    &{\Gamma}\,{\mathcal{U} \left( k_1 \right)} = {\mathcal{V} \left( k_1 \right)}\,{\Gamma}.
\end{align}
The resultant Hermitian gapped Hamiltonian ${\sf H} \left( {\bm k} \right)$ respects both chiral and nonsymmorphic symmetries,
\begin{align}
    &{\sf \Gamma}\,{\sf H} \left( {\bm k} \right) {\sf \Gamma}^{-1} 
    = - {\sf H} \left( {\bm k} \right),\quad
    {\sf \Gamma} \coloneqq 
    \begin{pmatrix}
        0 & \Gamma \\
	\Gamma & 0 \\
    \end{pmatrix},\\
    &{\sf \M} \left( k_1 \right)\,{\sf H} \left( {\bm k} \right) {\sf \M^{-1}}  \left( k_1 \right)
    = {\sf H} \left( \sigma{\bm k} \right),\quad
    {\sf \M} \left( k_1 \right) \coloneqq 
    \begin{pmatrix}
        {\mathcal{U} \left( k_1 \right)} & 0 \\
	0 & {\mathcal{V} \left( k_1 \right)} \\
    \end{pmatrix},\quad \left[{\sf \M} \left( k_1 \right)\right]^2 = e^{-ik_1},
\end{align}
as well as additional chiral symmetry,
\begin{equation}
    {\Sigma}\,{\sf H} \left( {\bm k} \right) {\Sigma}^{-1} 
    = - {\sf H} \left( {\bm k} \right),\quad
    {\Sigma} \coloneqq 
    \begin{pmatrix}
        1 & 0 \\
	0 & -1 \\
    \end{pmatrix}.
\end{equation}
Here, the symmetry operators satisfy the following relations:
\begin{equation}
    \left[ {\Gamma},\, {\sf \M} \left( k_1 \right) \right] = \left[ {\Sigma},\, {\sf \M} \left( k_1 \right) \right] = \left\{ {\sf \Gamma},\, {\Sigma} \right\} = 0.
\end{equation}

The presence of two chiral symmetries ${\sf \Gamma}$ and ${\Sigma}$ enables us to construct a unitary matrix $i {\sf \Gamma} {\Sigma}$ commuting with both ${\sf H} \left( {\bm k} \right)$ and ${\sf \M} \left( k_1 \right)$,
\begin{equation}
\left[ {\sf H} \left( {\bm k} \right),\,i {\sf \Gamma} {\Sigma} \right] = \left[ {\sf \M} \left( k_1 \right),\,i {\sf \Gamma} {\Sigma} \right] = 0.
\end{equation}
Then, the three matrices ${\sf H} \left( {\bm k} \right), {\sf \M} \left( k_1 \right)$, and $i {\sf \Gamma} {\Sigma}$ can be simultaneously diagonalized by a unitary matrix ${\sf Q}$, leading to
\begin{align}
    {\sf Q}^{\dagger} {\sf H} \left( {\bm k} \right) {\sf Q} &= 
    \begin{pmatrix}
        i H \left( {\bm k} \right) \Gamma & 0 \\
	   0 & -i H \left( {\bm k} \right) \Gamma \\
    \end{pmatrix}, \\
    {\sf Q}^{\dagger} {\sf \M} \left( k_1 \right) {\sf Q} &= 
    \begin{pmatrix}
        {\mathcal{U} \left( k_1 \right)} & 0 \\
	   0 & {\mathcal{U} \left( k_1 \right)} \\
    \end{pmatrix}, \\
    {\sf Q}^{\dagger} \left( i {\sf \Gamma} {\Sigma} \right) {\sf Q} &= \begin{pmatrix}
        1 & 0 \\
	   0 & -1 \\
    \end{pmatrix}; \\
    \quad {\sf Q} &\coloneqq \frac{1}{\sqrt{2}} \begin{pmatrix}
        1 & -i \\
	i \Gamma & -\Gamma \\
    \end{pmatrix}.
\end{align}
Consequently, the topological properties of non-Hermitian Hamiltonians $H \left( {\bm k} \right)$ are characterized by the Hermitian matrices $i H \left( {\bm k} \right) \Gamma$ with nonsymmorphic symmetry ${\mathcal{U} \left( k_1 \right)}$. 

The above unitary transformation further gives a general one-to-one correspondence between a non-Hermitian Hamiltonian $h_{{\rm AIII} + \mathcal{U}_{+}} \left( {\bm k} \right)$ in class AIII + $\mathcal{U}_{+}$ and a Hermitian Hamiltonian ${\sf h}_{{\rm A} + {\sf \M}} \left( {\bm k} \right)$ in class A + \M. 
From
\begin{align}
    {{\sf u} \left( k_1 \right)}\,{\sf h}_{{\rm A} + {\sf \M}} \left( {\bm k} \right) \,{{\sf u}^{-1} \left( k_1 \right)} 
    = {\sf h}_{{\rm A} + {\sf \M}} \left( \sigma{\bm k} \right)
    ,\quad \left[{{\sf u} \left( k_1 \right)}\right]^2 = e^{-ik_1},
\end{align}
one can systematically construct $h_{{\rm AIII} + \mathcal{U}_{+}} \left( {\bm k} \right)$ by $h_{{\rm AIII} + \mathcal{U}_{+}} \left( {\bm k} \right) \coloneqq -i {\sf h}_{{\rm A} + {\sf \M}} \left( {\bm k} \right)\gamma$ with a unitary matrix $\gamma$ satisfying $\gamma^2=1$. 
The obtained non-Hermitian Hamiltonian indeed meets the symmetry conditions for class AIII + $\mathcal{U}_{+}$:
\begin{align}
    &{u \left( k_1 \right)}\,h_{{\rm AIII} + \mathcal{U}_{+}} \left( {\bm k} \right) \,{v^{-1} \left( k_1 \right)} 
    = h_{{\rm AIII} + \mathcal{U}_{+}} \left( \sigma{\bm k} \right)
    ,\quad \left[{u \left( k_1 \right)}\right]^2 = \left[{v \left( k_1 \right)}\right]^2 = e^{-ik_1},\\
    &{\gamma}\,h_{{\rm AIII} + \mathcal{U}_{+}}^{\dagger} \left( {\bm k} \right) \,{\gamma}^{-1} 
    = - h_{{\rm AIII} + \mathcal{U}_{+}} \left( {\bm k} \right)
    ,\quad \gamma^2 = 1,\\
    &{\gamma}\,{u \left( k_1 \right)} = {v \left( k_1 \right)}\,{\gamma},
\end{align}
where ${u \left( k_1 \right)} \coloneqq {{\sf u} \left( k_1 \right)}$ and ${v \left( k_1 \right)} \coloneqq \gamma {{\sf u} \left( k_1 \right)} \gamma$.


\end{document}